%% file: main_aas.tex
\documentclass[twocolumn,twocolappendix]{aastex63}

\graphicspath{{./}{figures/}}
\usepackage{color}
\usepackage{soul}
\usepackage{natbib}
\usepackage{hyperref}
\usepackage{amsmath}
\usepackage{xspace}
\usepackage{graphicx}
\usepackage{enumitem}
\usepackage{amssymb}
\usepackage{xifthen}
\usepackage{hyperref}
\usepackage[normalem]{ulem}
\usepackage{comment}
\usepackage{float}
\usepackage{CJKutf8}
\usepackage{bbm}
\usepackage{supertabular}

\hypersetup{    
  colorlinks      = {true},
  linkcolor       = {blue},
  citecolor       = {blue},
  urlcolor        = {blue},
}

\newcommand{\code}[1]{\texttt{#1}\xspace}

\newcommand{\teff}{\ensuremath{T_\mathrm{eff}}\xspace}

\usepackage{graphicx,pifont}
\let\oldding\ding
\renewcommand{\ding}[2][1]{\scalebox{#1}{\oldding{#2}}}

\usepackage{array}
\newcolumntype{H}{>{\setbox0=\hbox\bgroup}c<{\egroup}@{}}

\shorttitle{The DESI Y1 RR Lyrae catalog}
\shortauthors{Medina, Li, Koposov, et al.}

\begin{document}

\title{The DESI Y1 RR Lyrae catalog I:\\
Empirical modeling of the cyclic variation of spectroscopic properties  \\ and a chemodynamical analysis of the outer halo
}

\input{authors.tex}


\begin{abstract}

We present the catalog of RR Lyrae stars (RRLs) observed in the first year of operations of the Dark Energy Spectroscopic Instrument (DESI) survey.
This catalog contains 6,240 RRLs out to $\sim120$\,kpc from the Galactic center and over 12,000 individual epochs with homogeneously-derived stellar atmospheric parameters. 
We introduce a novel methodology to model the cyclical variation of the spectroscopic properties of RRLs from single-epoch measurements. 
We employ this method to infer the systemic velocities and mean temperatures of fundamental and first-overtone mode RRLs in our sample (without distinguishing between individual spectral lines).
For fundamental mode pulsators, we obtain radial velocity curves with amplitudes of $\sim$30--80\,km\,s$^{-1}$ and effective temperature curves with 300--1,000\,K variations, whereas for first-overtone pulsators these amplitudes are $\sim20$\,km\,s$^{-1}$ and $\sim 600$\,K, respectively. 
We use our sample to study the metallicity distribution of the halo and its dependence on Galactocentric distance ($R_{\rm GC}$). Using a radius-dependent mixture model, we split the data into chemodynamically distinct components and find that our inner halo sample ($R_{\rm GC}\lesssim50$\,kpc) is predominantly composed of stars with [Fe/H] $\sim-1.5$ and largely radial orbits (with an anisotropy parameter $\beta\sim0.94$), that we associate with the Gaia-Sausage-Enceladus merger. 
Stars in the halo field exhibit a broader and more metal-poor [Fe/H] distribution with more circular orbits ($\beta\sim0.39$). 
The metallicity gradient of the metal-rich and the metal-poor components is found to be $0.005$ and $0.010$\,dex\,kpc$^{-1}$, respectively. 
Our catalog highlights DESI's tantalizing potential for studying the Milky Way and the pulsation properties of RRLs in the era of large spectroscopic surveys. 

\end{abstract}






\section{Introduction} \label{sec:intro}

RR Lyrae stars (RRLs, or RRL for single stars) are low-mass ($<1$\,M$_\odot$), short-period ($<1$\,d) pulsating variable stars that lie in the intersection between the horizontal branch (HB) and the instability strip in the Hertzsprung-Russell diagram \citep[see e.g.,][]{Walker1989,Catelan2015}.
Because of their ubiquity in old and metal-poor systems, RRLs are important tracers of the formation of the Galactic halo, globular clusters, and dwarf galaxies, providing unaltered insight on their early conditions and evolution \citep[e.g.,][and references therein]{Monelli2022}. 
Moreover, RRLs are excellent distance indicators (with uncertainties $<5$\,\%) due to their well-defined period-luminosity relation towards the infrared \citep[see e.g.,][]{Longmore1986,Bono2001,Catelan2004,Muraveva2015,Marconi2022,Mullen2023, Prudil2024a} and luminosity-metallicity relation in optical bands \citep[see e.g.,][]{Sandage1990,Caputo2000,Garofalo2022}.
This, together with their intrinsic brightness and their easily identifiable pulsation properties, makes them ideal targets to map the Galactic halo, where estimating distances to other stellar tracers is challenging (especially beyond $\sim20$\,kpc) and where they are found in large numbers 
\citep[e.g.,][]{Vivas2006,Mateu2018,Medina2018,Martinez-Vazquez2019,Vivas2020,Prudil2021,Cook2022,CabreraGarcia2024,Medina2024}.

Based on their modes of pulsation (and consequently, their periods, and light curve shapes), RRL variables are classified into three main groups: ab, c, and d-type RRLs (RRab, RRc, and RRd, respectively).
The first group (RRab) is the most numerous and corresponds to stars that pulsate in the radial fundamental mode and are characterized by their saw-tooth light curve shapes, and by the anticorrelation between their periods ($\gtrsim0.40$\,d) and pulsation amplitudes ($\sim0.3-1.5$\,mag in the $V-$band). 
Stars in the second group (RRc) pulsate in the radial first overtone and exhibit more sinusoidal light curves, with shorter periods than RRab ($\sim0.2$--$0.45$\,d) and smaller amplitudes ($\lesssim0.7$\,mag in $V$).
The less common third class (RRd) comprises stars in which the radial fundamental and the first-overtone mode are simultaneously excited (the latter being often the dominant mode), with relatively low amplitudes and roughly sinusoidal light curves with significant scatter around the mean curve \citep[see e.g.,][]{Jerzykiewicz1977,Nemec2024}.  

To take maximum advantage of RRLs as tracers of the outer halo, obtaining complete 6D phase-space information (positions and velocities) along with chemical abundances (or at least [Fe/H]) is imperative. 
Coupled with stellar atmosphere parameters, these data are critical for constraining the physics of RRL pulsation, as they can be used to investigate correlations between periods and amplitudes of RRLs, the dependence of the horizontal-branch morphology on [Fe/H] \citep[][]{Fabrizio2019,Savino2020}, 
the conditions of their formation \citep[e.g.,][]{Bono1997a,Kervella2019,Prudil2019,Bobrick2024}, 
the width of the RRL instability strip \citep[e.g.,][]{Preston2006,For2010,Marconi2015}, and the influence of metallicity on long-term period and amplitude modulations \citep[][]{Blazhko1907,Li2023}.
However, deriving the information from spectroscopic measurements is often challenging (particularly for faint, distant RRLs), hampered by the intrinsic short-timescale variation of their pulsation. 
In fact, conducting spectroscopic campaigns on RRLs must consider the variation of the aforementioned properties throughout the RRLs' pulsation cycle,  which restricts the ability to co-add multiple exposures to increase the signal-to-noise ratio due to the smearing of spectral features.
Moreover, while individual efforts are valuable for investigating small samples, conducting spectroscopic follow-ups, and identifying specific case studies
\citep[e.g.,][]{Kolenberg2010,For2011,Hansen2011,Nemec2013,Medina2023}, large catalogs of RRLs with spectroscopic measurements are often required in order to study the properties of RRLs as a population with high statistical significance.

A direct application of large RRL catalogs with chemodynamical information concerns the study of the metallicity and velocity distribution of the halo. 
Generally speaking, the metallicity distribution of halo stars and, in particular, the radial metallicity gradient of galaxies provide significant insights into the processes that govern the  evolution of galaxies. 
Indeed, it is expected that, as a consequence of the hierarchical formation of galaxies, massive (metal-rich) satellites sink into the host galaxy more efficiently than less massive (more metal-poor) satellites, resulting in a decrease in metallicity with increasing distance from the host \citep[see e.g.,][]{Larson1976,Tinsley1980}. 
Over the last decade, ample evidence of the presence of gradients in the metallicity distribution has been collected from simulations \citep[e.g.,][]{Tissera2014,Starkenburg2017,Buck2023,Cardona-Barrero2023,Buder2024} 
and observationally in both the Milky Way and external galaxies
\citep[e.g.,][] {Hayden2015,Anders2017,Dietz2020,Martinez-Vazquez2021,Taibi2022,Fu2024}.
Moreover, observations of Milky Way stars and simulations of Milky Way-like galaxies have shown that the velocity anisotropy of the halo can vary as a function of distance from the Galactic center \citep[e.g.,][]{Deason2013b,Cunningham2016,Bird2021,Iorio2021,Liu2022,Deason2024,He2024}.
Indeed, trends and features in the velocity anisotropy profile provide valuable insight on the accretion history of the Galaxy and the radialization of satellite orbits in the accretion process \citep[see e.g.,][]{Bird2015,Loebman2018}.
For instance, numerical simulations generally predict an increase of the velocity anisotropy with distance, from more isotropic near the Galactic center to more radial out to the virial radius \citep[see e.g.,][]{Sales2007,Rashkov2013,Emami2022,He2024,Mondal2024}. 
This, in turn, reflects the effects of dynamical friction on the accretion of multiple low-mass progenitors over long timescales \citep[see e.g,][]{Abadi2006,Debattista2008,Rashkov2013} and the orbital radialization of massive satellites \citep[e.g.,][]{Amorisco2017,Nipoti2017,Vasiliev2022}. 
Given their status as old and metal-poor halo tracers and the fact that they can be characterized out to large radii with precisely-derived distances \citep[e.g.,][]{Sesar2017,Stringer2021,Feng2024,Medina2024}, the existence of large catalogs of RRLs with homogeneously-derived spectroscopic properties are then also key for investigating the chemodynamics of the halo.

A significant number of medium to large-scale surveys with high potential for the spectroscopic study of RRLs have been conducted over the past few years. 
Examples of such surveys include 
the Radial Velocity Experiment \citep[RAVE;][]{Steinmetz2003,Steinmetz2020}, 
the Bulge Radial Velocity Assay \citep[BRAVA;][]{Rich2007}, through its Bulge RR Lyrae Radial Velocity Assay \citep[BRAVA-RR;][]{Kunder2020}, 
the Sloan Extension for Galactic Understanding and Exploration \citep[SEGUE;][]{Yanny2009}
the Apache Point Observatory Galactic Evolution Experiment \citep[APOGEE;][]{Majewski2017,Beaton2021}, 
the GALactic Archaeology with HERMES \citep[GALAH;][]{DeSilva2015,Buder2021},
and the Large sky Area Multi-Object fiber Spectroscopic Telescope \citep[LAMOST;][]{Cui2012,Zhao2012} survey. 
When combined with large photometric surveys and recent data releases from the {\it Gaia} mission \citep[][]{Gaia2016,Gaia2023}, these surveys have enabled the investigations of RRLs across different regions of the Galaxy in exquisite detail. 
Recent examples of these efforts are numerous. 
\citet{Ablimit2022}, for instance, used a combination of thousands of LAMOST, GALAH, APOGEE, and RAVE RRL spectra to study the assembly history of the Galaxy. 
A similar study was performed by \citet{Liu2022} to investigate the metallicity distribution of RRLs in the halo, combining LAMOST and SEGUE spectra. 
Similarly, single-epoch low and medium resolution spectra from the latest data release of the LAMOST survey, encompassing $\sim11,000$ individual RRL variables, were used by \citet{Wang2024} to determine their stellar atmospheric parameters, which can be used to shed new light on the study of stellar evolution and pulsation, as well as the structure of the Milky Way.
More recently, \citet{Prudil2024b} employed APOGEE spectra to develop model templates of radial velocity variations in RRLs within the Solar neighborhood, facilitating further studies of RRLs in other Galactic regions. 

The Dark Energy Spectroscopic Instrument (DESI) is one such large-scale spectroscopic survey, primarily designed to map the large-scale structure of the universe throughout its 5-year duration \citep[][]{DESI2016a,DESI2016b,DESI2022}.
In the last couple of years, DESI has successfully obtained medium-resolution (R$\sim$2,000-5,000) spectra of millions of sources in the northern hemisphere, consisting of mostly galaxies and quasars, but also Milky Way stars \citep[][]{Cooper2023}. 
The rich and homogeneous dataset provided by DESI offers a unique opportunity for the full characterization of thousands of RRLs, an opportunity that has not been explored in sufficient detail yet.

In this paper, we employ data from DESI's first year of operations (Y1) to build a catalog of RRLs with homogeneously-derived spectroscopic properties.
In Section~\ref{sec:data}, we present the dataset the DESI Y1 RRL sample is built upon, including RRL-relevant information of the DESI Milky Way Survey. 
Section~\ref{sec:analysis} provides an overview of the spectroscopic pipelines used to perform the processing of the DESI database, and details the methodology developed to derive the spectroscopic properties of RRLs accounting for their pulsations. 
We discuss the validation of such properties in Section~\ref{sec:validation}. 
In Section~\ref{sec:haloChemodyn}, as an example of usage of the DESI Y1 RRL catalog and to show the robustness of our data, we perform a study of the metallicity distribution, metallicity gradient, and velocity anisotropy of the halo out to large radii.
Lastly, we summarize the content of this paper and the main properties of the DESI Y1 RRL catalog in Section~\ref{sec:conclusions}.

\begin{figure}
\begin{center}
\includegraphics[angle=0,scale=.56]{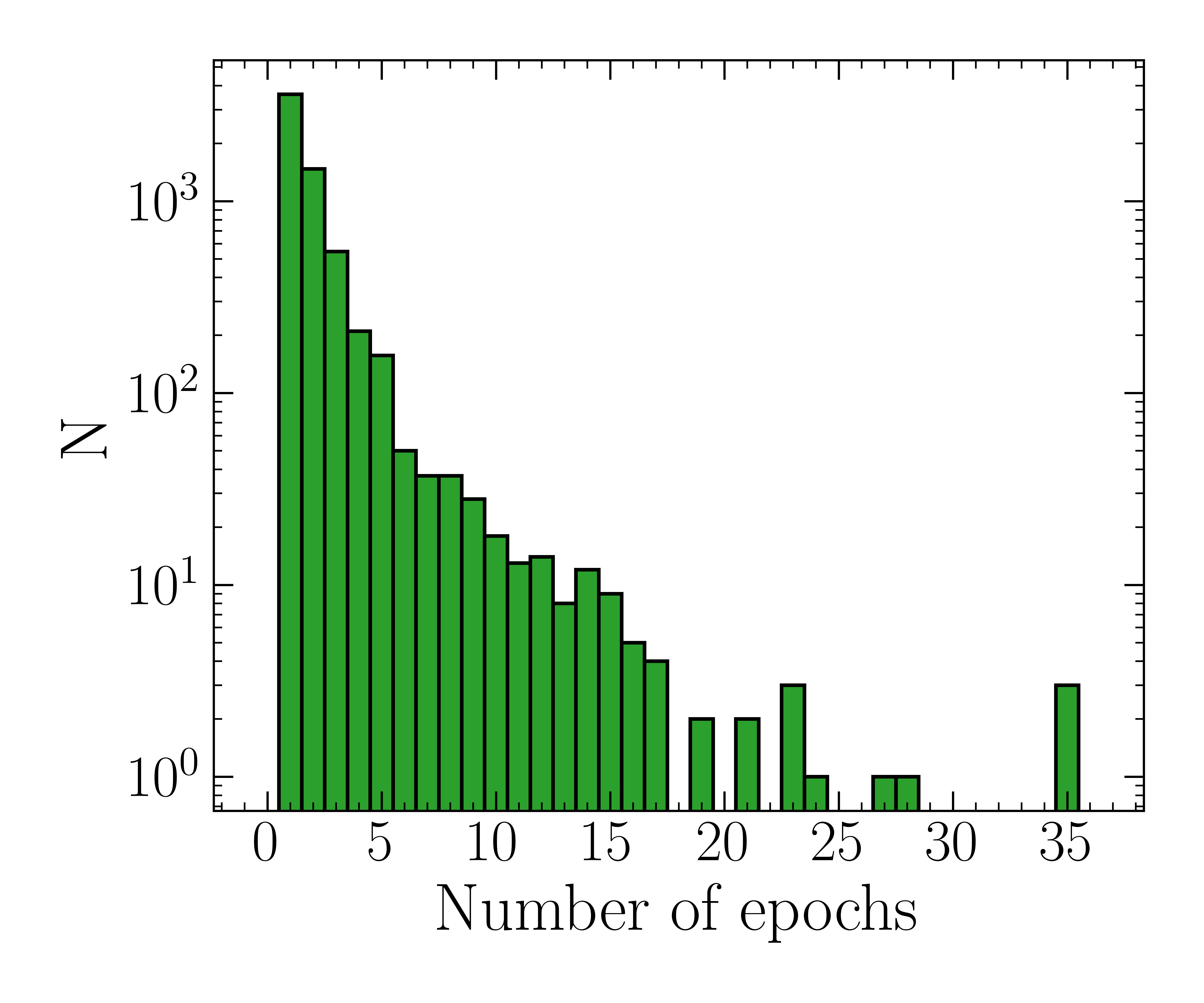}
\caption{
Number of spectroscopic epochs of the RRLs in the DESI Y1 catalog.
}
\label{fig:nepochs}
\end{center}
\end{figure}

\section{DESI data}
\label{sec:data}

The Dark Energy Spectroscopic Instrument \citep[DESI][]{DESI2016a,DESI2016b,DESI2022} is a multi-object spectroscopic instrument that deploys 5,000 fibers over a 3.2\,deg diameter field of view, mounted on the Mayall 4-m telescope at Kitt Peak National Observatory (KPNO). 
These fibers feed into 10 contiguous spectrographs (petals), each consisting of a blue arm (3,600-5,550\,\AA, with a resolution $R$ between 2,000 and 3,200), a red arm (5,550-6,560\,\AA, with $R$ between 3,200 and 4,100), and a near-infrared arm (6,560-9,800\,\AA, with $R$ between 4,100 and 5,000). 
The DESI processing and reduction pipeline \citep[][]{Guy2023} handles both the wavelength calibration and flux calibration of the acquired data.

The DESI survey follows a structure based on surveys and programs. 
The surveys are defined as the set of observations designed for the instrument's commissioning (CMX), three stages of science verification (SV), and the 5-year main survey conducted after completion of the SV phase.
The programs, defined as {\sc dark}, {\sc bright}, and {\sc backup}, correspond to the observing conditions in which the surveys are carried out. 
More specifically, they are selected using an effective time metric computed from the seeing, sky background, transparency, and airmass at the time of the observations \citep[i.e., the survey speed; see][]{Schlafly2023}.
A fourth program, {\sc other}, was used for the CMX and SV1 surveys. 

Although DESI is primarily designed to study the large-scale structure of the universe and to measure its expansion history using galaxies as targets \citep[see e.g.,][]{Levi2013}, about 7 million stars in the Galactic thick disk and halo have been observed as part of its Milky Way Survey (DESI-MWS).
This survey operates predominantly in bright-sky conditions and targets stars with magnitudes $16 < r < 19$. 
However, a significant number of stars (including RRLs) were observed in the {\sc dark} and {\sc other} programs. 
RR Lyrae variables are a prioritized target class for DESI-MWS. 
The target list of the DESI Y1 database includes sources with Galactic latitudes $|b|>20$\,deg classified as RRLs in the {\it Gaia} Data Release 2 (DR2) catalog, based on {\it Gaia}'s general variability classifier and the Special Object Studies (SOS) classifier \citep[][]{Holl2018,Clementini2019}, as well as faint RRLs in the Pan-STARRS1 catalog \citep[][]{Sesar2017}.
For a comprehensive description of the targeting strategy of DESI-MWS, we refer the reader to the overview provided by \citet{Cooper2023} and \citet{Koposov2024}. 
The DESI Y1 spectroscopic data can be accessed through the first public release of the DESI project \citep[DR1;][]{DESI_DR1_2025}, and the value added catalogs associated to DESI-MWS are presented in a separate paper \citep[][]{Koposov2025}.
The main data products presented in these catalogs, resulting from the processing pipelines of DESI-MWS, are detailed in Section~\ref{sec:obsProperties}.

\subsection{The DESI RR Lyrae catalog}

\begin{figure*}
        \centering    
        \includegraphics[width=0.66\textwidth]{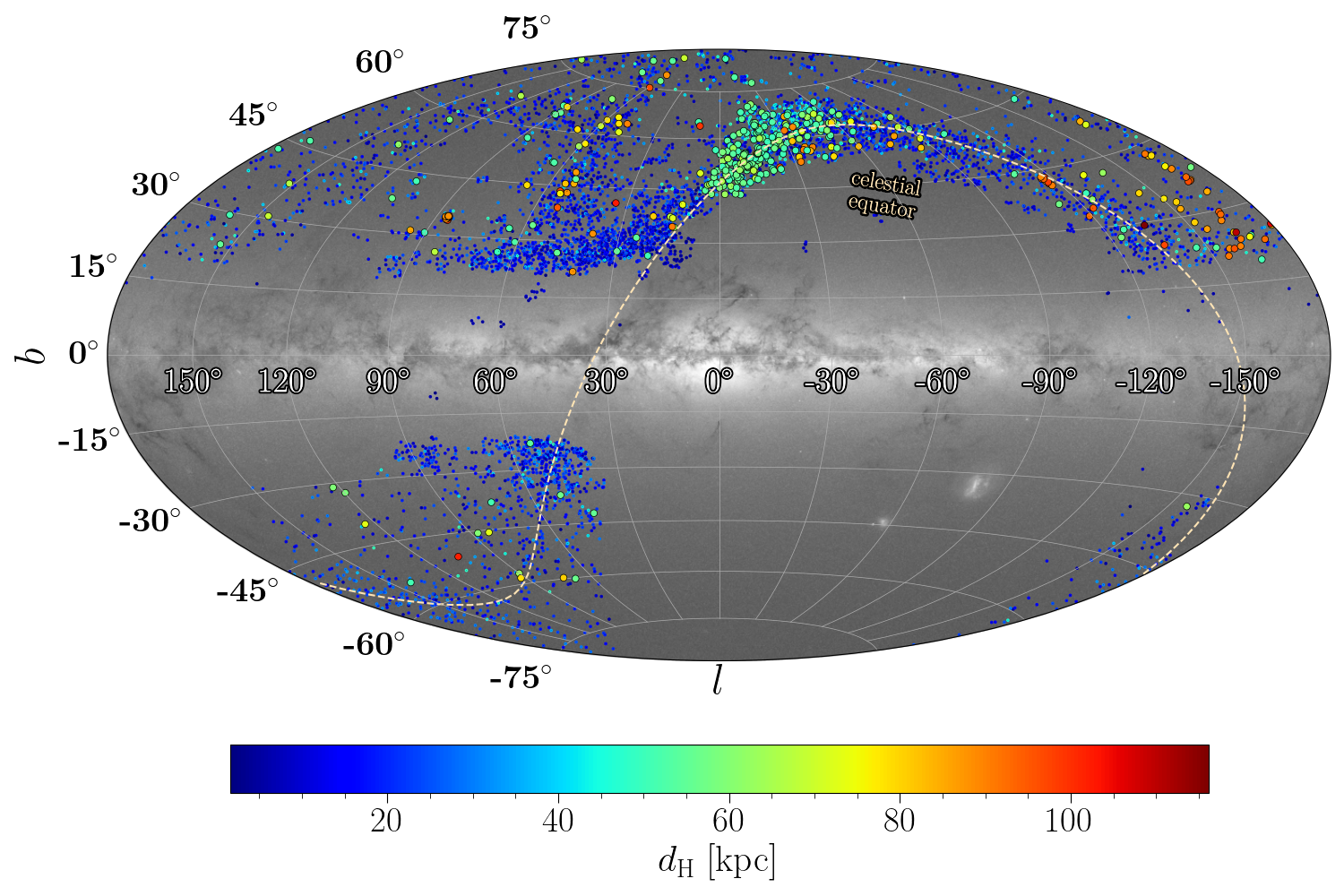}
        \includegraphics[width=0.30\textwidth]{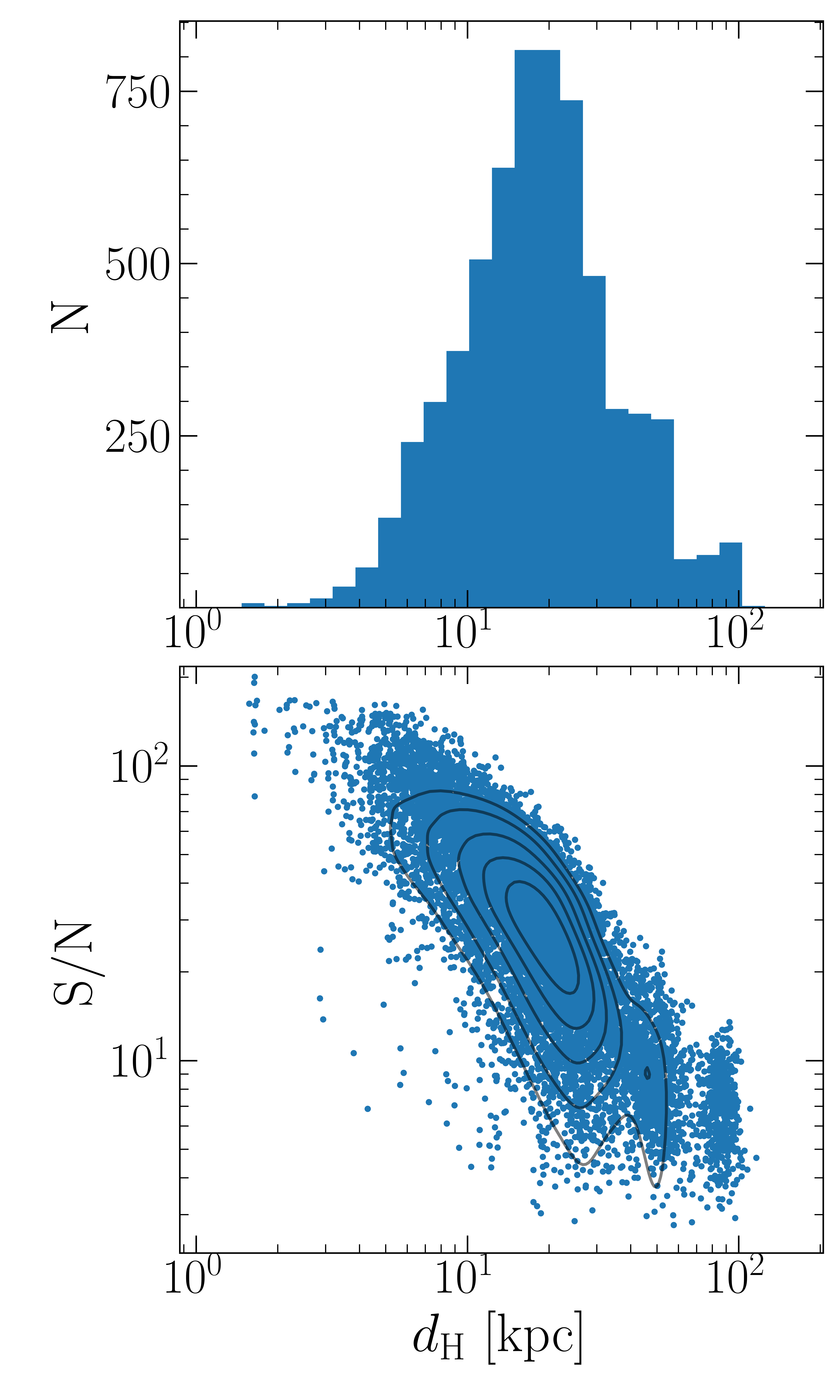}
    \caption{{\it Left}: Spatial distribution in Galactic coordinates of the DESI Y1 RRLs, color-coded by heliocentric distance ($d_{\rm H}$).
    The distances are computed following the methodology described in Section~\ref{sec:dist}. 
    The {\it Gaia} all-sky map is shown in the background as a reference. In this plot, the Sgr stream is visible as an extended overdensity at $d_{\rm H}\sim50$--$70$\,kpc, near the celestial equator. \textit{Background image credit: Gaia Data Processing and Analysis Consortium (DPAC); A. Moitinho / A. F. Silva / M. Barros / C. Barata, University of Lisbon, Portugal; H. Savietto, Fork Research, Portugal.} 
    {\it Top right}: Histogram of heliocentric distances of the DESI Y1 RRL catalog. This figure shows the natural decline of the number of RRLs observed as a function of distance. 
    {\it Bottom right}: Mean signal-to-noise ratio (S/N) of the DESI Y1 RRLs averaged from the three arms of DESI as a function of heliocentric distance. 
    The gap in the spatial distribution of RRLs at $d_{\rm H}\sim80$\,kpc is the result of our sample containing targets from DESI's {\sc bright} ($\lessapprox80$\,kpc) and {\sc dark} ($\gtrapprox80$\,kpc) programs. }
    \label{fig:map_and_histograms}
\end{figure*}

For this work, we considered single epoch observations from the main program and the science verification catalogs, observed as part of DESI Y1's {\sc bright}, {\sc dark}, and {\sc backup} programs. 
With this, the DESI RRL catalog contains stars down to $G\sim21$\,mag. 
These data are part of DESI's DR1 catalog and contains 9,760,208 single-epoch spectra of 4,873,914 sources with unique {\it Gaia} {\sc source\_id}. 
We perform basic cuts based on the spectroscopic properties of the DESI spectra (further described in Section~\ref{sec:analysis}) to ensure the quality of the final catalog. 
For this, only single-epoch observations with uncertainties in radial velocity smaller than 10\,km\,s$^{-1}$ and $\rm{[Fe/H]}>-4.0$\,dex are considered, and those for which no warnings are produced by DESI's processing pipeline RVSpecFit \citep[as indicated by the {\tt SUCCESS} flag set to True;][]{Koposov2019}.
This reduced the size of the catalog of usable exposures to 7,266,789. 

To construct the DESI Y1 RRL sample, we crossmatched the aforementioned filtered catalog with the recent {\it Gaia} DR3 RRL catalog \citep{Clementini2023}, which contains a total of 271,779 RRLs, using a crossmatch radius of 0.5\,arcsec. 
The result of this crossmatch is a catalog comprised of 12,301 epochs of 6,240 individual RRLs.
This makes the DESI Y1 RRL catalog one of the largest homogeneous spectroscopic RRL catalogs existing to date. 
Figure~\ref{fig:nepochs} depicts the number of spectroscopic epochs of the RRLs in our catalog.   
The number of epochs per star ranges from 1 to 35, and the bulk of our sample consists of stars observed only once ($\sim58$\%) or twice ($\sim24$\%).
For $\sim80$ RRLs DESI Y1 provides over 10 epochs, and 11 stars DESI possess over 20 epochs.

\begin{figure}
\begin{center}
\includegraphics[angle=0,scale=.32]{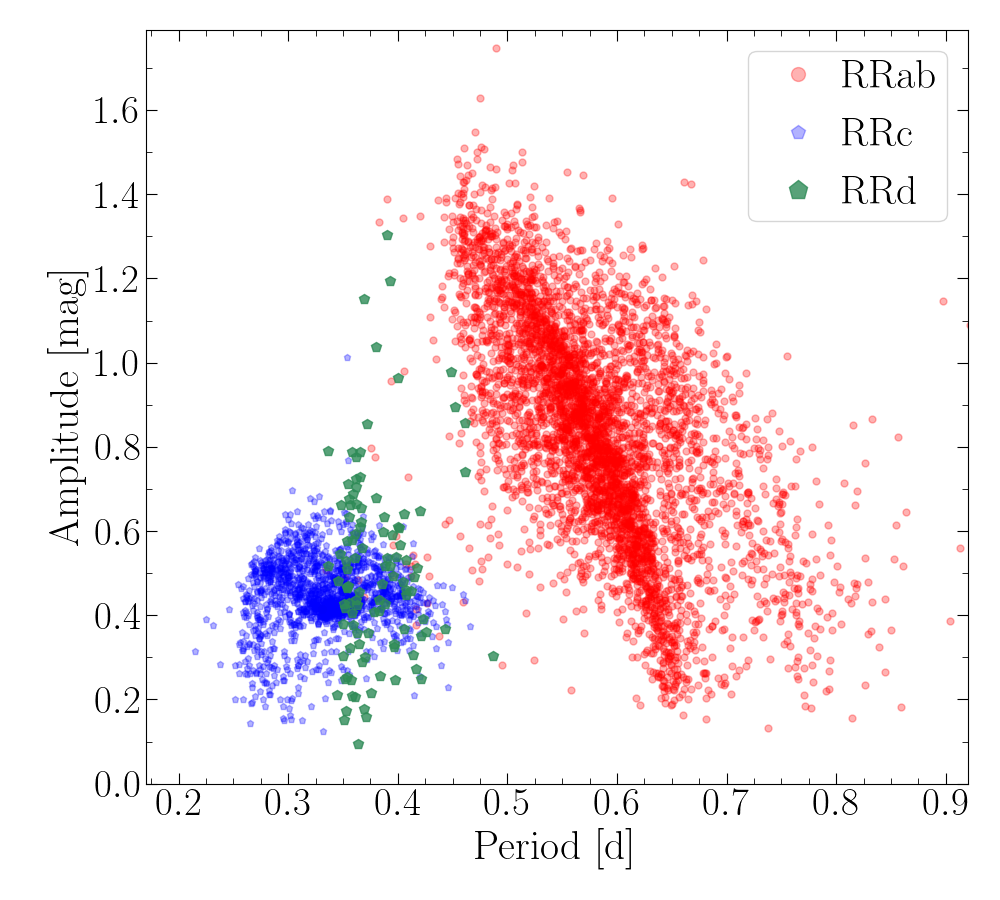}
\caption{
Bailey diagram of the DESI Y1 RRL sample. 
The depicted amplitudes of pulsation (in the $V-$band) are obtained from the peak-to-peak $G$ magnitudes in the {\it Gaia} catalog using the transformations described by \citet{Clementini2019}. 
The distinction between RRab, RRc, and RRd stars is based on the classification listed in the {\it Gaia} DR3 catalog \citep{Clementini2023}.
}
\label{fig:bailey}
\end{center}
\end{figure}

The spatial distribution of our sample is shown in Figure~\ref{fig:map_and_histograms}, in Galactic coordinates. 
In terms of classification, the DESI Y1 RRL sample consists of 4,524 RRab, 1,609 RRc, and 107 RRd stars. 
Figure~\ref{fig:bailey} shows their distribution in the so-called Bailey diagram (period vs. amplitude). 
Throughout this paper, we describe in detail the spectroscopic properties of the stars in our catalog making distinctions on their classification. 
A detailed analysis of the Bailey diagram and the dependence of its morphology on spectroscopic properties (mainly [Fe/H]) is presented in a separate paper \citep[][]{Medina2025b}.

\section{Spectroscopic analysis}
\label{sec:analysis}

\subsection{Observed spectroscopic properties}
\label{sec:obsProperties}

We base our analysis on the spectroscopic properties obtained by the main DESI-MWS data processing pipelines applied to DESI's DR1 catalog, namely the RVS and the SP pipelines. 
The RVS pipeline determines radial velocities and atmospheric parameters ($T_{\rm eff}$, log $g$, [Fe/H], and [$\alpha$/Fe]) of all stellar objects using the Python package RVSpecfit \citep{Koposov2019}, which was adapted to handle DESI data and is based on interpolated stellar templates \citep[PHOENIX models;][]{Husser2013}. 
The SP pipeline relies on the optimization and interpolation of model grids fitted to data, using the FORTRAN code FERRE\footnote{\href{https://github.com/callendeprieto/ferre}{https://github.com/callendeprieto/ferre}} \citep[][]{AllendePrieto2006}. 
These model grids have between two and five parameters (\teff, log $g$, [Fe/H], [$\alpha$/Fe], and microturbulence) and derives atmospheric parameters and covariance matrices for every spectrum, using PHOENIX models (same as for the RVS) for stars with $2300<\teff {\rm [K]}<5100$ and Kurucz ATLAS9 models \citep[][]{Kurucz1979,Kurucz1993} for stars with $3500<\teff {\rm [K]}<30000$ \citep[][]{Meszaros2012,AllendePrieto2018}.
This pipeline has been adapted for its application on DESI data through the Python code PIFERRE\footnote{\href{https://github.com/callendeprieto/piferre}{https://github.com/callendeprieto/piferre}}.
For more details of DESI's processing pipelines, we refer the reader to the overview of DESI-MWS by \citet{Cooper2023} and the early data release by \citet{Koposov2024}. 
We note that neither of these pipelines make special considerations for the analysis of short-period variable stars. 

To study the cyclical variation of spectroscopic properties of RRLs, accounting for their pulsating nature, as we do in this work, it is important to estimate when in the pulsation cycle (of each star) each epoch was taken. 
We estimate the phase of individual epochs $\phi$ from the reference epoch listed in the {\it Gaia} DR3 catalog and the time of the middle of DESI's exposures. 
These reference epochs correspond to the time of maximum light in the $G-$band light curve of each RRL, as measured by {\it Gaia}, and the resulting phases $\phi$ represent the moment in the pulsation of the star where a given observation was made. 
It is worth mentioning that time offsets between spectroscopic and photometric measurements (the latter used to build light curves and to infer reference epochs) can lead to shifts in the predicted phases, which may in turn result in inaccurately determined systemic properties. 
Empirical and theoretical evidence indicates that the evolution of RRLs off the zero-age horizontal branch and across the instability strip is associated with period changes \citep[hence, shifts in the predicted phases; see e.g.,][]{Jurcsik2001, Soszynski2011,Soszynski2014, Szeidl2011,ArellanoFerro2018,Bono2020}. 
One would expect that high period change rates lead to a low accuracy in phase determinations at a fixed epoch separation between spectroscopic and photometric measurements. 
Another effect that is worth mentioning is the amplitude, period, and phase modulation observed in 20\%–30\% of the RRab stars and 5-40\% of the RRc stars \citep[e.g.,][]{Szeidl1988,Moskalik2003,Nagy2006,Kolenberg2010, Catelan2015}, the so-called Blazhko effect \citep{Blazhko1907}. 
We note that recent works have reported higher incidence rates of the Blazhko effect among RRab stars ($\sim50$\%; see e.g., \citealt{Prudil2017,Netzel2018,Netzel2023,Molnar2022}).
The Blazhko effect occurs in timescales longer than the typical single-pulsation periods of RRLs \citep[typically weeks-months versus $\leq1.1$\,d; see e.g.,][]{Buchler2011,Prudil2017,Netzel2018} and with amplitude variations a few tenths of magnitudes smaller than those of a single cycle ($\leq1.0$\,mag).
The time difference between DESI's data and the reference epoch date from the light curves in {\it Gaia} ranges from $\sim2250$\,d (SV survey) to $\sim2880$\,d (main survey), with most of the DESI observations (74\%) being taken with an offset between $2500$\,d and said upper limit.
To quantify the effect of measurement offsets, period changes, and long timescales modulation in sizable samples is, however, a challenging task, as this requires long temporal baselines for measurements comparison. 
Thus, we recognize this limitation in the current analysis and identify this as a research avenue to be explored in the near future (e.g., combining future {\it Gaia} data releases with large collections of RRL light curves from the literature, such as OGLE). 

\subsection{Systemic velocity determination}
\label{sec:vsys}

For RRLs, neither radial velocities from coadded spectra  nor those from single exposures are suitable 
as estimates of their 
their center-of-mass velocities, due to their pulsating nature.  
In fact, not only can the amplitude of the observed radial velocity curve easily reach 50--60\,km\,s$^{-1}$ peak to peak, but its shape depends on the spectral lines measured to estimate the Doppler shift \citep[see e.g.,][]{Clementini1990,Jeffery2007}, as different lines are produced at different depths in the stellar atmosphere (metallic lines are formed deeper in the atmosphere than Balmer lines) and the motion of different atmospheric layers is not synchronous -- the so-called Van Hoof effect \citep[][]{Vanhoof1953}. 
Indeed, the radial velocity of RRab stars can exhibit variations of up to $\sim100$--$120$\,km\,s$^{-1}$ when measured using the H$_\alpha$ line, while those variations are approximately a factor of two smaller when using metallic lines \citep[][]{Sesar2012,Bono2020,Braga2021}. 
Thus, in order to estimate the center-of-mass velocity of RRLs (i.e., their {\it systemic} velocity), it is important to correct the observed velocity based on the phase of the pulsation at which the RRLs are observed.

Similar to the rest of the spectroscopic properties derived by the RVS pipeline, these observed RV estimates are available for single-epoch observations and for coadded spectra. 
For this work, we employ single and multi-epoch radial velocities to estimate the systemic velocity of our RRLs. 
In terms of uncertainties, 
$\sim85\%$
of our sample have observed (single-epoch) RVS radial velocity uncertainties $<3.5$\,km\,s$^{-1}$, with a median of $\sim1$\,km\,s$^{-1}$.
In the rest of this section, we describe two independent approaches followed to correct our observed velocities for the pulsation component affecting the systemic velocity estimates. 
For both approaches, we employ the radial velocities obtained with the DESI-MWS 
RVS pipeline applied to DESI's DR1 catalog, in conjunction with the available phase of observations $\phi$.

\begin{table*}[!tbp]
\caption{
Optimal parameters for each of the radial velocity curve models described in Section~\ref{sec:desiRVC} and derived using Cmdstanpy. 
We split our dataset into RRL subclasses and provide the number of observations used as input for the models ($N$).
For each parameter, we report the median of the posterior distributions and the confidence intervals (C.I.) defined from their 16th and 84th percentiles. 
}
\label{tab:rvc_params}
  \centering
\scriptsize
  \begin{minipage}{.4\linewidth}
    \begin{center}
     \begin{tabular}{Hcccc}
    \multicolumn{3}{l}{RRab ($N=8881$)}\\
\hline
& Parameter & Prior & Median & C.I. \\
\hline
      &                   $A_{11}$ &   $\mathcal{N} (0.3, 10.0)$ &    $-0.190$ &   [$-0.276$, $-0.101$] \\
      &                   $A_{12}$ &  $\mathcal{N} (-2.0, 10.0)$ &    $-0.967$ &   [$-1.124$, $-0.815$] \\
      &                   $A_{21}$ &   $\mathcal{N} (0.2, 10.0)$ &    $-0.062$ &    [$-0.138$, $0.013$] \\
      &                   $A_{22}$ &  $\mathcal{N} (-1.0, 10.0)$ &    $-0.516$ &   [$-0.645$, $-0.385$] \\
      &                   $A_{31}$ &   $\mathcal{N} (0.2, 10.0)$ &    $-0.095$ &   [$-0.164$, $-0.024$] \\
      &                   $A_{32}$ &  $\mathcal{N} (-0.7, 10.0)$ &    $-0.175$ &   [$-0.299$, $-0.053$] \\
      &                   $B_{11}$ &   $\mathcal{N} (0.3, 10.0)$ &    $-0.488$ &   [$-0.572$, $-0.404$] \\
      &                   $B_{12}$ &  $\mathcal{N} (-1.8, 10.0)$ &    $-0.603$ &   [$-0.750$, $-0.453$] \\
      &                   $B_{21}$ &   $\mathcal{N} (0.1, 10.0)$ &    $-0.219$ &   [$-0.294$, $-0.145$] \\
      &                   $B_{22}$ &  $\mathcal{N} (-0.4, 10.0)$ &     $0.100$ &    [$-0.027$, $0.232$] \\
      &                   $B_{31}$ &  $\mathcal{N} (-0.1, 10.0)$ &    $-0.281$ &   [$-0.354$, $-0.207$] \\
      &                   $B_{32}$ &   $\mathcal{N} (0.2, 10.0)$ &     $0.514$ &     [$0.388$, $0.638$] \\
      &              ${\rm Amp}_v$ &    $\mathcal{N} (1.0, 2.0)$ &     $0.782$ &     [$0.742$, $0.823$] \\
      &              $f_{\rm out}$ &    $\mathcal{U} (0.0, 1.0)$ &     $0.047$ &     [$0.042$, $0.051$] \\
      & $\ln\sigma_{v_{\rm out}}$ &    $\mathcal{U} (1.5, 3.0)$ &     $2.160$ &     [$2.141$, $2.179$] \\
       & $\ln\sigma_{v_{  }}$ &   $\mathcal{U} (-1.0, 1.5)$ &     $0.830$ &     [$0.820$, $0.841$] \\
      &             $\sigma_{v_{\rm sys}}$ & $\mathcal{U} (50.0, 500.0)$ &   $171.068$ & [$169.240$, $172.881$] \\
\hline
\end{tabular}
      \label{tab:left1}
    \end{center}
  \end{minipage}
  \quad
  \begin{minipage}{.4\linewidth}
    \begin{center}
      \begin{tabular}{Hcccc}
      \multicolumn{3}{l}{RRc ($N=3178$)}\\ 
\hline
& Parameter & Prior & Median & C.I. \\
\hline
      &                   $A_{11}$ &   $\mathcal{N} (0.3, 10.0)$ &    $-0.299$ &   [$-0.480$, $-0.137$] \\
      &                   $A_{12}$ &  $\mathcal{N} (-2.0, 10.0)$ &    $-0.859$ &   [$-1.356$, $-0.391$] \\
      &                   $A_{21}$ &   $\mathcal{N} (0.2, 10.0)$ &    $-0.013$ &    [$-0.163$, $0.131$] \\
      &                   $A_{22}$ &  $\mathcal{N} (-1.0, 10.0)$ &    $-0.232$ &    [$-0.670$, $0.201$] \\
      &                   $A_{31}$ &   $\mathcal{N} (0.2, 10.0)$ &     $0.217$ &     [$0.074$, $0.384$] \\
      &                   $A_{32}$ &  $\mathcal{N} (-0.7, 10.0)$ &    $-0.645$ &   [$-1.131$, $-0.225$] \\
      &                   $B_{11}$ &   $\mathcal{N} (0.3, 10.0)$ &    $-0.515$ &   [$-0.696$, $-0.353$] \\
      &                   $B_{12}$ &  $\mathcal{N} (-1.8, 10.0)$ &     $0.638$ &     [$0.202$, $1.103$] \\
      &                   $B_{21}$ &   $\mathcal{N} (0.1, 10.0)$ &    $-0.402$ &   [$-0.581$, $-0.245$] \\
      &                   $B_{22}$ &  $\mathcal{N} (-0.4, 10.0)$ &     $1.175$ &     [$0.710$, $1.704$] \\
      &                   $B_{31}$ &  $\mathcal{N} (-0.1, 10.0)$ &    $-0.112$ &    [$-0.255$, $0.029$] \\
      &                   $B_{32}$ &   $\mathcal{N} (0.2, 10.0)$ &     $0.380$ &    [$-0.042$, $0.799$] \\
      &              ${\rm Amp}_v$ &    $\mathcal{N} (1.0, 2.0)$ &     $0.909$ &     [$0.650$, $1.188$] \\
      &              $f_{\rm out}$ &    $\mathcal{U} (0.0, 1.0)$ &     $0.009$ &     [$0.006$, $0.012$] \\
      & $\ln\sigma_{v_{\rm out}}$ &    $\mathcal{U} (1.5, 3.0)$ &     $2.285$ &     [$2.224$, $2.354$] \\
      & $\ln\sigma_{v_{  }}$ &   $\mathcal{U} (-1.0, 1.5)$ &     $0.828$ &     [$0.815$, $0.842$] \\
      &             $\sigma_{v_{\rm sys}}$ & $\mathcal{U} (50.0, 500.0)$ &   $171.422$ & [$168.557$, $174.299$] \\
\hline
\end{tabular}
      \label{tab:right1}
    \end{center}
  \end{minipage}
\end{table*}

We highlight that, as described in Section~\ref{sec:obsProperties}, the observed velocities used in this work are determined using a template fitting methodology, essentially representing a weighted average velocity across all the lines used for the fit. 
Therefore, the Van Hoof effect is not accounted for to estimate observed velocities. 
We note that the weights of this average are captured by the inverse RV error, which quantifies the contribution of each spectral region to constraining the RV. 
Measured in 100\,\AA\ bins across DESI's spectral range for a constant S/N  spectrum, the inverse RV error shows that the fit is strongly dominated by spectral regions with wavelengths $\lambda \lesssim 4000$\,\AA.
The integrated weight from metallic lines in this region exceeds that of both H$_\beta$ and H$_\alpha$ by a factor of $\sim$3--4 (with H$_\alpha$ contributing only marginally to the overall fit). 
Since metallic lines display intrinsically smaller velocity variations than Balmer lines, this wavelength-dependent weighting implies that our observed velocities are predominantly sensitive to metallic line velocities.
Consequently, our RVC amplitudes are expected to be closer to those derived from metallic lines than from Balmer lines (as described in Section~\ref{sec:desiRVC}).
A detailed line-by-line analysis will be presented in a continuation of this work, employing a larger dataset from the first three years of DESI’s operations (DESI Y3).

\subsubsection{Radial velocity curve modeling}
\label{sec:desiRVC}

Our sample is composed of observations of individual stars that, when combined, cover homogeneously the full pulsation cycle of RRLs (in phase). 
Here, we construct a Bayesian inference model to determine the parameters describing the cyclic variation of the radial velocity of RRLs in our sample and the systemic velocity of each star.
For this, we assume that the sample is composed of single-epoch observations that are a good representation of the underlying phase-dependent radial velocity curve (RVC) variation, and observations that are not consistent with these models (i.e., outliers). 
The fraction of outlier observations for the model is represented by the parameter $f_{\rm out}$,  whereas the fraction of observed velocities used to construct a given RVC template is $(1-f_{\rm out})$. 
To account for correlations between the shape of the RVCs and the pulsation type,  we split the sample between fundamental mode and first overtone pulsators. 

For each star $i$, observed at phases $\phi_j$ (where the $j$ values depend on the total number of epochs per star), we model the systemic velocity $v_{{\rm sys}, i}$ obtained from its (multiple) $v_{{\rm obs}, i, j}$  observation(s) as: 

\begin{center}
\begin{align}
\begin{split}
P(v_{{\rm obs},i, j}|v_{{\rm mod},i}(\phi_{j})) = f_{\rm out} \mathcal{N}(v_{{\rm obs}, i, j}|0, \sigma_{v_{\rm out}})   \\
+ (1-f_{\rm out}) \mathcal{N}(v_{{\rm obs}, i,j}|v_{{\rm mod},i}(\phi_{j}), \sigma_{v_{{\rm RRL}, i, j}})
\end{split}
\end{align}
\end{center}
\label{eq:RVC_prob}

\noindent where the modeled velocity $v_{{\rm mod}, i}(\phi_{j})$ is defined as

\begin{align}
v_{{\rm mod},i}(\phi_{j}) & = v_{{\rm sys}, i} + K_i\cdot V(\phi_j),
\label{eq:RVcurves1}
\end{align}

\noindent with $v_{{\rm sys}, i}$ determined from a Gaussian prior centered at 0\,km\,s$^{-1}$ with scatter $\sigma_{v_{{\rm sys}}}$, $\mathcal{N}(0, \sigma_{v_{{\rm sys}}} )$, and 

\begin{align}
    K_i &= 25\cdot10^{{\rm Amp}_{v}\cdot\frac{(A_{G, i}-1)}{2.5}}  \label{eq:RVcurves2a}
\end{align}    
\begin{align}
V(\phi_j) &=  \sum_{k=1}^3 A_k(P) \cos(2\pi\phi_j k) + B_k(P) \sin(2\pi\phi_j k).  \label{eq:RVcurves2b}
\end{align}

\noindent In this representation, $A_{G,i}$ corresponds to the light-curve amplitude in the $G$ band of the $i-$th star and ${\rm Amp}_{v}$ is a model parameter. 
The two Fourier terms $A_k$ and $B_k$ are defined to account for the dependence of the light-curve shapes and the period $P$ through linear relations: 

\begin{align}
    A_k &= A_{k,1} + A_{k,2} P \\
B_k &= B_{k,1} + B_{k,2} P  \label{eq:RVcurves3}
\end{align}

\noindent where $A_{k,1}$, $A_{k,2}$, $B_{k,1}$, and $B_{k,2}$ are model parameters.
We note that the period of the RRLs is not a free parameter in the model and is taken from the {\it Gaia} catalog. 
Lastly, $\sigma_{v_{\rm out}}$ and $\sigma_{{v_{\rm RRL},i,j}}=\sqrt{\sigma_{v_{}}^2+\sigma_{v_{{{\rm obs},i,j}}}^2}$ represent the dispersion of the outlier and RRL velocity distribution, the latter encompassing a systematic velocity uncertainty $\sigma_{v_{ }}$ (in log scale), and the error of the observed radial velocities, $\sigma_{v_{{{\rm obs},i,j}}}$
Therefore, our model is defined by 
as many parameters as unique number of stars ($v_{{\rm sys}, i}$), plus 17 
parameters corresponding to the amplitude of the RVC ${\rm Amp}_{v}$, 
twelve Fourier coefficients, 
the fraction of outliers for the model ($f_{\rm out}$), 
the $\sigma$ of the outlier distribution $\sigma_{v_{\rm out}}$ (in log scale),  a systematic velocity uncertainty $\sigma_{v_{ }}$ (in log scale), 
and the weight of the velocity distribution 
($\sigma_{v_{{\rm sys}}}$).

We compute our models' parameters maximizing their likelihood using the probabilistic programming language for Bayesian inference Stan with its Python interface Cmdstanpy\footnote{\href{https://github.com/stan-dev/cmdstanpy}{https://github.com/stan-dev/cmdstanpy}}. 
The parameter space is explored with 36 chains of 2,000 steps each, which is sufficient to reach convergence and to compute robust results. 
We define the prior for each parameter as follows using uniform and normal distributions.
For the Fourier coefficients $A_k$ and $B_k$, in particular, we use normal distributions centered in values between $-2.0$ and $0.3$, chosen to ensure the convergence of the model when using the full period range of our sample.  
We estimate our model parameters from the median of the marginalized posterior distributions, and their uncertainties from the 16th/84th percentiles.
Table~\ref{tab:rvc_params} summarizes the priors and the resulting parameters for the RVC structural parameters, and includes the number of epochs used for each model. 

\begin{figure*}
\begin{center}
\includegraphics[angle=0,scale=.44]{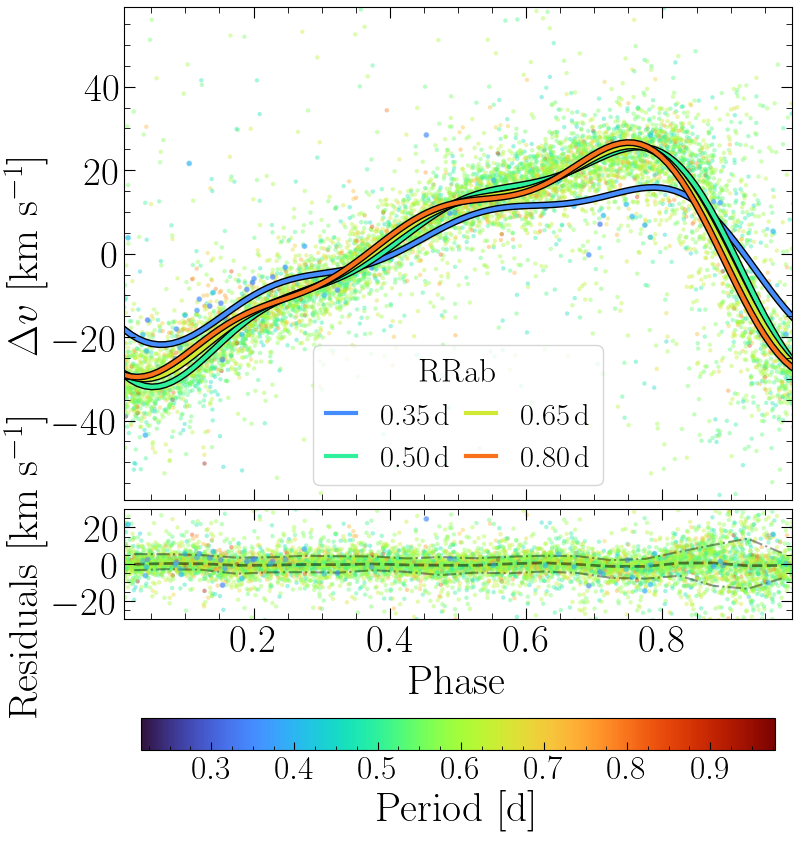}
\includegraphics[angle=0,scale=.44]{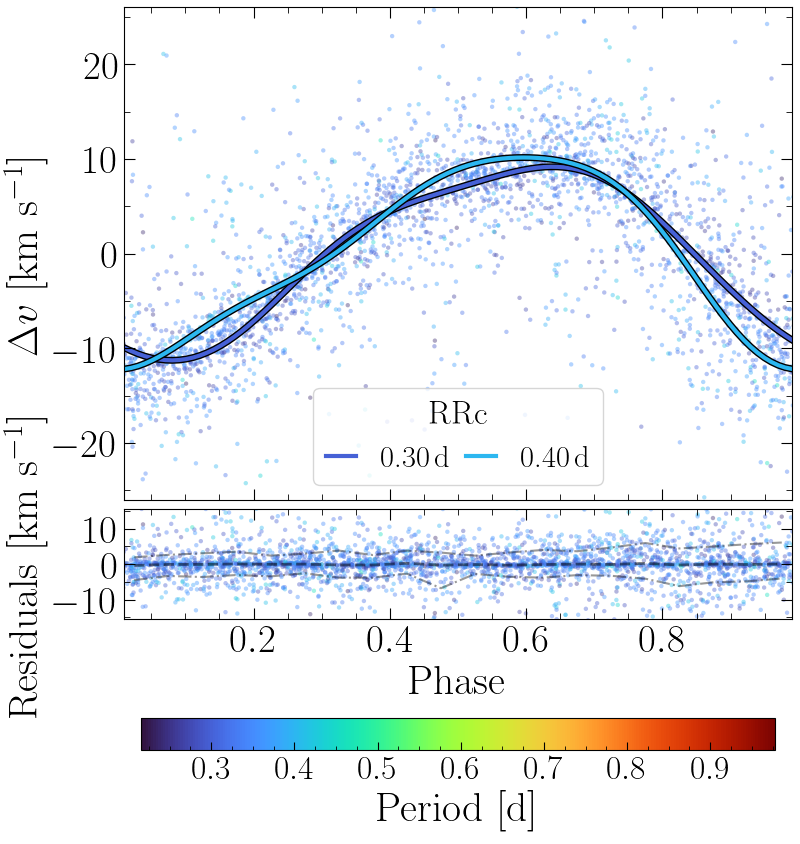}
\caption{
The resulting RVCs from the models described in Section~\ref{sec:desiRVC}, for RRab (left) and RRc (right). 
The observed velocities of individual observations relative to the RVCs are plotted with markers color-coded by period.  
For RRab variables, marginally larger dots are used for measurements of the significantly less numerous RRLs in the shorter period bin.
Example RVCs are displayed for fixed periods, in period bins centered at [0.35, 0.50, 0.65, 0.80]\,d for RRab, and [0.30, 0.40]\,d for RRc.
Each of the RVCs shown is scaled by a $G-$band amplitude $A_G$   corresponding to the mean {\sc peak\_to\_peak\_g} of the RRL sample that falls within the range of its corresponding period bin. 
The bottom panels in each plot display the deviations of the  observed velocities with respect to the velocities predicted by the model for each observation, depending on phase, pulsation period, and light curve amplitude. 
In the bottom panels, dashed lines represent the 16th, 50th, and 84th percentiles of the residual distribution, in phase bins of 0.05.  
}
\label{fig:rvcs}
\end{center}
\end{figure*}

Figure~\ref{fig:rvcs} displays examples of RVC templates constructed from the parameters resulting from this methodology.
The figure shows RVCs as the change in velocity relative to the RRL systemic velocity ($\Delta v$), as a function of phase for four periods of RRab stars (0.35, 0.50, 0.65, and 0.80\,d) and for two periods of RRc stars (0.30 and 0.40\,d).
The figure also shows the corrected velocity (observed velocity minus derived systemic velocity) for all DESI Y1 RRLs, split by RRL classification and color-coded by period.
From the RVCs derived for RRab stars, we observe radial velocity variations (amplitudes) of 
$\sim50$\,km\,s$^{-1}$ throughout their pulsation cycle, with a slight dependence on the period of pulsation. 
The RVCs of RRab are characterized by a smooth increase in observed velocity for most of the pulsation and a steep decrease at $\phi\sim0.80$, close to the phase of maximum contraction (at the start of the rising branch of the RRL luminosity), where abrupt changes in the stars' atmospheric kinetic energy are expected \citep[see e.g.,][]{Kolenberg2010}.
For RRc stars, the RVCs are in general more sinusoidal with radial velocity variations of the order of $\sim20$\,km\,s$^{-1}$ for both of the periods shown, in agreement with previously reported RVC measurements \citep[see e.g.,][]{Braga2021, Prudil2024b}. 
We note that Figure~\ref{fig:rvcs} displays a RVC with a distinctly small line-of-sight amplitude for short-period RRab at 0.35 d (of the order of $\sim30$\,km\,s$^{-1}$), and we consider this observable partially attributable to low number statistics. 
In fact, 92\% of our RRab sample lies in the [0.45, 0.70]\,d period range and only 51 RRab have periods shorter than 0.45\,d, in contrast to RRc stars, which have a much more compact period distribution (92\% of the RRc stars have periods within [0.25, 0.40]\,d). 
Consequently, our model does not provide robust estimates for the velocity correction (and the amplitude of RVCs) of RRab stars with periods $<0.45$\,d.
We note that excluding these stars from the model yields parameters and systemic velocities consistent well within uncertainties for the bulk of our sample, indicating that they do not affect the global results significantly.

\begin{figure*}
\begin{center}
\includegraphics[angle=0,scale=.44]{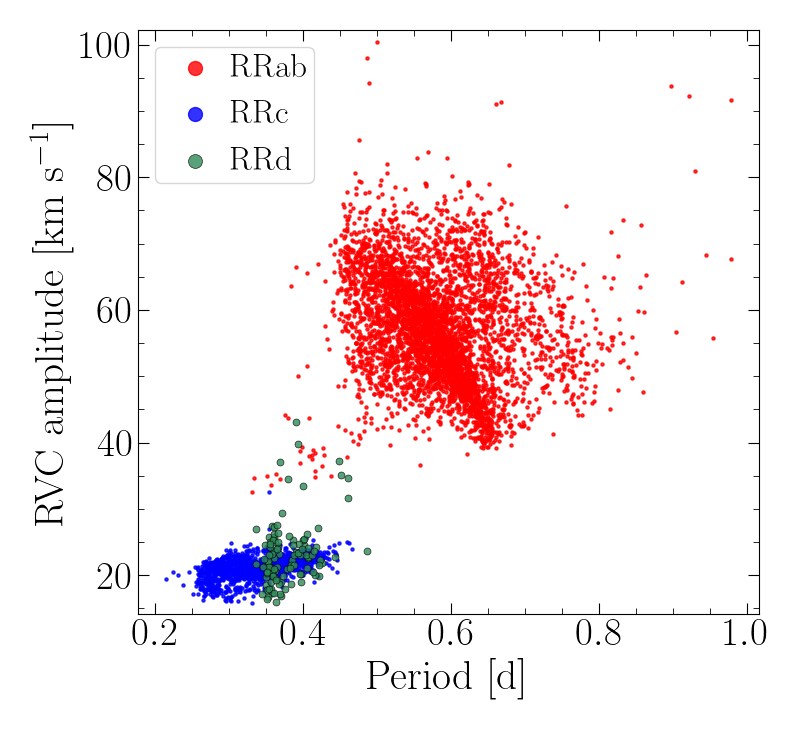}
\includegraphics[angle=0,scale=.44]{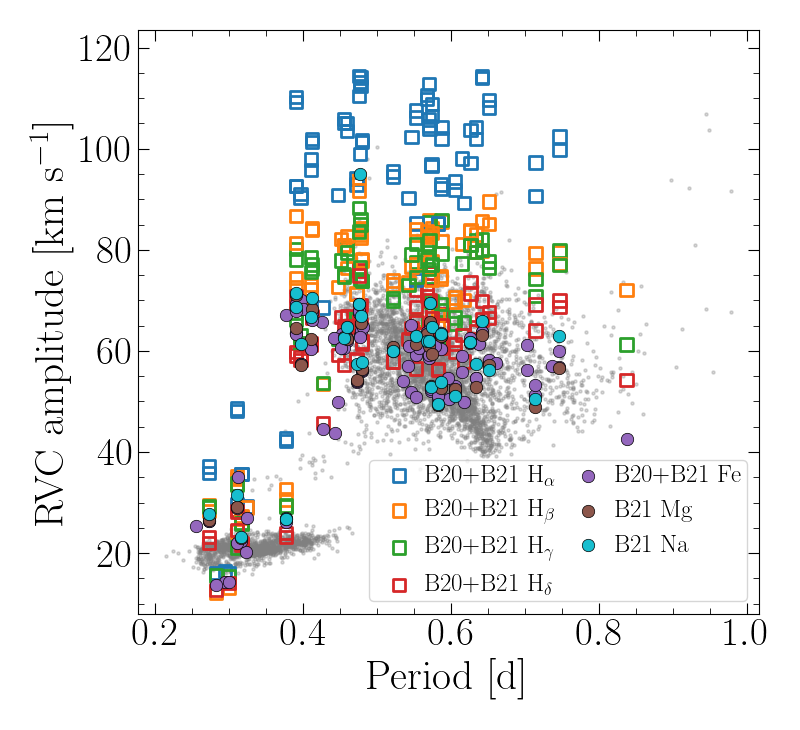}
\caption{
{\it Left}: Amplitude of the RVCs obtained with the model described in Section~\ref{sec:desiRVC} for the RRLs in our sample, as a function of their pulsation periods. 
{\it Right}: Comparison of our results with those from \citet{Bono2020} (B20) and \citet{Braga2021} (B21) using Balmer lines (H$_{\alpha}$, H$_{\beta}$, H$_{\gamma}$, H$_{\delta}$) and metallic lines (Fe, Mg, Na). 
The RVC amplitudes derived with our method resemble those from H$_\delta$ and metallic lines in both works. 
}
\label{fig:rvc_amplitudes}
\end{center}
\end{figure*}

Figure~\ref{fig:rvc_amplitudes} displays the inferred RVC amplitude for every RRL in our sample, computed using Equations~\ref{eq:RVcurves2a} and \ref{eq:RVcurves2b} and the best-fit parameters of our model (Table~\ref{tab:rvc_params}). 
The figure shows that, on a star-by-star basis, the amplitude of RVCs ranges between 40-80\,km\,s$^{-1}$ for RRab stars and 15-25\,km\,s$^{-1}$ for RRc stars. 
In the case of RRab stars, in particular, the RVC amplitude distribution follows closely the shape of their position in the Bailey diagram -- a consequence of our definition of the RVC shapes relying on their lightcurve amplitudes (Equations~\ref{eq:RVcurves2a} and \ref{eq:RVcurves2b}).

The trends shown in Figure~\ref{fig:rvc_amplitudes} represent the general expectation for the period dependency of RVC amplitudes, that is, decreasing  amplitudes for increasing period in RRab stars. 
This is in agreement with previous empirical studies of radial velocity variations of RRLs, which are based on measurements fully covering the pulsation cycle of individual stars. 
The figure also shows a comparison with the empirical RVC amplitudes measured for RRab and RRc stars by \citet{Bono2020} and \citet{Braga2021} using Balmer lines (H$_{\alpha}$, H$_{\beta}$, H$_{\gamma}$, H$_{\delta}$) and metallic lines (Fe, Mg, Na). 
We find that our results are consistent with those obtained from metallic lines in their works, an expected result given that the code used in our work to measure observed radial velocities  (RVSpecfit) does not treat Balmer lines preferentially and metallic lines have an overall higher weight in the template fitting (especially for wavelengths shorter than 4500\AA). 
We also note that the RVC amplitudes in our work reach lower amplitudes than those from  \citet{Bono2020} and \citet{Braga2021}. 
We attribute this to selection effects, considering that the samples selected in their works are composed (predominantly) of RRab stars with $V-$band amplitudes larger than 0.5\,mag, and RRc stars with large amplitudes ($>0.5$\,mag) or with a narrow period distribution around $\sim0.3$\,d.

The systemic velocity uncertainties are derived from the $1\sigma$ intervals of the velocity posterior distribution of every star in our sample that fulfills our adopted quality cuts, regardless of the number of available epochs. 
These uncertainties range from 1 to $\sim150$\,km\,s$^{-1}$ for RRab stars ($\sim75$\% with uncertainties $<30$\,km\,s$^{-1}$), and from 1 to $\sim130$\,km\,s$^{-1}$ for RRc variables ($\sim85$\% with uncertainties $<10$\,km\,s$^{-1}$), and are mostly driven by the number of epochs per star. 
In fact, only stars with fewer than three epochs display large systemic velocity uncertainties, with RRab being more affected by the constraint on the number of epochs than RRc (as the median of their distributions are 30\,km\,s$^{-1}$ and 8\,km\,s$^{-1}$, respectively).
On the other hand, the vast majority (99\%) of stars with three or more than three epochs have systemic velocity uncertainties $<10$\,km\,s$^{-1}$, and 87\% of them have uncertainties $<5$\,km\,s$^{-1}$. 
The decrease in velocity uncertainty as a function of number of epochs $N$ can be characterized as $10.94^{+6.21}_{-3.89}\ N^{ -0.61^{+0.12}_{-0.15} }$\,km\,s$^{-1}$ for RRab stars, and $7.09^{+3.35}_{-0.41}\ N^{ -0.49^{+0.13}_{-0.01} }$\,km\,s$^{-1}$ for RRc stars (where the error of the coefficients represent the 1$\sigma$ confidence regions). 
Thus, stars with large velocity uncertainties are outliers in our sample. 
We note that testing our methodology on a similar dataset but twice as large, namely the DESI second data release, does not significantly affect the aforementioned decrease in velocity uncertainty as a function of $N$ (as this requires a more complex modeling that takes into account, e.g., outlier models that depend on phase). 
This, however, results in a factor of two increase in the precision of the RVC parameters, and resolves the limitations of the model for short-period RRab stars, confirming the importance of the sample size in constructing the RVCs.

In terms of the scatter of single epoch measurements with respect to the velocity predicted by the RVCs at a given phase (measured as observed minus predicted radial velocity; see Figure~\ref{fig:rvcs}),  
we observe a slight variation as a function of period and, more notably, pulsation type. 
Indeed, the median difference between the observed and predicted velocity is $\leq1$\,km\,s$^{-1}$ across the period ranges analyzed, with the more significant differences for RRab in the 0.40-0.60\,d period range.
As for the standard deviation of the velocity difference distributions, they are of the order of $\sim10$\,km\,s$^{-1}$ for RRab ($\leq12$\,km\,s$^{-1}$ across period ranges) and $\sim6$\,km\,s$^{-1}$ for RRab ($\leq8$\,km\,s$^{-1}$ across period ranges), but are highly affected by a few outliers.
Figure~\ref{fig:rvcs} also shows the variation of the 16th, 50th, and 84th percentiles of these distributions, in phase bins of 0.05. 
For RRab, the 1$\sigma$ regions of the residuals vary 
from 3-6\,km\,s$^{-1}$ for $\phi$ in the range [0.00, 0.75] and increase to $\sim15$\,km\,s$^{-1}$ in [0.75, 1.00]. 
For RRc, the 1$\sigma$ region is found between 3 and 6\,km\,s$^{-1}$ consistently across the entire RRL pulsation cycle.

Lastly, we highlight that no distinction between velocities measured from different absorption lines is done for this analysis to account for the Van Hoof effect. 
This is likely the reason for the presence of bumps in the RVCs of both RRab and RRc stars (the latter being less affected overall), and a tentative explanation for the aforementioned high systemic velocity uncertainties. 
Larger samples of RRLs with measured velocities from individual lines, obtained from DESI Y3, will enable a detailed derivation of line-by-line RVCs.  
The results of such analysis will be presented in a follow-up study (Medina et al. in prep.).

\subsubsection{Literature RVCs}
\label{sec:B21RVC}

Line-of-sight velocity templates for RRLs are available in the literature and are constructed from visual and/or near infrared spectra of single-mode RRLs, using catalogs of different sizes \citep[e.g.,][]{Sesar2012,Braga2021,Prudil2024b}. 
Here, we describe our approach to estimate the systemic velocity of the DESI RRLs accounting for their pulsating component employing the radial velocity curve models of \citet{Braga2021} (hereafter B21), for which we use the epoch of maximum light as a phase zero reference. 
These templates were built using 31 RRab stars (split into period bins) and five RRc stars with well-sampled radial velocity curves. 
The velocities described above represent an ideal use-case for the models of \citet{Braga2021}, as the latter are available not only for different RRL types, but also for different sets of lines, namely Balmer and metallic lines. 

To obtain RRLs' velocities on a line-by-line basis, for each star, we first generated a radial velocity template using the spectral analysis tool {\sc iSpec} \citep[][]{Blanco-Cuaresma2014a,Blanco-Cuaresma2019}. 
These spectra are produced adopting ATLAS9 model atmospheres \citep[][]{Castelli2003} and employing a SPECTRUM radial transfer code, with the Vienna Atomic Line Database (VALD) atomic line list in the range 3000–11000\,\AA\ \citep[][]{Piskunov1995,Ryabchikova2015}.
To generate the spectra for each star, we adopt the atmospheric parameters obtained by the RVS pipeline ($T_{\rm eff}$, log $g$, [Fe/H], $v \sin i$), assuming a limb darkening coefficient of 0.6 (as recommended by \citealt{Blanco-Cuaresma2014a}) and a fixed resolution for each arm of the spectrograph, mimicking those of DESI. 
Then, for each epoch of a given star, we compute the mean velocity shift (and its standard deviation) comparing with the template for H$_\alpha$ (at rest wavelength 6563\,\AA), H$_\beta$ (4861\,\AA), H$_\gamma$ (4340\,\AA), H$_\delta$ (4102\,\AA), and the calcium-triplet region (at 8498--8662\,\AA) using a $\chi^2$ minimization routine and a Markov Chain Monte Carlo (MCMC) sampling. 
We perform the $\chi^2$ computation in a $\pm65$\,\AA\ from the center of each line, and in the range 8460--8680\,\AA\ for the calcium-triplet region. 
We use the velocity obtained by the RVS pipeline as a first guess for the individual lines' velocities, sampling the velocity space within a 30\,km\,s$^{-1}$ window from these values. 
This value is sufficient to account for velocity differences caused by different line-to-line measurements for most of the RRLs' pulsation cycle \citep[see e.g.,][]{Sesar2012}.
To obtain a single systemic velocity for a given RRL, we then applied the radial velocity curve correction from B21 for each line and then averaged them, weighted by their uncertainties.

\subsection{Iron abundances}
\label{sec:feh}

\begin{figure}
\begin{center}
\includegraphics[angle=0,scale=.48]{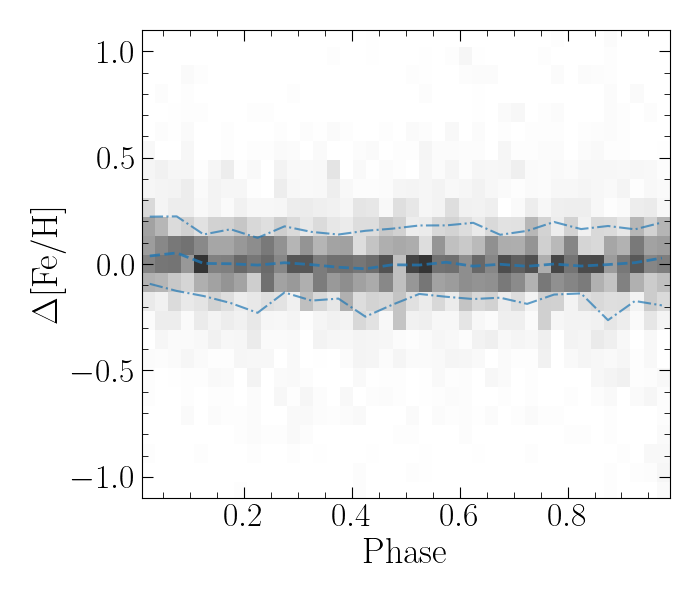}
\caption{
Column normalized 2D histogram depicting the variation of iron abundance with respect to the mean [Fe/H] ($\Delta$[Fe/H]) of each star in our catalog with multi-epoch spectra, as a function of phase. 
Dashed lines represent the 16th, 50th, and 84th percentiles of the $\Delta$[Fe/H] distribution in phase bins of width 0.05, and show a median difference of $\sim0.00$\,dex with $1\sigma$ variation of up to $\sim0.20$\,dex across the RRLs' pulsation.
}
\label{fig:FEHvsPhase}
\end{center}
\end{figure}

We derive the [Fe/H] of the stars in our sample using the RVS pipeline on the single epoch catalog and those from SP pipeline on DESI's coadded spectra. 
For the former, if more than one epoch is available, we adopt their weighted average [Fe/H] as a reasonable estimate of the stars' metallicities. 
We note, however, that the presence and strength of metallic lines might be subject to variations throughout the pulsation of RRLs.
In fact, the recent work by \citet{Wang2024} showed that the [Fe/H] of RRL might exhibit variations of about 0.25-0.30\,dex near $\phi\sim0.90$. 
Figure~\ref{fig:FEHvsPhase} depicts the differences between single epoch [Fe/H] and the mean [Fe/H] ($\Delta$[Fe/H]) for every star in our sample with multi-epoch spectra as a function of phase.
When splitting the pulsation cycle in 0.05 phase bins, we find no clear correlation between $\Delta$[Fe/H] and $\phi$. 
Furthermore, our results show a median difference of $\sim0.00$\,dex with a $1\sigma$ variation of up to $\sim0.20$\,dex consistently throughout the RRL pulsation cycle. 
Thus, we do not expect our RVS-based multi-epoch iron abundances to present strong biases with phase.

In addition to the iron abundances measured by the RVS and the SP pipelines (described in Section~\ref{sec:obsProperties}), we estimate the metallicity of our sample using an alternative approach that relies on existing [Fe/H] calibrations and that is suitable for DESI's medium-resolution spectra.
More specifically, we derived the [Fe/H] of our sample using the recent calibration of the $\Delta$S method \citep[][]{Preston1959,Layden1994} 
done by \citet{Crestani2021a}. 
This method estimates [Fe/H] of an RRL based on its correlation with the equivalent width of the Ca K line (3933\,\AA) and different combinations of Balmer lines (more specifically, H$_\beta$, H$_\gamma$, and H$_\delta$, at 4861, 4340, and 4102\,\AA, respectively). 
We modeled these absorption lines using the Python code \textsc{mpfit}\footnote{\href{https://github.com/segasai/astrolibpy}{https://github.com/segasai/astrolibpy}}
and measured their equivalent widths using a combination of Gaussian and Lorentzian profiles, 
with a width of 20\,\AA\ around the lines central wavelengths.

\subsection{Distance determination}
\label{sec:dist}

\begin{table*}[!tbp]
\caption{
Same as Table~\ref{tab:rvc_params} but for the model parameters of the \teff\ curves. }
\label{tab:teffc_params}
  \centering
\scriptsize
  \begin{minipage}{.5\linewidth}
    \begin{center}
     \begin{tabular}{Hcccc}
    \multicolumn{3}{l}{RRab ($N=8881$)}\\
\hline
& Parameter & Prior & Median & C.I. \\
\hline
      &                               $A_{11}$ &    $\mathcal{N} (-2.0, 10.0)$ &     $5.267$ &     [$2.989$, $7.551$] \\
      &                               $A_{12}$ &    $\mathcal{N} (40.0, 20.0)$ &    $25.408$ &   [$21.237$, $29.531$] \\
      &                               $A_{31}$ &     $\mathcal{N} (0.2, 10.0)$ &     $6.001$ &     [$4.041$, $7.973$] \\
      &                               $A_{12}$ &    $\mathcal{N} (40.0, 20.0)$ &    $25.408$ &   [$21.237$, $29.531$] \\
      &                               $A_{22}$ &     $\mathcal{N} (7.0, 20.0)$ &     $6.716$ &    [$2.991$, $10.419$] \\
      &                               $A_{32}$ &    $\mathcal{N} (-0.7, 20.0)$ &    $-3.701$ &   [$-7.211$, $-0.236$] \\
      &                               $B_{11}$ &    $\mathcal{N} (-5.0, 10.0)$ &    $-0.093$ &    [$-2.148$, $1.955$] \\
      &                               $B_{12}$ &     $\mathcal{N} (6.0, 20.0)$ &     $8.605$ &    [$5.029$, $12.207$] \\
      &                               $B_{31}$ &    $\mathcal{N} (-0.1, 10.0)$ &    $-0.190$ &    [$-2.126$, $1.710$] \\
      &                               $B_{21}$ &    $\mathcal{N} (-1.0, 10.0)$ &    $-2.165$ &   [$-4.023$, $-0.210$] \\
      &                               $B_{22}$ &    $\mathcal{N} (15.0, 20.0)$ &     $5.559$ &     [$2.102$, $8.880$] \\
      &                               $B_{32}$ &     $\mathcal{N} (0.2, 20.0)$ &    $-2.408$ &    [$-5.772$, $1.027$] \\
      &              ${\rm Amp}_{T_{\rm eff}}$ &      $\mathcal{N} (1.0, 2.0)$ &     $1.435$ &     [$1.376$, $1.495$] \\
      &                          $f_{\rm out}$ &      $\mathcal{U} (0.0, 1.0)$ &     $0.010$ &     [$0.009$, $0.012$] \\
      & $\ln\sigma_{{T_{\rm eff}}_{\rm out}}$ &      $\mathcal{U} (1.5, 5.0)$ &     $3.853$ &     [$3.821$, $3.887$] \\
      & $\ln\sigma_{{T_{\rm eff}}} $ &     $\mathcal{U} (-1.0, 2.5)$ &     $2.333$ &     [$2.326$, $2.341$] \\
      &             $\sigma_{{T_{\rm eff}\ {\rm sys} } }$ & $\mathcal{U} (50.0, 2,000.0)$ &   $424.116$ & [$418.6$, $429.8$] \\

\hline
\end{tabular}
      \label{tab:left2}
    \end{center}
  \end{minipage}
  \quad
  \begin{minipage}{.4\linewidth}
    \begin{center}
      \begin{tabular}{Hcccc}
      \multicolumn{3}{l}{RRc ($N=3178$)}\\ 
\hline
& Parameter & Prior & Median & C.I. \\
\hline
      &                               $A_{11}$ &    $\mathcal{N} (-2.0, 10.0)$ &    $22.143$ &     [$18.803$, $25.575$] \\
      &                               $A_{12}$ &    $\mathcal{N} (40.0, 20.0)$ &   $-28.380$ &   [$-36.215$, $-20.601$] \\
      &                               $A_{31}$ &     $\mathcal{N} (0.2, 10.0)$ &    $-5.164$ &     [$-7.779$, $-2.699$] \\
      &                               $A_{12}$ &    $\mathcal{N} (40.0, 20.0)$ &   $-28.380$ &   [$-36.215$, $-20.601$] \\
      &                               $A_{22}$ &     $\mathcal{N} (7.0, 20.0)$ &    $-0.550$ &      [$-7.893$, $6.650$] \\
      &                               $A_{32}$ &    $\mathcal{N} (-0.7, 20.0)$ &    $16.384$ &      [$9.142$, $24.074$] \\
      &                               $B_{11}$ &    $\mathcal{N} (-5.0, 10.0)$ &    $-3.482$ &     [$-6.072$, $-0.981$] \\
      &                               $B_{12}$ &     $\mathcal{N} (6.0, 20.0)$ &     $8.138$ &      [$0.713$, $15.531$] \\
      &                               $B_{31}$ &    $\mathcal{N} (-0.1, 10.0)$ &     $2.252$ &      [$-0.133$, $4.716$] \\
      &                               $B_{21}$ &    $\mathcal{N} (-1.0, 10.0)$ &    $-6.779$ &     [$-9.412$, $-4.240$] \\
      &                               $B_{22}$ &    $\mathcal{N} (15.0, 20.0)$ &    $17.917$ &     [$10.457$, $25.622$] \\
      &                               $B_{32}$ &     $\mathcal{N} (0.2, 20.0)$ &    $-5.467$ &     [$-12.621$, $1.600$] \\
      &              ${\rm Amp}_{T_{\rm eff}}$ &      $\mathcal{N} (1.0, 2.0)$ &     $0.429$ &       [$0.198$, $0.654$] \\
      &                          $f_{\rm out}$ &      $\mathcal{U} (0.0, 1.0)$ &     $0.002$ &       [$0.001$, $0.003$] \\
      & $\ln\sigma_{{T_{\rm eff}}_{\rm out}}$ &      $\mathcal{U} (1.5, 5.0)$ &     $3.888$ &       [$3.759$, $4.058$] \\
      & $\ln\sigma_{{T_{\rm eff}} } $ &     $\mathcal{U} (-1.0, 2.5)$ &     $2.261$ &       [$2.251$, $2.270$] \\
      &             $\sigma_{{T_{\rm eff}\ {\rm sys}} }$ & $\mathcal{U} (50.0, 2,000.0)$ &   $989.075$ & [$971.9$, $1,006.8$] \\
\hline
\end{tabular}
      \label{tab:right2}
    \end{center}
  \end{minipage}
\end{table*}

The heliocentric distance ($d_{\rm H}$) of our RRL sample is determined using the absolute magnitude-metallicity relation inferred by \citet{Garofalo2022} in the {\it Gaia} $G$ band, using the intensity-averaged $G$ magnitude provided in the {\it Gaia} catalog ({\sc int\_average\_g}).
That is, we employ the following relation:

\begin{equation}
\begin{array}{lc}
\mathbf{M_G = (0.33^{+0.02}_{-0.02}) {\rm [Fe/H]} + (1.05^{+0.03}_{-0.03})},  
\end{array}
\label{eq:magfeh_relation_rrls}
\end{equation}

where $M_G$ represents the absolute magnitude in $G$.
Thus, knowledge of the RRLs iron abundances is required to compute their distances, due to the linear dependency of $M_G$ on [Fe/H] in Equation~\ref{eq:magfeh_relation_rrls}.

For our sample of halo field RRLs, we employ the iron abundances estimated by DESI's RVS pipeline (Section~\ref{sec:feh}).
To obtain reliable $d_{\rm H}$ estimates for our RRLs in known stellar systems, we crossmatched our sample with existing catalogs of high probability stars in globular cluster and dwarf galaxies. 
Further details of the crossmatch process are provided in Section~\ref{sec:field_gc_dwarf_stream}.
For every star in any of these globular clusters or dwarf galaxies, we adopt the mean metallicity of the corresponding host.
Under this assumption, metallicity deviations at a 0.25\,dex level correspond to distance differences of $\sim$2\%,
which fall within the typical distance uncertainty of our sample ($\sim$5\%, which corresponds to [Fe/H] differences of $\sim$0.4\,dex).

\begin{figure*}
\begin{center}
\includegraphics[angle=0,scale=.40]{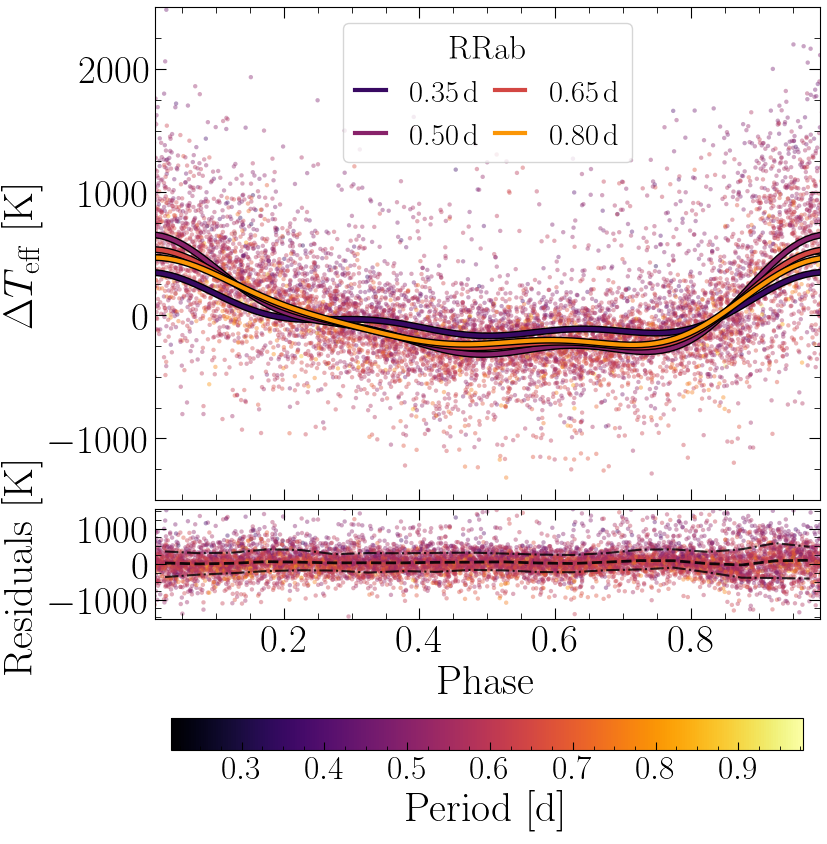}
\includegraphics[angle=0,scale=.40]{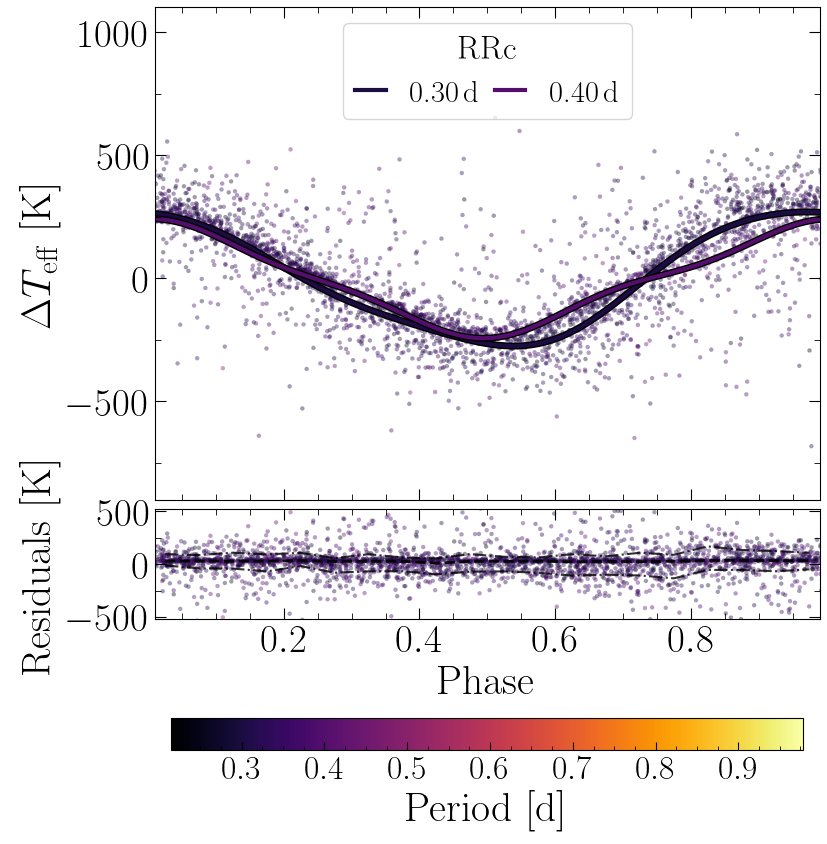}
\caption{
Same as Figure~\ref{fig:rvcs} but for the \teff\ variation of the RRLs in our sample throughout their pulsation periods, for RRab (left) and RRc (right). 
The \teff\ variation curves described in Section~\ref{sec:teff} are plotted with solid lines, and the observed \teff\ of individual observations relative to those curves are depicted with markers color-coded by period.  
The \teff\ curves shown in these panels are computed for the same period bins as for Figure~\ref{fig:rvcs} (i.e., centered at [0.35, 0.50, 0.65, 0.80]\,d for RRab, and [0.30, 0.40]\,d for RRc), and scaled by the $G-$band amplitude of the RRLs in their corresponding period bins.}
\label{fig:teffcs}
\end{center}
\end{figure*}

Distances are computed via distance modulus using the absolute magnitudes from Equation~\ref{eq:magfeh_relation_rrls} and extinction-corrected intensity-averaged $G$ magnitude from {\it Gaia}. 
We employ the dust maps from \citet{Schlafly2011}
and adopt the traditional value of relative visibility ($R_V$) in the diffuse interstellar medium 
$R_V = 3.1$ \citep{Schultz1975,Cardelli1989} with $A_G/A_V = 0.85926$\footnote{\href{http://stev.oapd.inaf.it/cgi-bin/cmd_3.3}{http://stev.oapd.inaf.it/cgi-bin/cmd\_3.3}}.
Figure~\ref{fig:map_and_histograms} depicts the $d_{\rm H}$ distribution of the RRLs in kpc, showing that our sample covers a wide range of distances, out to 
$\sim115$\,kpc. 
Additionally, Figure~\ref{fig:map_and_histograms} shows the DESI three-arm averaged signal-to-noise ratio (S/N) as a function of $d_{\rm H}$, which is  $\leq$10 for most stars beyond 50\,kpc. 

We note the presence of an extended overdensity near the celestial equator in Figure~\ref{fig:map_and_histograms}, with $d_{\rm H}$ ranging from $\sim50$--$70$\,kpc. 
This overdensity corresponds to the Sagittarius (Sgr) stream. 
A detailed description of the memebership determination of DESI RRLs to the Sgr stream is provided in Section~\ref{sec:field_gc_dwarf_stream}.

\subsection{Effective temperatures}
\label{sec:teff}

The \teff estimates for 
the RRLs in our sample relies largely on the temperature  measurements from the RVS pipeline and our knowledge of the ephemeris of each observation. 
It is important to recognize that the single-epoch \teff observations are not good estimates of the RRLs' mean \teff\ for a significant fraction of the RRL pulsation cycle, and their variation in temperature can easily reach 1000-2000\,K peak to peak \citep[see e.g.,][]{For2011,Pancino2015}.

\begin{figure*}
\includegraphics[angle=0,scale=.4]{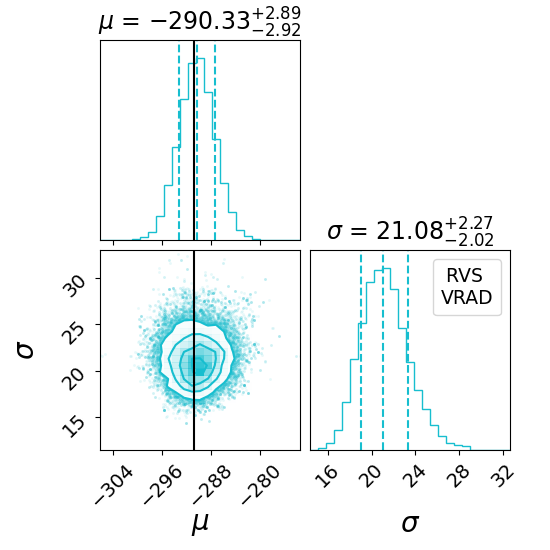}
\includegraphics[angle=0,scale=.4]{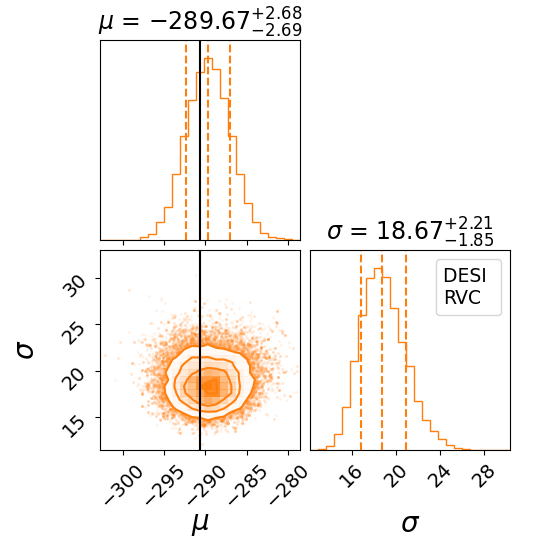}
\includegraphics[angle=0,scale=.4]{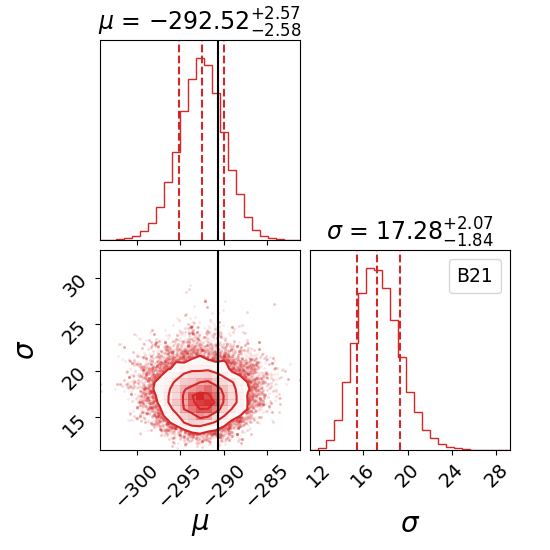}
\caption{
Posterior distribution from MCMC sampling of the mean velocity and velocity dispersion of RRLs in Draco dSph.  
This figure displays the sampling space obtained from applying our methodology to the uncorrected velocities from the RVS pipeline (in turquoise), to the velocities corrected by our Bayesian approach  (orange), and to those corrected using the radial velocity templates from B21 (red).
In each panel, a vertical line represents the line of sight velocity reported by \citet{Walker2015} ($-290.7\pm0.8$\,km\,s$^{-1}$). 
We note that a bias of 0.93\,km\,s$^{-1}$ exists in the determination of velocities with DESI \citep[][]{Koposov2024}, which affects the leftmost and middle panels.  
Considering this bias, our RVC-modeled velocities provides the best agreement with the results of \citet{Walker2015}.
}
\label{fig:vdisp_pal5_draco1}
\end{figure*}

\begin{figure}
\begin{center}
\includegraphics[angle=0,scale=.3]{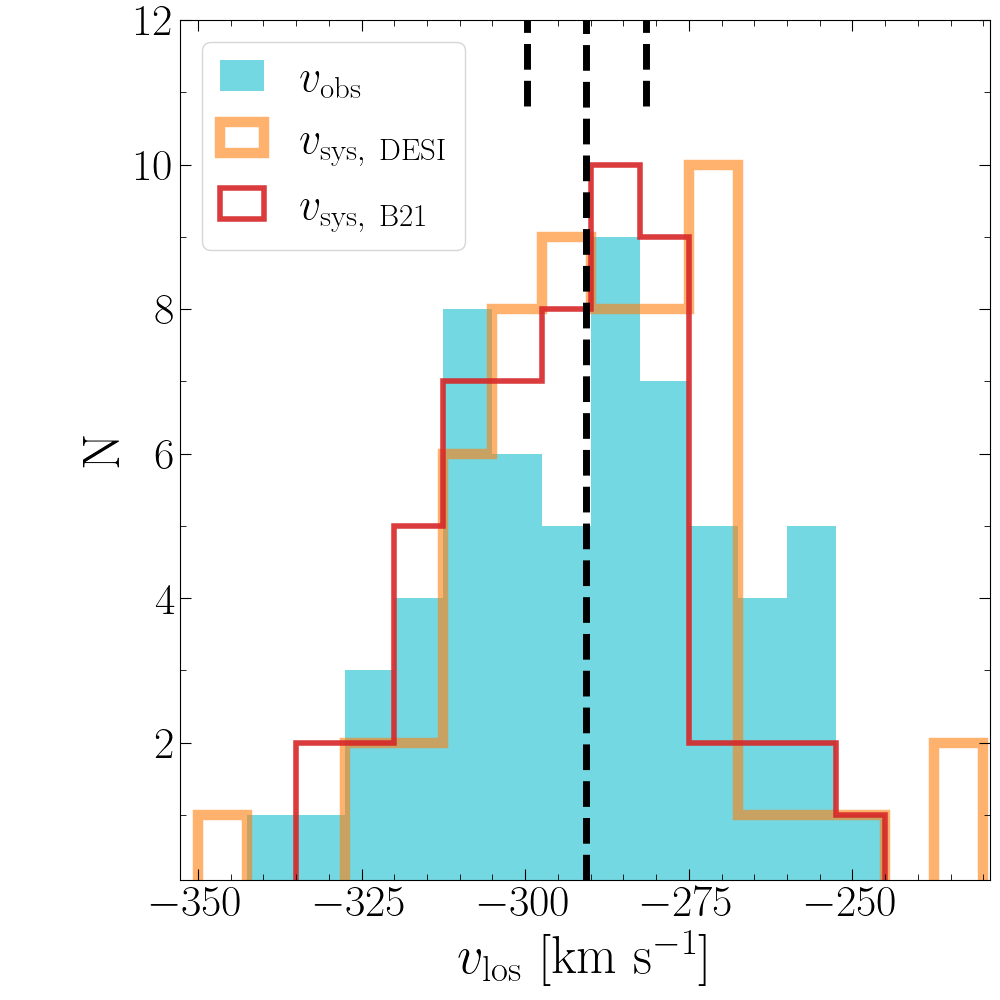}
\caption{
Line-of-sight velocity distribution of RRLs in Draco, for the RRLs' velocities taken directly from the RVSpec pipeline (turquoise filled bars), the velocities corrected using our RVC modeling (orange filled bars), and those corrected using the B21 RVC models (red unfilled bars).
A long dashed vertical line represents the $v_{\rm los}$ of Draco reported by \citet{Walker2015} ($-290.7\pm0.8$\,km\,s$^{-1}$), and short vertical lines depict their reported velocity dispersion ($-9.1\pm1.2$\,km\,s$^{-1}$). 
This plot shows that not accounting for the pulsation component of our sample results in a wide dispersion around the expected value, whereas narrower distributions (overall) are observed when using the corrected velocities. 
}
\label{fig:vdisp_draco2}
\end{center}
\end{figure}

We modeled the \teff variation of our sample in a similar manner as done for the systemic velocity estimates (Section~\ref{sec:desiRVC}), based on the derivation of variation curves. 
That is, we build the model

\begin{center}
\begin{align}
\begin{split}
P(T_{{\rm eff\ obs},i, j}|T_{{\rm eff\  mod, i}}(\phi_{j})) = f_{\rm out} \mathcal{N}(T_{{\rm eff\ obs}, i,j}|0, \sigma_{T_{{\rm eff,  out}}})   \\
 + (1-f_{\rm out}) \mathcal{N}(T_{{\rm eff\  obs}, i,j}|T_{{\rm eff\  mod}, i}(\phi_{j}), \sigma_{T_{{\rm eff,  RRL}, i, j  }} )
\end{split}
\end{align}
\end{center}
\label{eq:Teff_prob}

\noindent where $T_{\rm eff\ obs, i,j}$ is the temperature of a star $i$ observed at phases $\phi_j$ and 

\begin{align}
T_{{\rm eff\ mod},i}(\phi_{j}) = T_{{\rm eff\ sys}, i} + K_i\cdot V_{T_{\rm eff}}(\phi_j),  \label{eq:Teffcurves0}
\end{align}

\noindent with $T_{{\rm eff\ }{\rm sys}, i}$ defined from a Gaussian prior centered at 6000\,K with scatter $\sigma_{T_{{\rm eff}\ {\rm sys}}}$, $\mathcal{N}(6000, \sigma_{T_{{\rm eff}\ {\rm sys}}} )$, and

\begin{align}
    K_i &= 25\cdot10^{{\rm Amp}_{T_{\rm eff} }\cdot\frac{(A_{G, i}-1)}{2.5}} \\
V(\phi_j) &=  \sum_{k=1}^3 A_k(P) \cos(2\pi\phi_j k) + B_k(P) \sin(2\pi\phi_j k) . \label{eq:Teffcurves}
\end{align}

\noindent In this equation, we define the Fourier parameters $A_k$ and $B_k$ as a function of period $P$, so that $A_k = A_{k,1} + A_{k,2} P $ and $B_k = B_{k,1} + B_{k,2} P $. As in the case of the RVC modeling, in the previous equation $A_{G, i}$ represents the peak-to-peak $G-$band amplitude of the $i$-th RRL, and ${\rm Amp}_{T_{\rm eff}}$, $A_{k,1}$, $A_{k,2}$, $B_{k,1}$, and $B_{k,2}$ are the model parameters (i.e., these parameters are shared by the RRLs in the sample used to model the \teff curve). 
This means that the number of parameters of a model is equal to 
17 
(those in Equation~\ref{eq:Teffcurves}) plus the number of stars used for the model (that is, a single `systemic' \teff\ is estimated for each star). 
Similar to the procedure followed to estimate the RRLs' center-of-mass velocities (Section~\ref{sec:desiRVC}), we compute the best model parameters following a maximum likelihood approach using the Python interface to CmdStan, CmdStanPy.

\begin{table*}
\scriptsize
\caption{
Line of sight velocity ($v_x$) and velocity dispersion ($\sigma_{v_x}$) of the systems with RRLs analyzed in Section~\ref{sec:validation}. 
Here, $v_{\rm lit}$ is used to represent these values as listed in the literature. 
The references to previous works correspond to \citet{Walker2015} (W15), \citet{Baumgardt2018MNRAS.478.1520B} (BH18), and \citet{Simon2007ApJ...670..313S} (SG07). 
The velocity obtained from applying our MCMC methodology without correcting the velocity of our RRLs by their pulsating component is denoted by $v_{\rm RVS}$.  
The columns $v_0$ and $v_{\rm B21}$ represent our results from MCMC sampling the systems' velocity after correcting for the pulsation of the RRLs using our Bayesian model and the radial velocity curves of B21, respectively. 
We also include the number of stars $N$ used to compute the statistics of the DESI RRLs, and we highlight that only the statistics of Draco are robust. 
We note, however, the remarkable agreement between the mean metallicity estimates obtained from our RVC modeling ($v_0$) and those from the literature for the systems with $\geq4$ RRLs when the RVS bias ($-0.93$\,km\,s$^{-1}$) is considered.  
}
\label{tab:vdisp_comparison}
\begin{center}
\begin{tabular}{|c|c|c|c|c|c|c|cHH|c|c|c|}
\toprule
         Name & $v_{\rm lit}$  & $\sigma_{v_{\rm lit}}$ &  Reference &        $v_{\rm RVS}$ & $\sigma_{v_{\rm RVS}}$ &                  $v_0$ &      $\sigma_{v_0}$ &       $v_{\rm Balmer}$ & $\sigma_{v_{\rm Balmer}}$ &   $v_{\rm B21}$ & $\sigma_{v_{\rm B21}}$ &    N \\
          & (km\,s$^{-1}$) & (km\,s$^{-1}$)  &      &   (km\,s$^{-1}$)  & (km\,s$^{-1}$)  &                  (km\,s$^{-1}$)  &     (km\,s$^{-1}$)  &      (km\,s$^{-1}$) & km\,s$^{-1}$  &   (km\,s$^{-1}$)  & (km\,s$^{-1}$)  &     \\         
\hline
  Draco &      $-290.7\pm0.8$ &      $9.1\pm1.2$ &    W15 & $-290.3^{+2.9}_{-2.9}$ &   $21.1^{+2.3}_{-2.0}$ & $-289.7^{+2.7}_{-2.7}$ & $18.7^{+2.2}_{-1.8}$ & $-292.9^{+2.7}_{-2.7}$ &      $17.8^{+2.1}_{-1.9}$ & $-292.5^{+2.6}_{-2.6}$ &          $17.3^{+2.1}_{-1.8}$ & $59$ \\
       Pal~5 &       $-58.6\pm0.1$ &          $0.7\ $ & BH18 &  $-56.6^{+4.8}_{-4.8}$ &   $11.0^{+5.4}_{-3.1}$ &  $-57.9^{+2.0}_{-1.9}$ &  $0.2^{+1.5}_{-0.2}$ &  $-59.0^{+3.9}_{-4.2}$ &       $6.6^{+5.7}_{-4.3}$ &  $-55.5^{+2.8}_{-3.2}$ &           $1.2^{+5.6}_{-1.2}$ &  $5$ \\
    NGC~5904 &        $53.5\pm0.2$ &          $7.8\ $ &                                  BH18 &   $36.8^{+4.7}_{-4.5}$ &    $9.7^{+5.5}_{-3.0}$ &   $55.6^{+4.0}_{-3.3}$ &  $7.8^{+4.8}_{-2.6}$ &   $56.1^{+3.0}_{-2.5}$ &       $4.9^{+3.6}_{-1.8}$ &   $55.5^{+3.3}_{-2.2}$ &           $4.1^{+3.9}_{-2.2}$ &  $4$ \\
    NGC~5466 &       $106.8\pm0.2$ &          $1.6\ $ & BH18 &   $99.9^{+8.6}_{-9.5}$ &  $19.6^{+13.4}_{-6.6}$ &  $109.8^{+5.1}_{-2.9}$ &  $0.6^{+8.0}_{-0.6}$ & $112.4^{+11.2}_{-7.3}$ &    $14.8^{+15.7}_{-13.6}$ & $113.2^{+11.7}_{-9.0}$ &        $18.7^{+16.3}_{-11.0}$ &  $3$ \\
    NGC~5024 &       $-63.4\pm0.2$ &          $5.6\ $ &                                  BH18 &  $-55.0^{+9.0}_{-8.0}$ &  $17.8^{+11.5}_{-6.2}$ &  $-62.2^{+1.8}_{-2.1}$ &  $2.4^{+3.5}_{-2.3}$ &  $-63.6^{+1.3}_{-1.4}$ &       $0.2^{+1.2}_{-0.1}$ &  $-63.5^{+1.4}_{-1.5}$ &           $0.2^{+1.3}_{-0.2}$ &  $3$ \\
Ursa~Major~II &      $-116.5\pm1.9$ &      $6.7\pm1.4$ &     SG07 & $-104.7^{+9.1}_{-8.9}$ &  $19.4^{+17.5}_{-7.7}$ & $-122.6^{+2.2}_{-2.2}$ &  $0.2^{+1.7}_{-0.2}$ & $-119.8^{+1.6}_{-1.0}$ &       $0.9^{+2.7}_{-0.9}$ & $-119.7^{+1.6}_{-1.3}$ &           $0.9^{+3.4}_{-0.9}$ &  $2$ \\
    NGC~6341 &      $-120.5\pm0.3$ &          $8.7\ $ &                                  BH18 & $-138.5^{+8.0}_{-8.2}$ &  $17.3^{+15.1}_{-6.5}$ & $-121.2^{+2.3}_{-2.0}$ &  $2.0^{+4.5}_{-1.9}$ & $-122.0^{+3.8}_{-3.0}$ &       $0.5^{+4.6}_{-0.4}$ & $-119.4^{+3.1}_{-3.1}$ &           $0.3^{+3.0}_{-0.3}$ &  $2$ \\
    NGC~5053 &        $42.8\pm0.2$ &          $1.6\ $ & BH18 & $45.9^{+16.6}_{-16.8}$ & $37.2^{+25.8}_{-14.1}$ &   $43.9^{+1.6}_{-1.6}$ &  $0.2^{+1.3}_{-0.2}$ &   $38.9^{+1.7}_{-1.7}$ &       $0.2^{+1.5}_{-0.2}$ &   $40.0^{+2.2}_{-2.2}$ &           $0.2^{+1.9}_{-0.2}$ &  $2$ \\

\hline 
\end{tabular}
\end{center}
\end{table*}

To compute \teff\ curves, we split our sample into fundamental mode and first-overtone pulsators. 
Furthermore, for the modeling process we only consider observations with no warnings from the RVS pipeline, with $5500< $ \teff\ (K) $<8500$ to avoid outliers, and with  \teff\ uncertainties $<300$\,K.   
The adopted priors for our model and the resulting best parameters are provided in Table~\ref{tab:teffc_params}.

Similar to Figure~\ref{fig:rvcs}, Figure~\ref{fig:teffcs} shows RRab \teff\ curve templates for four periods, and RRc \teff\ curve templates for two periods, together with the relative \teff\ of every measurement in our sample
(with respect to the mean \teff, for each star).
The figure depicts \teff\ variations of $\sim300$--$1000$\,K throughout the pulsation cycle of RRab stars, and a more consistent variation of $\sim600$\,K for RRc stars.
We note that, for RRab, the derived curves fail at modeling the \teff\ increase immediately after the phase of maximum compression, or minimum atmospheric kinetic energy, where the stars rapidly approach the phase of maximum light (at $\phi\gtrsim0.9$).
Regarding the level of scatter of \teff\, variation of individual stars, the bulk of the $\Delta\teff$ distribution  reaches $\sim700$\,K below the mean temperature for both for RRab and RRc stars, and $\sim1000$\,K and $\sim500$\,K above it for RRab and RRc stars, respectively (that is, the \teff\ variations are larger for the former).

As for the mean \teff\ uncertainties, unlike our systemic velocity estimates, we do not find strong correlations with the number of epochs for both RRab and RRc stars.
In fact, the median \teff\ uncertainty is of $\sim30$--$35$\,K in both cases, regardless of the number of epochs.
We do, however, observe that stars with less than 4 epochs are more prone to outliers in the \teff\ uncertainty distribution, and the likelihood of obtaining outliers in this distribution decreases as the number of epoch increases. 

In terms of residuals (observed vs. predicted \teff), 
we observe that the shape of the distribution is a function of $\phi$, with a median of $\sim45$\,K.
If we split the pulsation cycle in phase bins of width 0.05, we find a clear correlation between the residuals and $\phi$. 
Indeed, the median of the $\phi$-dependent residual distributions is predominantly $>0$\,K and varies between $\sim-25$ and $\sim100$\,K (the latter observed near $\phi=0.95$). 
The residual dispersion ($1\sigma$ region) is also a function of $\phi$ for RRab, and reaches a maximum of close to 500\,K in the $\phi$ range [0.90-1.00]. 
For RRc stars, the median is consistently between 10 and 35\,K across their pulsation cycle, with $1\sigma$ dispersion regions between 30 and 150\,K (depending on $\phi$). Both the median and the dispersion of the residuals reach a maximum at $\phi\sim0.85$. 
We thus conclude that our mean \teff\ estimates are reliable, although caution should be considered in the $0.85<\phi<1.00$ range (for both single-epoch-corrected \teff\ and for our modelled \teff\ curves).

\begin{figure*}
\includegraphics[angle=0,scale=.4]{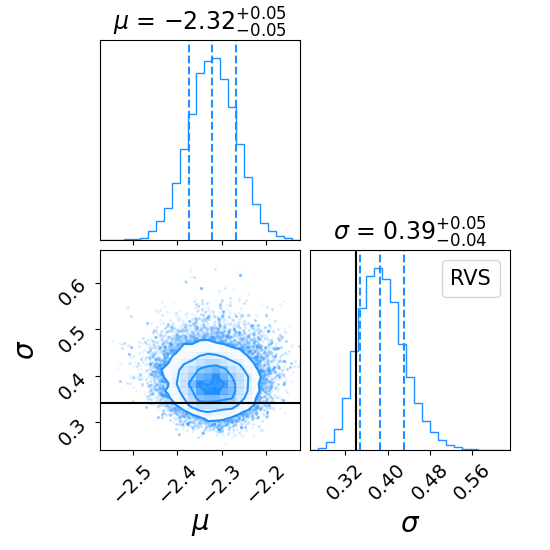}
\includegraphics[angle=0,scale=.4]{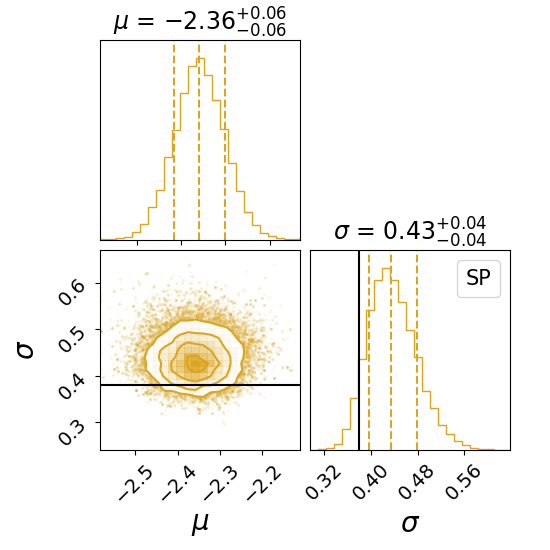}
\includegraphics[angle=0,scale=.4]{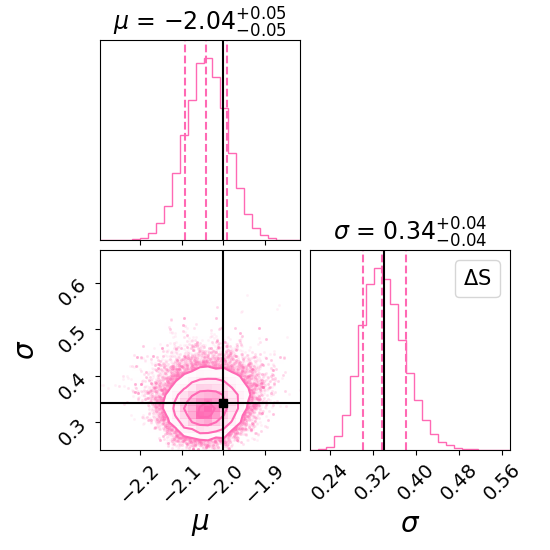}
\caption{
Same as Figure~\ref{fig:vdisp_pal5_draco1} but showing the posterior of the metallicity distribution (mean $\mu$ and dispersion $\sigma$) of RRLs in Draco. 
The metallicities shown correspond to those obtained from the RVS and the SP pipelines (left and center panels, respectively), and from the $\Delta$S method (right). 
The literature value for the mean metallicity of Draco ($-2.00\pm0.02$\,dex) and its dispersion ($0.34\pm0.02$\,dex) are displayed with solid lines, showing that the $\Delta$S metallicity estimates are the most consistent with these values. 
For the left and center panels, the literature value falls outside the displayed region for $\mu$. 
}
\label{fig:metdisp_pal5_draco1}
\end{figure*}

\section{Validation of spectroscopic properties}
\label{sec:validation}

\subsection{Selection of RRLs in the Galactic field and in substructures}
\label{sec:field_gc_dwarf_stream}

To identify RRLs in the Galactic halo field, and to distinguish them from RRLs belonging to known globular clusters and dwarf galaxies, we perform a crossmatch with the catalogs of stars in globular clusters and in dwarf galaxies compiled by \citet{vasiliev_GCs} and \citet{Pace2022}, respectively. 
Considering that the astrometric information in these catalogs are also based on {\it Gaia} data, the crossmatch is performed using a 0.5\,arcsec search radius, ensuring a pure sample of RRLs dwarf galaxies and clusters (although this might slightly increase the contamination in our Galactic field sample). 
We should note, however, that no special distinctions for RRLs were made in the compilation of the catalog (e.g., considering their pulsating nature), and therefore membership to these systems is tentative. 
From this crossmatch, we find a total of 22 RRLs in nine Milky Way star clusters (Pal~5, NGC~5904, NGC~5466, NGC~5024, NGC~5053, NGC~6341, NGC~2419, NGC~4147, and NGC~5634) and 62 RRLs in three dwarf galaxies (Draco, Ursa Major II, and Bo{\"o}tes III).
Of these 62 RRLs, 59 are found in the Draco dwarf spheroidal galaxy (dSph). 

The DESI Y1 footprint significantly overlaps with that of the Sgr stream \citep{Cooper2023}. 
Similar to \citet{Erkal2021} and \citet{Bystroem2024}, we label stars as candidate Sgr stream members based on their position relative to the stream, their heliocentric distance, and their radial velocity. 
For the positions, 
we transform the stars' equatorial coordinates into a Sgr stream coordinate system $\tilde{\Lambda}_{\odot}$, $\tilde{B}_{\odot}$. 
These coordinates represent stream longitude and latitude, respectively. 
We follow the convention of \citet{Belokurov2014} and adopt 
a stream longitude $\tilde{\Lambda}_{\odot}$ that increases in the direction of motion of Sgr.
Therefore, $\tilde{\Lambda}_{\odot} {\rm [deg]} = 360-{\Lambda}_{\odot}$, where ${\Lambda}_{\odot}$ is the Sgr stream longitude defined by \citet{Majewski2003}.  
For the distances, we use the values defined in Section~\ref{sec:dist} and label stars within three standard deviations from the distance splines of \citet{Hernitschek2017} (which reach distances out to $\sim100$\,kpc) and with $|\tilde{B}_{\odot}|<15$\,deg. 
For the radial velocities, we use the center-of-mass velocities obtained in Section~\ref{sec:desiRVC} and select stars within 3.5 standard deviations from the values reported by \citet{Vasiliev2021} as candidate Sgr stream stars.
Following these definitions, we find a total of 600 RRLs that are candidate members of the Sgr stream. 
We note that 246 of these stars (and 292 of our full sample) display membership probabilities larger than 0.5 in the catalog of Sgr members of \citet{Ramos2022}, when using {\it Gaia}'s {\sc source\_id} for the crossmatch.

Consequently, the list of RRLs that are not in the aforementioned clusters or galaxies, nor in the Sgr stream (from our definition of Sgr membership), contains 5554 stars.
This corresponds to 89\% of the full sample. 
We refer to this sample as the DESI field RRLs from hereon. 

\subsection{Draco as a testbed for parameter comparisons}

In order to validate our results and to test the performance of our methodology quantifying the effect of our assumptions, we compare our velocity and [Fe/H] estimates with literature values for the Draco dSph. 
We selected this galaxy as it contains the largest number of RRLs of the globular clusters and dwarf galaxies observed by DESI (see Section~\ref{sec:field_gc_dwarf_stream}). 
The catalog resulting from the crossmatch with the compilation of dwarf galaxies properties by \citet{Pace2022}, contains their main chemo-dynamical properties.
This includes proper motions, line-of-sight velocities, velocity dispersion, metallicities, and metallicity dispersion.
It is worth mentioning that the sample of Draco stars in DESI extends further in radius from its center than previous studies \citep[e.g.,][]{Walker2015}. 
Therefore, special considerations should be made given the location of our RRL sample in the outskirts of Draco (as discussed in Section~\ref{sec:dracofeh}).
Additionally, the catalog by \citet{Pace2022} lists likely member stars with individually-measured membership probabilities ($P_{\rm memb}$). 
Here, we select RRLs with $P_{\rm memb}>0.90$. 
We note in passing that variable stars are not considered special cases for the computation of these probabilities (e.g., to correct their observed velocities for their pulsation), and therefore, their true membership is not guaranteed.

\begin{table*}
\scriptsize
\caption{
Similar to Table~\ref{tab:vdisp_comparison}, but comparing the metallicity and metallicity dispersion $\sigma$ obtained in this work (with subscripts ${\rm RVS}$, ${\rm SP}$, and ${\Delta\rm S}$) for RRLs in Draco and in globular clusters with those from the literature (${\rm [Fe/H]}_{\rm Lit}$ and ${\rm [Fe/H]}_{\rm Lit}$). 
In this table, we compare with the metallicities reported by \citet{Kirby:2013} (K13), \citet{Bailin2019} (B19), and \citet{Bailin2022} (B22).
}
\label{tab:metdisp_comparison}
\begin{center}
\begin{tabular}{|c|c|c|c|c|c|c|c|c|c|c|}
\toprule
Name & [Fe/H]$_{\rm Lit}$ & $\sigma_{{\rm [Fe/H]}_{\rm Lit}}$ &                                  ref &      [Fe/H]$_{\rm RVS}$ & $\sigma_{{\rm [Fe/H]}_{\rm RVS}}$ &       [Fe/H]$_{\rm SP}$ & $\sigma_{{\rm [Fe/H]}_{\rm SP}}$ & [Fe/H]$_{\Delta {\rm S} }$ & $\sigma_{{\rm [Fe/H]}_{\Delta {\rm S} }}$ &    N \\       
 & (dex) & (dex) &              &      (dex) & (dex) &      (dex) & (dex) & (dex) & (dex) &   \\      
\hline
     Draco &     $-2.00^{+0.02}_{-0.02}$ &                     $0.34\pm0.02$ &  K13 & $-2.32^{+0.05}_{-0.05}$ &            $0.39^{+0.05}_{-0.04}$ & $-2.36^{+0.06}_{-0.06}$ &           $0.43^{+0.04}_{-0.04}$ &    $-2.04^{+0.05}_{-0.05}$ &                   $0.34^{+0.04}_{-0.04}$ & $59$ \\
       Pal~5 &            $-1.32\pm0.10$ &                                -- & B22 & $-1.45^{+0.05}_{-0.04}$ &            $0.09^{+0.05}_{-0.03}$ & $-1.48^{+0.03}_{-0.03}$ &           $0.07^{+0.04}_{-0.02}$ &    $-1.28^{+0.07}_{-0.08}$ &                   $0.15^{+0.10}_{-0.06}$ &  $5$ \\
    NGC~5904 &            $-1.25\pm0.10$ &                                $0.03^{+0.00}_{-0.00}$ & B22 & $-1.44^{+0.02}_{-0.02}$ &            $0.02^{+0.02}_{-0.01}$ & $-1.50^{+0.03}_{-0.03}$ &           $0.06^{+0.04}_{-0.02}$ &    $-1.22^{+0.11}_{-0.11}$ &                   $0.21^{+0.17}_{-0.09}$ &  $4$ \\
    NGC~5466 &            $-2.20\pm0.10$ &                                -- & B22 & $-2.09^{+0.07}_{-0.07}$ &            $0.11^{+0.11}_{-0.04}$ & $-2.06^{+0.06}_{-0.06}$ &           $0.10^{+0.11}_{-0.04}$ &    $-1.80^{+0.14}_{-0.12}$ &                   $0.17^{+0.20}_{-0.09}$ &  $3$ \\
    NGC~5024 &            $-1.97\pm0.10$ &                               $0.07^{+0.01}_{-0.01}$ & B22 & $-2.28^{+0.03}_{-0.03}$ &            $0.03^{+0.04}_{-0.01}$ & $-2.32^{+0.02}_{-0.02}$ &           $0.03^{+0.03}_{-0.01}$ &    $-2.08^{+0.09}_{-0.09}$ &                   $0.06^{+0.14}_{-0.04}$ &  $3$ \\
    NGC~6341 &            $-2.24\pm0.03$ &                                $0.08^{+0.03}_{-0.01}$ & B19 & $-2.50^{+0.04}_{-0.04}$ &            $0.04^{+0.11}_{-0.03}$ & $-2.58^{+0.14}_{-0.15}$ &           $0.17^{+0.39}_{-0.09}$ &    $-2.12^{+0.10}_{-0.09}$ &                   $0.09^{+0.25}_{-0.07}$ &  $2$ \\
    NGC~5053 &            $-2.23\pm0.10$ &                                $0.04^{+0.02}_{-0.00}$ & B22 & $-2.55^{+0.10}_{-0.08}$ &            $0.10^{+0.23}_{-0.06}$ & $-2.60^{+0.18}_{-0.16}$ &           $0.19^{+0.45}_{-0.10}$ &    $-2.12^{+0.08}_{-0.08}$ &                   $0.05^{+0.17}_{-0.04}$ &  $2$ \\
\hline 
\end{tabular}
\end{center}
\end{table*}

\subsubsection{Line-of-sight velocity}
\label{sec:dracovlos}

To compute the mean observed velocity of a group of RRLs in Draco 
(its systemic velocity) and its velocity dispersion, we employ the MCMC Python parameter-space sampler \code{emcee}\footnote{\href{https://github.com/dfm/emcee}{https://github.com/dfm/emcee}} \citep{emcee}. 
This computation relies on the assumption that the velocities in the system follow a Gaussian distribution, which can be described by a mean systemic velocity $\mu$ and an intrinsic dispersion $\sigma_v$. 
Then, following \citet{Walker2006}, we adopt a Gaussian likelihood function of the form:

\begin{equation}
\log\ \mathcal{L} = -\frac{1}{2} \left( 
\sum_{i=1}^{N} {\rm log}(\sigma_v^2+\sigma_{v_i}^2) + \sum_{i=1}^{N} \frac{ (v_i-\mu)^2 }{\sigma_v^2+\sigma_{v_i}^2}
\right),
\label{eq:vdispmcmc}
\end{equation}

\noindent where $v_i$ and $\sigma_{v_i}$ refer to the velocity of a given star in the system and its measured uncertainty. 
For this process, we use 50 walkers, a burn-in period of 100 steps, and a total number of steps of 1000.
These choices ensure the convergence of the MCMC routine. 
Finally, we report the values corresponding to the median of the posterior distributions, adopting the 16th and 84th percentile limits as lower and upper uncertainties, respectively. 
We perform this analysis for 
a) our uncorrected RVS velocities, 
b) our velocities corrected by the Bayesian approach described in Section~\ref{sec:desiRVC}, and
c) our velocities corrected by the RVCs of B21 (Section~\ref{sec:B21RVC}). 
Our results are summarized in Table~\ref{tab:vdisp_comparison}. 
Figure~\ref{fig:vdisp_pal5_draco1} shows the mean velocity and the velocity dispersion obtained with this method for Draco.  
For this analysis, we do not take into account the bias caused by the spatial distribution of RRLs in Draco, as stars located in its outskirts might exhibit higher velocity dispersions than stars closer to its center \citep[see e.g.,][]{Walker2007}.

From the values in Table~\ref{tab:vdisp_comparison}, we conclude that correcting for the pulsation of the RRLs plays a significant role in correctly estimating the mean velocity of a system, regardless of the number of RRLs used (although we should highlight the impact of low number statistics on this conclusion). 
A notable result is that we are able to recover the line of sight velocity of Draco ($-290.7\pm0.8$\,km\,s$^{-1}$, \citealt{Walker2015}) without needing to apply the pulsation correction. 
At first sight, this estimate is even better than those resulting from the corrected velocities. 
We note, however, that a 0.93\,km\,s$^{-1}$ bias was detected by \citet{Koposov2024}, when comparing RVS-based velocities in DESI with APOGEE spectra (DESI's velocities are 0.93\,km\,s$^{-1}$ larger than those in APOGEE). 
This bias does not affect our velocities inferred with the B21 RVCs, as the line-by-line velocities used for this correction are computed independently from RVS.
When applying the velocity bias to our estimates, we find that the systemic velocities of Draco are $-291.12$\,km\,s$^{-1}$, $-290.59$\,km\,s$^{-1}$, and $-292.52$\,km\,s$^{-1}$ when using uncorrected, DESI RVS-corrected, and B21-corrected RRL velocities, respectively.
Thus, our RVS-corrected velocities are in remarkable agreement with the reported velocity of Draco.
This is the case for most of the systems shown in Table~\ref{tab:vdisp_comparison} as well, when the 0.93\,km\,s$^{-1}$ bias is considered.  
We also note that we obtain the largest velocity dispersion when using uncorrected velocities. 
Lastly, the velocity dispersion observed in our corrected samples is consistent with the velocity dispersion  found in Draco by \citet{Walker2015} ($9.1\pm1.2$\,km\,s$^{-1}$) and the uncertainties of our systemic velocities (typically $<15$\,km\,s$^{-1}$ for the stars in Draco), added in quadrature.

\begin{figure*}
\includegraphics[angle=0,scale=.42]{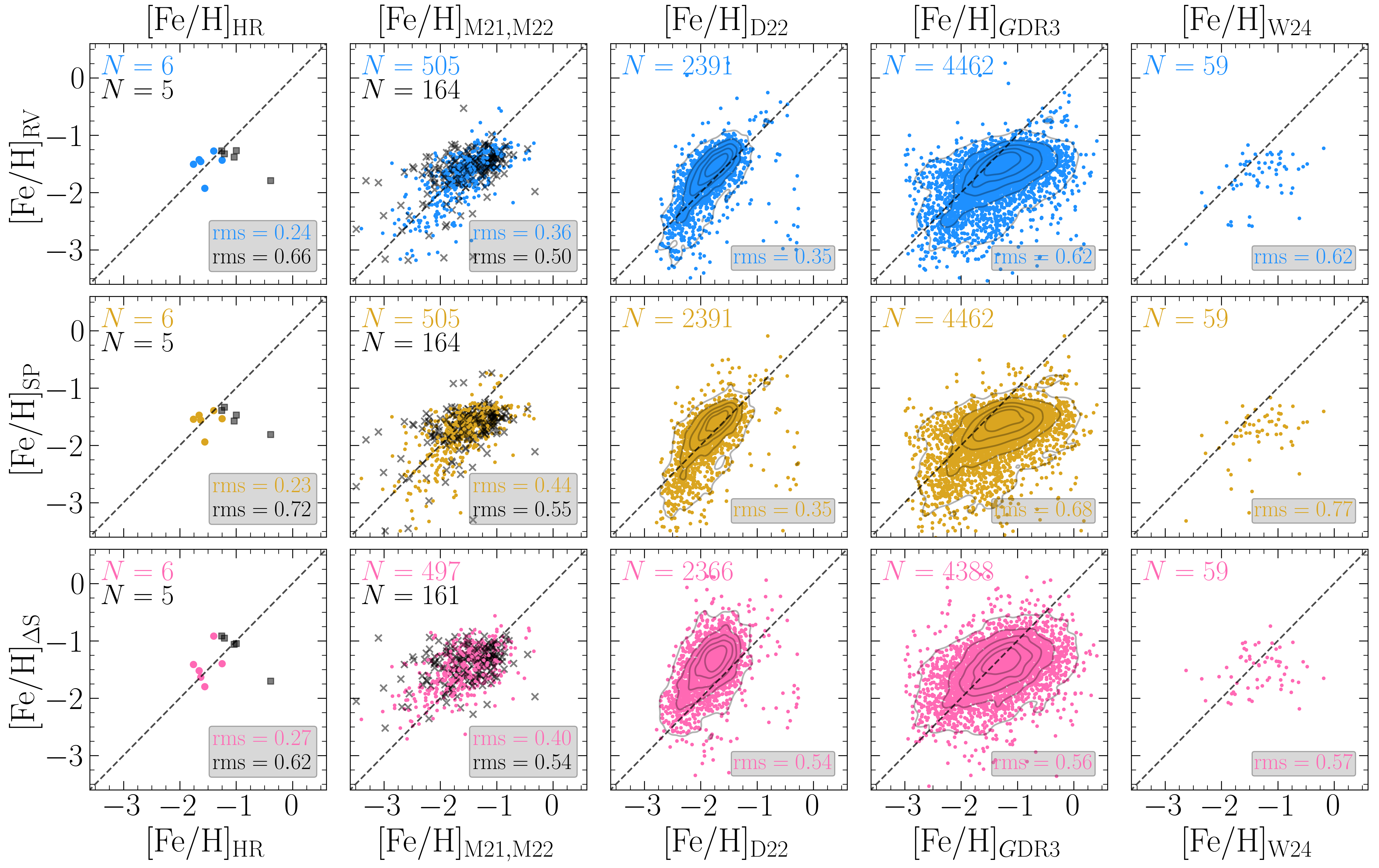}
\caption{
Comparison of the [Fe/H] derived by the DESI pipelines for our sample of RRLs with those from the literature. 
In the leftmost column,  we show the comparisons between our [Fe/H] and those from high-resolution studies including those tabulated by \citet{Crestani2021a} and  \citet{Crestani2021b} (filled circles, CHR calibration sample), and APOGEE (squares). 
The second column (from left to right) shows the result of a crossmatch with the photometric metallicities derived by \citet{Mullen2021} (for RRab, M21) and \citet{Mullen2022} (for RRc, M22), respectively. 
In these panels, filled circles represent stars with [Fe/H] computed from optical light curves by these authors, whereas crosses are used for their metallicities from infrared light curves. 
The middle column displays a comparison with the photometric metallicity predictions of RRab using {\it Gaia} optical and $K_S$ infrared bands from \citet{Dekany2022} (D22).
The fourth column shows the results from the comparison with the metallicities in the {\it Gaia} DR3 catalog. 
The rightmost column depicts the comparison with the mean [M/H] derived by \citet{Wang2024} (W24) using a template-matching method on LAMOST low- and medium-resolution spectra. 
A diagonal dashed line is used in each case to represent the identity relation. 
}
\label{fig:met_comparisons}
\end{figure*}

The velocity spread of our measurements, as observed in the distribution of individual stars' velocities, also represents a benchmark for assessing the precision and accuracy of our line-of-sight velocity corrections. 
Since this metric is highly dependent on the number of stars, we only analyze the velocity dispersion of RRLs in Draco. 
Figure~\ref{fig:vdisp_draco2} depicts the distribution of our three samples of RRL velocities. 
As it can be inferred from Table~\ref{tab:vdisp_comparison} and directly observed in Figure~\ref{fig:vdisp_draco2}, the velocity distributions from both of our estimates of (corrected) systemic velocities are narrower than for uncorrected RRLs, although in all cases the median of the distributions are in good agreement with each other. 
For the uncorrected velocities, the width of the distribution measured from its 
16th and 84th percentiles, corresponds to 21.8 and 22.0\,km\,s$^{-1}$, respectively.  
For the velocities inferred from our Bayesian approach, the velocity spread is reduced to 
15.1 and 20.2\,km\,s$^{-1}$, equivalent to a 10-30\% decrease. 
It is noteworthy, however, that we find two stars in the distribution of DESI RVC-corrected velocities (with {\it Gaia} IDs 1433146680194267520 and 1433157675310524288) for which the correction results in larger deviations with respect to the bulk of the distribution. 
For these stars, observed five times each, most of the epochs were taken near $\phi\sim1.0$ and $0.75$, respectively. 
The case of the B21-template-corrected velocities is similar to the previous one, as we also observe a significant reduction of the width of the distribution (to 14.5 and 18.6\,km\,s$^{-1}$, or 15-30\% decrease).

\subsubsection{Metallicity}
\label{sec:dracofeh}

Similar to the line of sight velocities comparison, we estimate the mean metallicity and the metallicity dispersion of the RRLs in Draco using MCMC. 
For this exercise, we employ the iron abundances obtained with DESI's RVS and SP pipelines, and from the $\Delta$S method. 
The resulting posterior distributions are depicted in Figure~\ref{fig:metdisp_pal5_draco1}.

From the figure, it is clear that the mean of the [Fe/H] distribution of the RRLs in Draco is systematically more metal poor than the value reported by \citet{Kirby:2013} (${\rm [Fe/H]}=-2.00\pm0.02$\,dex) for the RVS and SP-derived metallicities (by 0.32 to 0.36\,dex, respectively). 
The metallicity dispersion is also larger for the values derived by these pipelines, as compared with that of \citet{Kirby:2013} although the RVS metallicity dispersion is just marginally higher. 
We find the best agreement with the \citet{Kirby:2013} values when using the $\Delta$S values, both for the mean of the metallicity distribution (which is marginally more metal poor, by $0.04$\,dex) and for the metallicity dispersion (which is the same as that literature value). 

This is also the case if we compare with the mean metallicity of Draco found by \citet{Kirby2011} ([Fe/H] $=-1.93\pm0.01$\,dex). 
A tentative explanation for the discrepancy between the DESI (RVS and SP pipeline) metallicities is the metallicity gradient found in Draco by \citet{Kirby2011}, of $-0.73$\,dex\,deg$^{-1}$.
Indeed, the mean distance of our RRLs to the dwarf's center (RA, DEC) $=(260.0684$\,deg, $57.9185$\,deg$)$ \citep{Munoz2018} 
is 0.18\,deg, whereas the sample of \citet{Kirby2011} is constrained to a region within $\sim0.1$\,deg.
Adopting their metallicity gradient results in a decline of 0.13\,dex in metallicity at 0.18\,deg from the center of Draco, which is consistent with our $\Delta$S measurement. 
Moreover, the median [Fe/H] from the RVS, SP, and $\Delta$S method when only considering RRLs within 0.1\,deg are $-2.18$, $-2.17$, and $-1.87$\,dex, respectively, which shows a $\sim0.25$\,dex difference for the two first cases when compared with the results by \citet{Kirby2011}.
These values can also be compared with the RVS metallicities of the Draco members detected in DESI analyzed by \citet{Koposov2024} (i.e., not only RRLs), $-2.25$\,dex (or $-1.97$\,dex within 0.1\,deg).  
Thus, the disagreement between our RVS and SP metallicities and those from the literature are likely not only explained by the metallicity gradient of Draco, but also by the fact that the short-period pulsation of RRLs is not accounted for in these pipelines.

Following the same methodology, we compare the metallicity and metallicity dispersion of globular clusters with at least two RRLs in our sample with literature values (similar to Table~\ref{tab:vdisp_comparison}), taking into account that our estimations are affected by low number statistics.
As shown in Table~\ref{tab:metdisp_comparison}, 
both RVS and SP metallicities are underestimated at a $\sim0.15$-$0.20$\,dex level at ${\rm [Fe/H]}\sim -1.30$\,dex \citep[for systems like Pal~5 and NGC~5904;][]{Bailin2022}, and $\sim0.20$-$0.40$\,dex for ${\rm [Fe/H]}\leq -1.8$\,dex (for systems like NGC~6341 and NGC~5053 at ${\rm [Fe/H]}\sim -2.25$\,dex; \citealt{Bailin2019,Bailin2022}).
Conversely, for the [Fe/H] obtained with the $\Delta$S method we find an overestimation of the system's metallicities at a $\sim0.05$\,dex level at ${\rm [Fe/H]}\sim -1.30$\,dex, and $\sim0.10$-$0.40$\,dex for the more metal-poor systems.

\subsection{Comparison with [Fe/H] in literature catalogs}
\label{sec:metallicity_comparison}

To evaluate the performance of our determined iron abundances and to find any existing trends in our data, we compare our metallicity estimates with those derived in the literature for RRLs from low-, medium-, and high-resolution spectra. 
For this comparison, we again use the [Fe/H] obtained with the DESI MWS processing pipelines (RVS and SP) and those from the $\Delta$S method. 
We crossmatch the DESI Y1 RRL catalog with existing catalogs containing large numbers of halo RRLs.
Here we consider those derived from high-resolution spectra from \citet{Crestani2021a} and from APOGEE \citep[][]{Majewski2017,Beaton2021}, 
those from the photometrically-derived (from optical to infrared light curves) metallicities of \citet{Mullen2021} (RRab), \citet{Mullen2022} (RRc), and \citet{Dekany2022} (RRab), and the medium- to low-resolution-based metallicities from {\it Gaia} DR3 \citep[][]{Gaia2023,Clementini2023} and LAMOST \citep[][]{Wang2024}. 
We note that, for the 59 stars in common with \citet{Wang2024}, the median number of epochs is three, with a maximum of 10 epochs and nine stars having more than 7 epochs. 
For the stars in this sample, we use the mean of the provided [M/H] for our comparison. 
The result of these comparisons is shown in Figure~\ref{fig:met_comparisons}, along with the root mean square (rms) of the stellar metallicities used in the plots. 

Figure~\ref{fig:met_comparisons} shows that the rms of the comparisons lie typically between 0.3 and 0.7\,dex. 
From the comparison with the high-resolution sample, we find that the rms of each of our metallicity estimates is $<0.30$\,dex (without counting the APOGEE star with {\it Gaia} {\sc source\_id} 4420736655826086400, which is an outlier), although the very small number statistics used for this computation should be considered.  
From the comparison with photometric metallicities of \citet{Mullen2021} and \citet{Mullen2022}, we observe the rms to vary between $\sim0.35$ and $\sim0.45$\,dex, with a consistently larger scatter for metallicities from mid-infrared light curves (in the range 0.50-0.55\,dex). 
Our [Fe/H] estimates are systematically higher than those of \citet{Dekany2022} (0.15, 0.40, and 0.10\,dex more metal-rich, as measured by the median differences), regardless of the method used. 
We note that a similar trend was observed by these authors when comparing their metallicities with those estimated by \citet{Hajdu2018} from $K_{S}-$band light curves (a $\sim0.4$\,dex offset in the same direction as our results) and \citet{Iorio2021} from {\it Gaia} $G-$band light curves (0.2–0.3\,dex, also in the same direction). 
The correlation between our measurements are clearer for RV and SP (with rms $\sim0.35$\,dex), with a larger rms for our $\Delta$S estimates ($\sim0.55$\,dex). 
Both our RV and SP metallicities are systematically more metal-poor than those from {\it Gaia} (with median [Fe/H] differences of 0.30, 0.10, and 0.40\,dex), and the data shows large rms in each case ($>0.5$\,dex). 
This is also seen for the comparison with the metallicities from \citet{Wang2024} (with median [Fe/H] differences of 0.35, 0.15, 0.45\,dex), and in this case we observe marginal evidence of a correlation with our $\Delta$S values. 
In short, our derived [Fe/H] are consistent overall (at a $\lesssim0.25$\,dex level) with high-resolution metallicities (subject to low number statistics), and our RVS and SP metallicities show the largest differences with those from {\it Gaia} DR3 and the values from LAMOST based on template fitting.  
For photometric metallicities, we find a better agreement with those derived from optical and near-infrared band light curves than those from mid-infrared light curves.

\begin{figure}
\begin{center}
\includegraphics[angle=0,scale=.37]{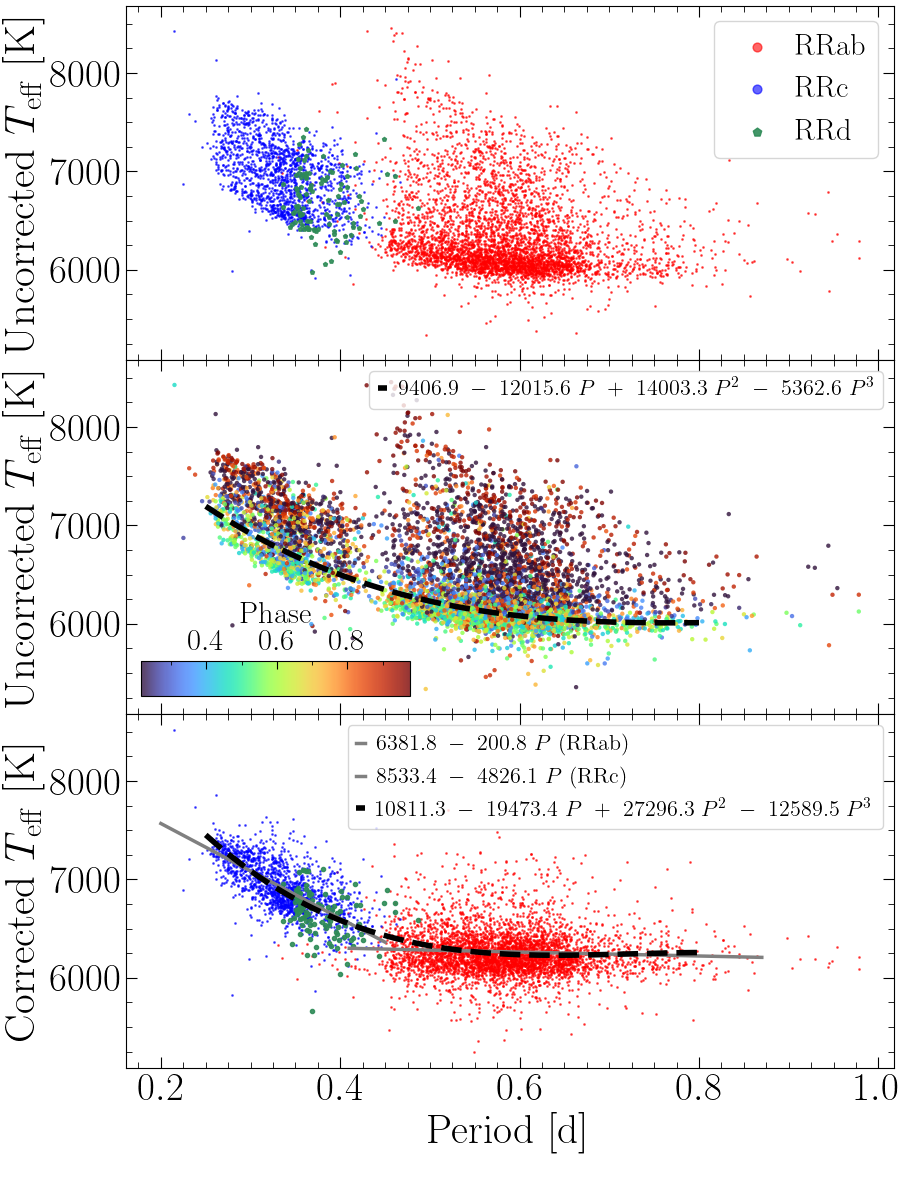}
\caption{
The period-\teff\ anticorrelations of RRLs observed in the DESI catalog. 
The {\it top} (color-coded by their pulsation modes) and {\it middle} (color-coded by their phases) panels depict these quantities for uncorrected temperature estimates, and the {\it bottom} panel shows the temperatures corrected by our \teff\ variation model. 
Applying our model significantly reduces the observed scatter in temperature for both fundamental and first-overtone pulsators, making the observations consistent with the anticorrelations predicted by pulsation theory, although the \teff\ decrease in RRab results in a less clear decreasing trend with period.  
In the {\it bottom} panel, linear functions are used to fit the observed trend for RRab and RRc separately.  
The full dataset, without classification distinctions, is successfully represented by a single third-degree polynomial.
}
\label{fig:perteff}
\end{center}
\end{figure}

\subsection{The Period-\teff\ relation}

According to stellar evolution theory, not only do RRLs occur in a narrow region of the Hertzprung-Russel diagram (the instability strip), but also their sizes (density) and periods evolve as they move along the horizontal branch \citep[according to the pulsation equation; see][and references therein]{Catelan2015}. 
As a consequence, a correlation between periods of pulsation and \teff\ would be expected for stars along the horizontal branch, and checking for such correlation is a suitable test for our \teff\ variation models. 

Figure~\ref{fig:perteff} shows the position of our sample in the $P$-\teff\ space, for temperatures directly taken from the RVS pipeline (uncorrected) and for temperatures corrected by our \teff\ modeling (Section~\ref{sec:teff}). 
The figure shows that the vast majority of \teff\ measurements lying above the expected polynomial decrease of \teff\ with period are caused by epochs at phases $\lesssim0.4$ and $\gtrsim0.65$. 
The figure also illustrates a clear improvement in the distinctness of the $P$-\teff\ sequences for our sample, resulting from narrower \teff\ distributions as a function of $P$. 
Indeed, for RRc stars, implementing the \teff\ correction makes a significant difference in reducing the scatter of the anticorrelation. 
For RRab, on the other hand, the correction comes at a cost, as after applying it the quadratic-like $P$-\teff\ part of the sequence becomes less clear, even though the presence of high temperature estimates is significantly reduced.  
The broadening of the RRab region can be attributed to the intrinsic scatter in our \teff\ corrections. 
Furthermore, we observe that the feature of decreasing \teff\ for short period RRab ($<0.4$\,d), visible in the top and bottom panels of Figure~\ref{fig:perteff}, is lost after applying the correction, and is likely as a consequence of the
larger range of observed \teff\ and the low number of RRab in these period ranges. 

To characterize these anticorrelations, we employ {\sc emcee} adopting a third-order polynomial to model the period-\teff\ relation for RRab and RRc stars. 
We find that 32 walkers and chains with 400 steps are sufficient to appropriately sample the parameters' posterior distributions. 
For the uncorrected sample, if only \teff\ measurements for phases between 0.4 and 0.65 are considered, the resulting relation is

\begin{equation}
\begin{array}{lc}
T_{\rm eff} {\rm [K]}=9406.9_{-299.9}^{+1072.6} \ -\ 12015.6_{-6627.4}^{+1841.7}\ P\ \\ \ \ \ \ \ \  +\ 14003.3_{-3523.0}^{+12567.3}\ P^2\ -\ 5362.6_{-7469.3}^{+2228.8}\ P^3.
\end{array}
\label{eq:perteff_all_uncor}
\end{equation}

\noindent If instead we employ our entire dataset of corrected \teff\ estimates, we obtain 

\begin{equation}
\begin{array}{lc}
T_{\rm eff} {\rm [K]}=10811.3_{-79.5}^{+99.6} \ -\ 19473.4_{-618.0}^{+469.3}\ P\ \\ \ \ \ \ \ \ +\ 27296.3_{-818.1}^{+1233.6}\ P^2\ -\ 12589.5_{-766.3}^{+508.8}\ P^3
\end{array}
\label{eq:perteff_all}
\end{equation}

A similar empirical study was performed by \citet{Li2020} using $\sim1700$ RRLs in the LAMOST catalog.  
In that work, the data was used to identify RRLs with abnormal temperatures, a signature interpreted as stemming from the misclassification of the RRLs or their association with companions in binary systems.
Therefore, we recognize this as a direct application of our catalog, but its exploration is beyond the scope of our work. 
A more detailed analysis of the \teff\ of the stars in our sample, and their use to constrain the shape of the instability strip as a function of [Fe/H] is presented in a companion paper \citep[][]{Medina2025b}.

\begin{figure}
\begin{center}
\includegraphics[angle=0,scale=.42]{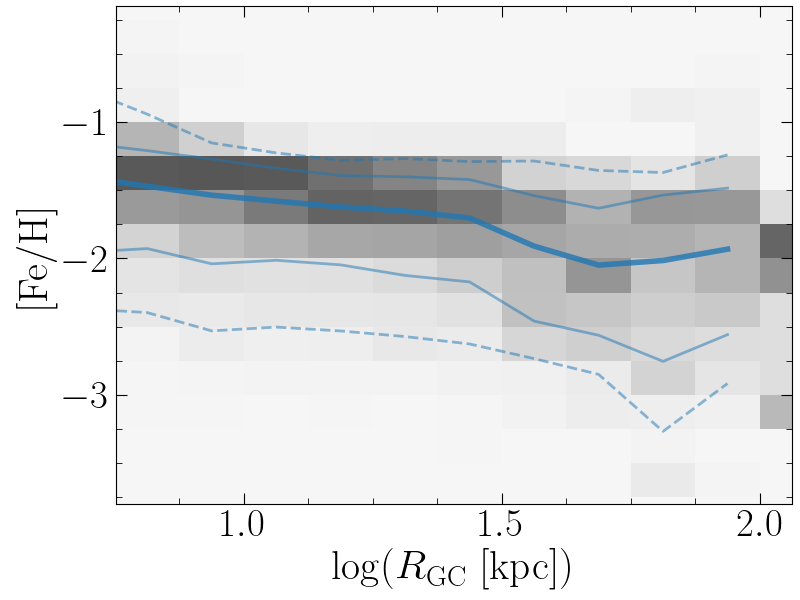}
\includegraphics[angle=0,scale=.42]{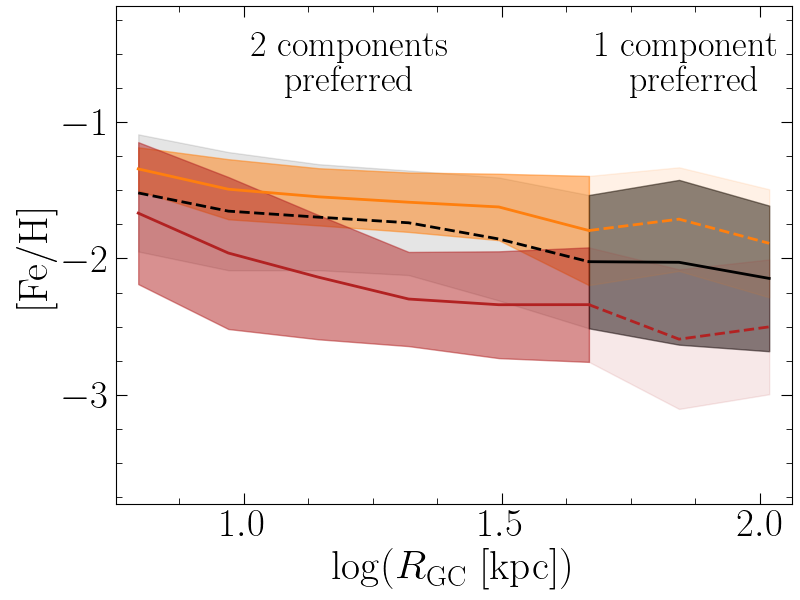}
\caption{
{\it Top}: Column normalized 2-D histogram showing the iron abundance of DESI Y1 field RRLs as a function of Galactocentric distance. 
A thick solid line represents the moving median across the $R_{\rm GC}$ bins, whereas thin solid (dashed) lines are used to depict the [16,84]th ([5,95]th) interval regions. 
The binned data indicates two distinct regions: one characterized by a significant and relatively metal-rich overdensity with low dispersion in [Fe/H] within 40-50\,kpc, and a region with large scatter beyond that limit. 
{\it Bottom}: 
Results of our Gaussian mixture modeling for the RRL [Fe/H] distribution in log($R_{\rm GC}$) bins.  
The mean of the Gaussians is shown with either dashed or solid lines, and the shaded regions represent the width $\sigma$ of these Gaussians. 
Solid lines are used to depict if either the one two components model is preferred over the one component model, whereas dashed lines represent the disfavored model (as measured by the BIC metric). 
The transition between these two regimes occurs at $\sim$45\,kpc.
}
\label{fig:MDF_1}
\end{center}
\end{figure}

\section{The chemodynamics of the halo}
\label{sec:haloChemodyn}

Here, we analyze the bulk chemodynamical properties of the $\sim5500$ RRLs in the halo field (i.e., those not in globular clusters, dwarf galaxies, or the Sagittarius stream).
Specifically, we use these properties to investigate the metallicity distribution function (MDF) and the gradient of metallicity as a function of Galactocentric distance ($R_{\rm GC}$) in the halo, and the kinematics of halo RRLs.
For this, we compute $R_{\rm GC}$ 
adopting a a spherical halo 
and a distance of the Sun to the Galactic center of $8.2$\,kpc \citep{Gravity2021}.

\subsection{Halo metallicity distribution}
\label{sec:haloMDF}

Figure~\ref{fig:MDF_1} (top panel) depicts the [Fe/H] of our sample as a function of Galactocentric distance in a column-normalized 2-D histogram. 
In the figure, we also show the median and the [16,84]th and [5,95]th percentile intervals of the overall metallicity distribution as a function of the binned $\log(R_{\rm GC})$. 
Two distinct regimes are visible in the figure: one dominated 
by a prominent metal-rich component (with [Fe/H] $\sim-1.5$\,dex), which extends out to $\log(R_{\rm GC})\sim1.6$ ($\sim40$\,kpc), and one that displays a larger scatter in metallicity ($\sim0.25$\,dex larger, as measured by the percentile intervals), which is more metal-poor overall ([Fe/H] $\leq-2.0$\,dex).

\begin{figure}
\begin{center}
\includegraphics[angle=0,scale=.42]{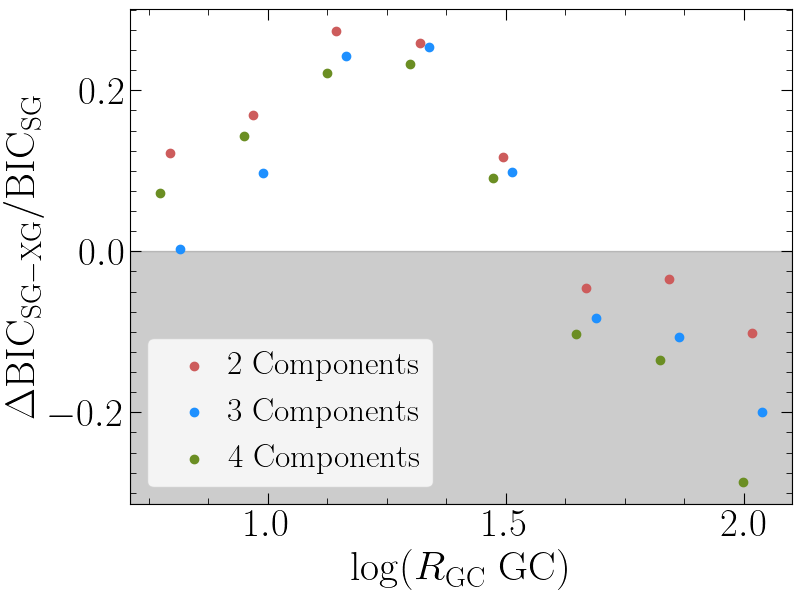}
\caption{
Relative BIC score of multi-component Gaussian mixture models of the halo MDF with respect to a single Gaussian fit.
At a given radius, a model with a significantly larger (relative) BIC score depicts a preference of such model over others. 
The shaded region is used to highlight relative BIC values $<0$, i.e., where a 1-component model is preferred over a multi-component Gaussian decomposition. 
In each distance bin, a small offset in the x-axis is added to the data for a better comparison of the values shown. 
}
\label{fig:AICBIC_vs_dist}
\end{center}
\end{figure}

\begin{figure*}
\begin{center}
\includegraphics[angle=0,scale=.42]{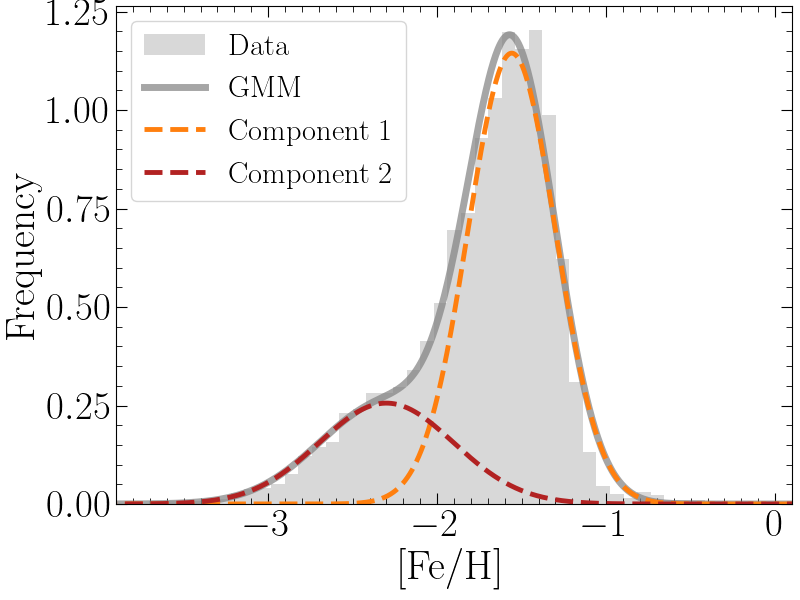}
\includegraphics[angle=0,scale=.42]{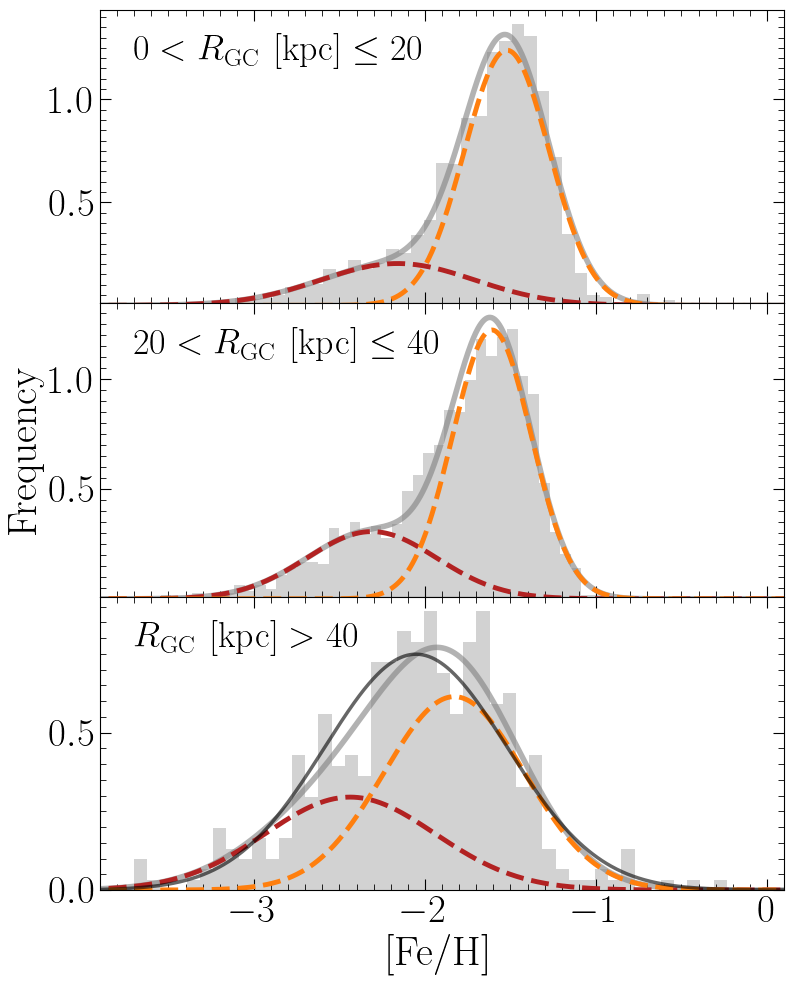}
\caption{
{\it Left}: [Fe/H] distribution of the entire DESI Y1 sample. This distribution can be modeled by two Gaussian components: one thin component centered at ${\rm [Fe/H]}\sim-1.6$\,dex and a wider and more metal-poor component with a mean at ${\rm [Fe/H]}\sim-2.3$\,dex.
{\it Right}: Same as the left panel, but for the RRLs in different halo regions.  
These panels show the dependency of the number of Gaussian components needed to model the data, as (at least) two components are observed in the inner halo, whereas a single Gaussian is able to reproduce the metallicity distribution of the outer halo. 
The central position of these distributions become more metal-poor as the RRL distances increase. 
}
\label{fig:MDF_GMM}
\end{center}
\end{figure*}

To understand the properties of these two components, we split our RRLs in distance bins in logarithmic space and model the MDF of the RRLs in each bin following a Gaussian mixture modeling approach. 
The chosen binning ensures a reasonable number of stars to be used for the MDF analysis in each bin.  
For each MDF, we adopt a model consisting of one Gaussian (single component model) and one using two Gaussians (two component model). 
The result of this analysis is displayed in Figure~\ref{fig:MDF_1} (bottom panel).  
In this figure, the central position and variance of the Gaussian components used to model the MDFs (per bin) are shown.
In each bin, we compute the Bayesian information criterion \citep[BIC;][]{Schwarz1978} metric to assess which model is preferred. 
Figure~\ref{fig:MDF_1} shows that, based on the BIC metric, within $\log(R_{\rm GC}\ [{\rm kpc}])\sim1.65$ ($R_{\rm GC}\sim45$\,kpc) a 2-component Gaussian is preferred over a single Gaussian fit. 
Beyond that radius, a single Gaussian model is preferred.  
We highlight that adopting these models and using the aforementioned metric to select a better model is only valid under the assumption that the underlying distributions are, in fact, Gaussians. 
In that regard, we note that populations covering a wide range of metallicites can exhibit considerably non-Gaussian chemical distributions \citep[see e.g.,][]{Naidu2020,Myeong2022}, and this approach is a strong oversimplification
to study the chemodynamical distribution of halo stars \citep[see e.g.,][]{An2021gse,Liang2021,CabreraGarcia2024}. 
Indeed, there is growing evidence showing that the MDF of the components of the halo have strongly non-Gaussian shapes
\citep[e.g.,][]{Naidu2021,Liu2022,Donlon2023,Khoperskov2023,Zhu2024,Mori2024,Viswanathan2024}, which is supported by expectations from the products of galaxy chemical evolution models \citep[e.g.,][]{Leaman2012,Sanders2021}.

As described above, we assume that the halo MDF can be represented by one or two Gaussian components only, which is an oversimplification of the true nature of the halo. 
Indeed, several authors have reported the presence of multiple components in the MDF of the halo as a function of $R_{\rm GC}$ \citep[see e.g.,][]{Naidu2020,Viswanathan2024}. 
To test this with our dataset, we repeated the aforementioned analysis incorporating up to two additional components to the Gaussian mixture model. 
In Table~\ref{tab:aicbic}, we report the BIC score of multiple Gaussian mixture fits (${\rm XG}$) relative to the BIC score of a single Gaussian model (SG), i.e., $\Delta{\rm BIC}_{{\rm SG} - {\rm XG}}=({\rm BIC}_{\rm SG}-{\rm BIC}_{\rm XG})$.
Figure~\ref{fig:AICBIC_vs_dist} shows the same metric but normalized by ${\rm BIC}_{\rm SG}$ for a better visualization. 
Thus, the table and the figure show where a single Gaussian suffices/fails to describe the MDF (relative BIC $<0$ and $>0$, respectively), and where a given multiple component model is preferred over different number of components (inferred from comparatively larger relative BIC, at a given $R_{\rm GC}$ bin).

\begin{table}
\scriptsize
\caption{
Comparison of the BIC scores of multigaussian mixture models with respect to a single component model. 
Values $>0$ indicate the preference of a multicomponent mixture model over a single Gaussian. 
}
\label{tab:aicbic}
\begin{center}
\begin{tabular}{cHHHHHcccHHHHH}
\toprule
 $R_{\rm GC}$ & BIC$_{\rm SG}$ & BIC$_{\rm 2G}$ & BIC$_{\rm 3G}$ & BIC$_{\rm 4G}$ & BIC$_{\rm 5G}$ & $\Delta$BIC$_{{\rm SG}-{\rm 2G}}$ & $\Delta$BIC$_{{\rm SG}-{\rm 3G}}$ & $\Delta$BIC$_{{\rm SG}-{\rm 4G}}$ & $\Delta$BIC$_{{\rm SG}-{\rm 5G}}$ & Relative BIC$_{\rm 2G}$ & Relative BIC$_{\rm 3G}$ & Relative BIC$_{\rm 4G}$ & Relative BIC$_{\rm 5G}$ \\
{\rm [kpc]} &  &  &  &  & &  &  &  &  &  &  &  &  \\
\hline
      $6.2$ &        $478.8$ &        $420.5$ &        $477.4$ &        $444.2$ &        $450.9$ &                            $58.3$ &                             $1.4$ &                            $34.6$ &                            $27.9$ &                   $0.1$ &                   $0.0$ &                   $0.1$ &                   $0.1$ \\
       $9.3$ &        $919.4$ &        $763.8$ &        $830.4$ &        $787.6$ &        $802.4$ &                           $155.5$ &                            $89.0$ &                           $131.8$ &                           $117.0$ &                   $0.2$ &                   $0.1$ &                   $0.1$ &                   $0.1$ \\
      $13.9$ &      $1,343.4$ &        $976.7$ &      $1,018.4$ &      $1,046.1$ &      $1,007.9$ &                           $366.7$ &                           $325.1$ &                           $297.3$ &                           $335.5$ &                   $0.3$ &                   $0.2$ &                   $0.2$ &                   $0.2$ \\
      $20.9$ &      $1,511.1$ &      $1,120.4$ &      $1,127.9$ &      $1,159.0$ &      $1,182.1$ &                           $390.7$ &                           $383.2$ &                           $352.1$ &                           $329.0$ &                   $0.3$ &                   $0.3$ &                   $0.2$ &                   $0.2$ \\
      $31.2$ &      $1,039.5$ &        $918.3$ &        $937.5$ &        $945.4$ &        $961.4$ &                           $121.3$ &                           $102.1$ &                            $94.2$ &                            $78.1$ &                   $0.1$ &                   $0.1$ &                   $0.1$ &                   $0.1$ \\
      $46.6$ &        $414.6$ &        $433.4$ &        $448.9$ &        $457.0$ &        $459.9$ &                           $-18.8$ &                           $-34.3$ &                           $-42.3$ &                           $-45.2$ &                  $-0.0$ &                  $-0.1$ &                  $-0.1$ &                  $-0.1$ \\
      $69.7$ &        $203.2$ &        $210.3$ &        $224.9$ &        $230.7$ &        $238.3$ &                            $-7.1$ &                           $-21.7$ &                           $-27.5$ &                           $-35.1$ &                  $-0.0$ &                  $-0.1$ &                  $-0.1$ &                  $-0.2$ \\
     $104.3$ &        $117.6$ &        $129.6$ &        $141.1$ &        $151.3$ &        $152.7$ &                           $-11.9$ &                           $-23.4$ &                           $-33.6$ &                           $-35.1$ &                  $-0.1$ &                  $-0.2$ &                  $-0.3$ &                  $-0.3$ \\

\hline
\end{tabular}
\end{center}
\end{table}

From this model evaluation and from an inspection of the mixture model MDF fits, we find that describing our data with two Gaussians is a reasonable assumption for a good fraction of the region $\log(R_{\rm GC})\lesssim1.6$ ($\sim40$\,kpc), whereas beyond that radius a single Gaussian model is always preferred. 
This is also the case when adopting a different evaluation metric, namely the Akaike information criterion \citep[AIC;][]{Akaike1981}. 
Our result contrasts with the description of the outer halo ($R_{\rm GC}>45$\,kpc) using photometric metallicities from the Pristine Survey \citep[][]{starkenburg_pristine_2019} recently reported by \citet{Viswanathan2024} (Figure 13 in their work), where three Gaussian 
components are required to characterize the MDF.
This discrepancy should be attributed, at least partially, to low number statistics affecting our sample of RRLs at $R_{\rm GC}>45$\,kpc (a factor of 10 smaller than that used by \citealt{Viswanathan2024}), which limits our ability to draw firm conclusions at these distances.  
Larger samples of outer halo RRLs with homogeneously-derived [Fe/H] (e.g., from upcoming DESI data releases) are required to inspect the number of components of the MDF at $R_{\rm GC}>50$\,kpc in greater detail.

Figure~\ref{fig:MDF_GMM} shows the MDF of our entire sample modeled by the multiple component approach, and a similar visualization for the RRLs in three representative regions, namely $R_{\rm GC}<20$\,kpc (the inner halo), $20<R_{\rm GC}\ {\rm [kpc]}\leq40$\,kpc (``extended'' inner halo, or transition region), and $R_{\rm GC}>40$\,kpc (the outer halo).
The MDFs in this figure show how, for $R_{\rm GC}<20$\,kpc, the data is well represented by two Gaussians, centered at ${\rm [Fe/H]}=-1.52$\,dex (with width $\sigma=0.25$\,dex) and ${\rm [Fe/H]}=-2.15$\,dex ($\sigma=0.45$\,dex). 
As the distance to the Galactic center increases, the position of these Gaussians shifts towards the more metal-poor regime, at ${\rm [Fe/H]}=-1.61$\,dex ($\sigma=0.23$\,dex) and ${\rm [Fe/H]}=-2.31$\,dex ($\sigma=0.38$\,dex). 
Beyond 40\,kpc, the two components become less distinct, and the data can be represented by a single Gaussian (${\rm [Fe/H]}=-2.05$\,dex with $\sigma=0.53$\,dex).

Our findings are consistent with the current knowledge of the dual origin of the Galactic halo, where its inner regions are composed of a combination of stars accreted from massive satellites and stars formed in-situ, whereas its outskirts are predominantly composed of stars accreted from satellites.
These formation scenarios and the dual component halo have been thoroughly investigated in the past \citep[e.g.,][]{White1978,Chiba2001,Naidu2020,Limberg2022}. 
Indeed, the radial metallicity gradient has been interpreted to be the outcome of massive radial mergers \citep[e.g.,][]{Carollo2007,Carollo2010}, which includes the accretion of the massive and metal-rich dwarf galaxy responsible for the Gaia-Sausage-Enceladus (GSE) merger event \citep[][]{Belokurov2018,Haywood2018,Helmi2018}.
The presence of this metal-rich component is supported by our results. 
Moreover, the limit between the two aforementioned regimes occurs between 40 and 50\,kpc, roughly consistent with the break in the halo number density profile reported by several authors using different stellar tracers across the entire sky \citep[see e.g.,][]{Keller2008,Deason2013,Das2016,Deason2018,Medina2018,Thomas2018,Stringer2021,Yu2024,Amarante2024,Medina2024}. 
Recently, \citet{Liu2022} studied the MDF of a sample of $\sim4300$ RRLs with SDSS \citep{Yanny2009} and LAMOST \citep{Deng2012,Liu2020} spectroscopic measurements. 
We note that, although the transition region found by these authors is closer to $30$\,kpc, we find an overall agreement with their results.

\subsection{The metallicity gradient of the metal-poor halo and GSE}

To characterize the halo metallicity gradient with our RRL sample, we extend the methodology employed above and follow a Bayesian mixture model approach. 
We assume that the halo can be effectively described by two components, and that each component (thin and broad, from their [Fe/H] distribution) can be described by the mean and the variance of a Gaussian in radial velocity ($v_r$), azimuthal velocity ($v_\phi$), and [Fe/H]. 
The inclusion of the stars' velocities in the model is supported by studies that confirm
existing correlations between the Galactocentric velocity and iron abundance of the two observed overdensities in the inner halo \citep[see e.g.,][]{Belokurov2018}. 
We emphasize that, although this approach allows us to capture observable chemodynamical trends in the halo, adopting a model with Gaussian [Fe/H] and velocity distributions for the halo components is a caveat in our analysis \citep[see e.g.,][]{Lancaster2019,Sanders2021} 
with room for improvement in future work. 
Furthermore, previous works have reported the presence of more than two components in the chemodynamical distribution of halo stars (and of halo RRLs in particular).  
\citet{Iorio2021}, for instance, identified four distinct halo components using the {\it Gaia} DR2 RRL catalog, corresponding to the disk, the GSE merger, the quasi isotropic accreted halo, and a metal-rich centrally concentrated in-situ component. 
Although we expect our sample to be dominated by the GSE and the quasi isotropic halo, given the footprint of DESI Y1 and the analysis presented in Section~\ref{sec:haloMDF}, a detailed evaluation of the contribution of each of these components to our sample remains to be done and will be the focus of a forthcoming work.  

We allow the mean metallicity to vary as a linear function of Galactocentric distance (in kpc) in the model, as $\mu_{\rm [Fe/H]}=a+b\ R_{\rm GC}$. 
Moreover, we assume that the fraction of stars in the metal-rich component $p_{\rm GSE}$ varies with radius as

\begin{equation}
\begin{array}{ll}
p_{\rm GSE}(R_{\rm GC}) = p_{\rm inner} + (p_{\rm outer} - p_{\rm inner})/(1+e^{-(R_{\rm GC}-\mu)/k})
\end{array}
\label{eq:p_gse}
\end{equation}

where $p_{\rm inner}$, $p_{\rm outer}$, $\mu$, and $k$ are model parameters. 
Using this representation, the model is allowed to constrain the extent in distance of the metal-rich component (as defined by the parameter $\mu$), and the fraction of stars associated with that component depends on whether $R_{\rm GC}<\mu$ (where $p_{\rm GSE}$ tends to $p_{\rm inner}$) or $R_{\rm GC}>\mu$ (where $p_{\rm GSE}$ tends to $p_{\rm outer}$).
Then, the likelihood for a star $i$ belonging to the GSE and metal-poor component ($\mathcal{L}_i^{{\rm GSE}}$ and $\mathcal{L}_i^{{\rm MP}}$, respectively), defined from multivariate Gaussians in $v_R$, $v_{\phi}$, and [Fe/H], and the likelihood function ($\mathcal{L}^{{\rm tot}}$) are defined as:

\begin{center}
\begin{align}
\mathcal{L}_i^{\rm GSE}& = \mathcal{L}_{v_r, i}^{\rm GSE} \times \mathcal{L}_{v_\phi, i}^{\rm GSE} \times \mathcal{L}_{\rm [Fe/H], i}^{\rm GSE},\\
\mathcal{L}_i^{\rm MP}& = \mathcal{L}_{v_r, i}^{\rm MP} \times \mathcal{L}_{v_\phi, i}^{\rm MP} \times \mathcal{L}_{\rm [Fe/H], i}^{\rm MP},\\
\ln(\mathcal{L}^{\rm tot})& = \sum_{i=1}^{N} \ln(p_{\rm GSE}\ \mathcal{L}_i^{\rm GSE}+ (1-p_{\rm GSE})\ \mathcal{L}_i^{\rm MP}), 
\end{align}
\end{center}
\label{eq:MDF_likelihood}

Following Bayes' rule, the posterior probability of a star belonging to the GSE-like component is:

\begin{equation}
\begin{array}{ll}
P_i = \dfrac{p_{\rm GSE} \mathcal{L}_i^{\rm GSE}}{p_{\rm GSE} \mathcal{L}_i^{\rm GSE}+ (1-p_{\rm GSE}) \mathcal{L}_i^{\rm MP}}.
\end{array}
\label{eq:MDF_prob}
\end{equation}

Therefore, our model is described by 18 parameters: four to define the fraction of stars belonging to the GSE component ($p_{\rm inner}$, $p_{\rm outer}$, $\mu$, and $k$), four to trace the dependence of [Fe/H] on $R_{\rm GC}$ (two per component, $a_{{\rm [Fe/H]} }^{\rm GSE}$, $b_{{\rm [Fe/H]}}^{\rm GSE}$, $a_{{\rm [Fe/H]}}^{\rm MP}$, and $b_{{\rm [Fe/H]}}^{\rm MP}$, where the superindices ${\rm GSE}$ and ${\rm MP}$ refer to the GSE-like and metal-poor components, respectively), two for the width of the [Fe/H] Gaussians (one per component, $\sigma_{{\rm [Fe/H]}}^{\rm GSE}$ and $\sigma_{{\rm [Fe/H]}}^{\rm MP}$), four for the means of the velocity Gaussians ($v_\phi^{\rm GSE}$, $v_r^{\rm GSE}$, $v_\phi^{\rm MP}$, $v_r^{\rm MP}$), and four for their variance ($\sigma_{v_\phi}^{\rm GSE}$, $\sigma_{v_r}^{\rm GSE}$, $\sigma_{v_\phi}^{\rm MP}$, $\sigma_{v_r}^{\rm MP}$).  
The adopted flat priors for each parameter are provided in Table~\ref{tab:gmm_mcmc}.

\begin{figure}
\begin{center}
\includegraphics[angle=0,scale=.44]{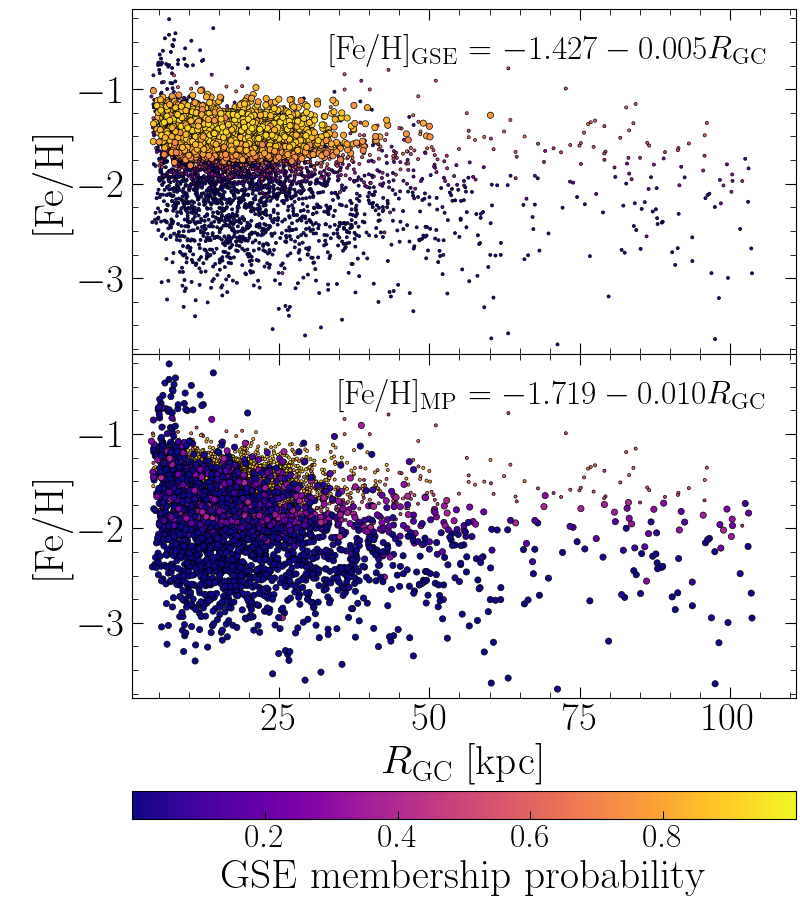}
\caption{
Iron abundance as a function of Galactocentric distance for our RRL sample, color-coded by the GSE membership probability computed by our Bayesian mixture model.
The top panel shows, with large circles, the stars likely associated with the GSE merger (with $p_{\rm GSE}>0.7$), whereas the bottom panel uses large circles for the stars with $p_{\rm GSE}<0.4$.
The color scheme shows that the extent of the stars likely associated with the GSE accretion event is predominantly $<40$\,kpc.
}
\label{fig:MDF_GMM_wtracks}
\end{center}
\end{figure}

\begin{figure}
\begin{center}
\includegraphics[angle=0,scale=.42]{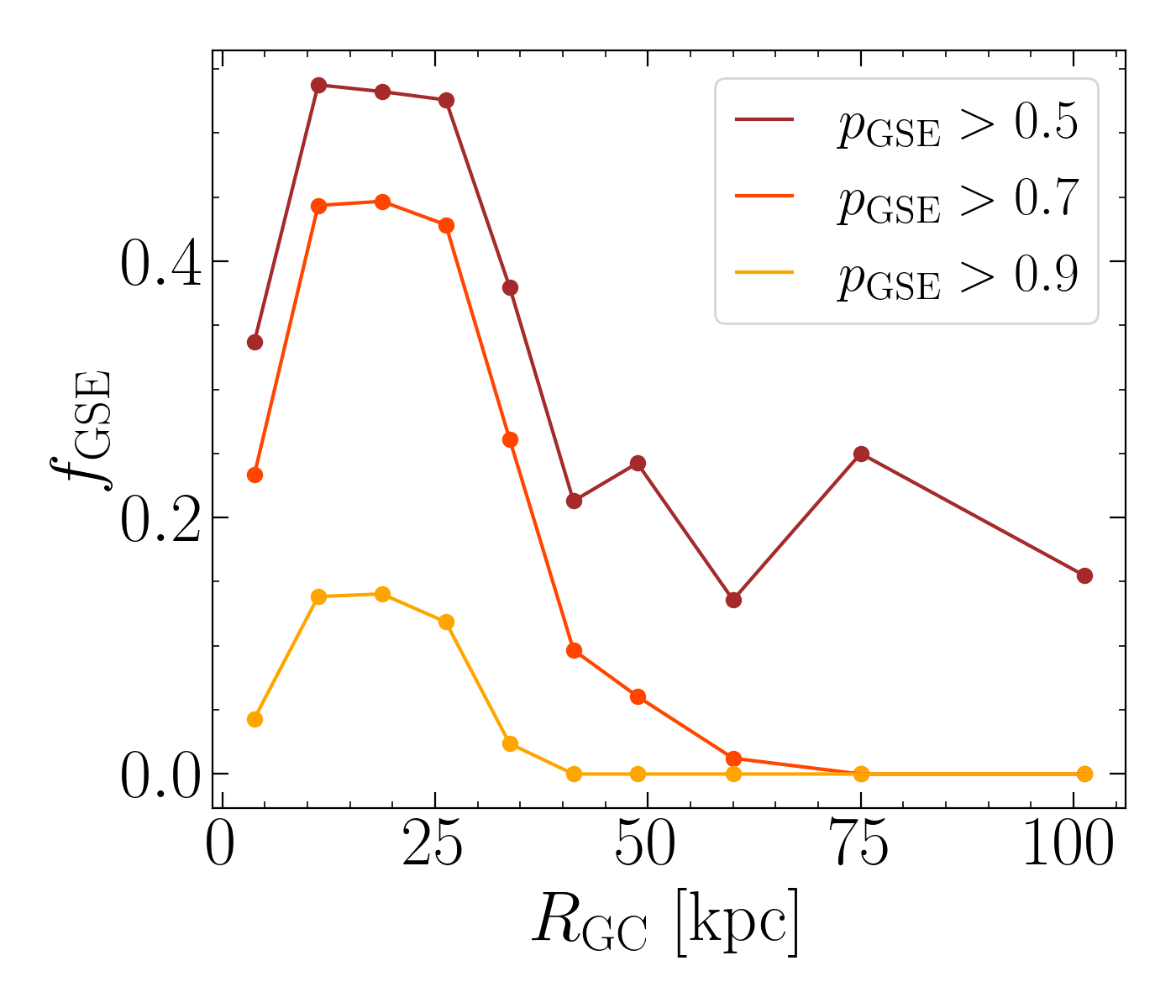}
\caption{
Fraction of RRLs likely associated to GSE as a function of Galactocentric distance, using different cuts in $p_{\rm GSE}$. 
}
\label{fig:gsefrac}
\end{center}
\end{figure}

To determine the best fit parameters, 
we explore the parameter space following an MCMC methodology and the Python package {\sc emcee} \citep[][]{emcee}. 
We find that 32 walkers, each taking 2000 MCMC steps, suffice to obtain robust results. 
We define the optimal parameters as the median of the marginalized distributions, and their uncertainties as their 16th and 84th percentiles.  
Our results are summarized in Table~\ref{tab:gmm_mcmc} and depicted in Figure~\ref{fig:MDF_GMM_wtracks}.

Our analysis indicates that the mean metallicity of the Gaussians composing the RRLs MDF can be described as:

\begin{equation}
\begin{array}{ll}
{\rm [Fe/H]_{\rm GSE}}& = -1.427^{+0.010}_{-0.012} - 0.005^{+0.000}_{-0.001}\ R_{\rm GC},\\ 
{\rm [Fe/H]_{\rm MP}}& = -1.719^{+0.012}_{-0.011} - 0.010^{+0.001}_{-0.000}\ R_{\rm GC} 
\end{array}
\label{eq:MDF_GMM_gradient}
\end{equation}

where the constant and multiplicative coefficients are provided in units of dex and dex\,kpc$^{-1}$.
The GSE component is characterized by a scatter of $\sigma_{\rm [Fe/H]}^{\rm GSE}=0.184^{+0.004}_{-0.005}$\,dex, whereas the metal-poor component displays $\sigma_{\rm [Fe/H]}^{\rm MP}=0.474^{+0.006}_{-0.007}$\,dex. 
From Equation~\ref{eq:MDF_GMM_gradient}, our sample shows a mild decrease in iron abundance as a function of $R_{\rm GC}$ (of $0.005$ and $0.010$\,dex\,kpc$^{-1}$ for the GSE and metal-poor component, respectively). 
We note that the gradient of the metal-poor component is larger than that found by \citet{Liu2022} ($0.003$\,dex\,kpc$^{-1}$) in the halo field using LAMOST RRLs.
Our results also confirm the slightly negative metallicity gradient detected by these authors in GSE.
However, the metallicity gradient found for the metal-rich component is about a factor of two smaller than the value reported by these authors (0.009\,dex\,kpc$^{-1}$). 
We also note that our GSE metallicity gradient is about three times shallower than the one measured by \citet{Khoperskov2023} using APOGEE DR17.

\begin{table}
\scriptsize
\caption{
Resulting parameters of our $R_{\rm GC}$-dependent mixture modeling of the [Fe/H] distribution in the halo. 
}
\label{tab:gmm_mcmc}
\begin{center}
\begin{tabular}{cccc}
\hline
Parameter & Prior & Median & C.I. \\
\hline
$p_{\rm inner}$ & $\mathcal{U}$($0.000$, $1.000$) &  $0.498$ & [$0.486$, $0.510$] \\
$p_{\rm outer}$ & $\mathcal{U}$($0.000$, $1.000$) & $0.379$ & [$0.336$, $0.420$] \\
$\mu$ [kpc] & $\mathcal{U}$($30.0$, $60.0$) & $37.1$ & [$35.08$, $40.6$] \\
$k$ & log-$\mathcal{U}$($-1.000$, $0.900$) & $-0.239$ & [$-0.741$, $0.243$] \\
$v_\phi^{\rm GSE}$\ [km\,s$^{-1}$] & $\mathcal{U}$($-100.000$, $150.000$) & $4.580$ & [$3.867$, $5.296$] \\
$\sigma_{v_\phi}^{\rm GSE}$\ [km\,s$^{-1}$] & log-$\mathcal{U}$($1.000$, $2.500$) & $25.194$ & [$24.337$, $26.024$] \\
$v_r^{\rm GSE}$\ [km\,s$^{-1}$] & $\mathcal{U}$($-100.000$, $150.000$) & $3.689$ & [$0.455$, $6.716$] \\
$\sigma_{v_r}^{\rm GSE}$\ [km\,s$^{-1}$] & log-$\mathcal{U}$($1.000$, $2.500$) & $135.952$ & [$133.929$, $138.141$] \\
$a^{\rm GSE}$\ [dex] & $\mathcal{U}$($-3.000$, $-1.000$) & $-1.427$ & [$-1.439$, $-1.417$] \\
$b^{\rm GSE}$\ [dex\,kpc$^{-1}$] & $\mathcal{U}$($-0.010$, $0.005$) & $-0.005$ & [$-0.006$, $-0.005$] \\
$\sigma_{\rm [Fe/H]}^{\rm GSE}$\ [dex] & $\mathcal{U}$($0.000$, $4.000$) & $0.184$ & [$0.179$, $0.188$] \\
$v_\phi^{\rm MP}$\ [km\,s$^{-1}$] & $\mathcal{U}$($-100.000$, $300.000$) & $-2.438$ & [$-4.443$, $-0.390$] \\
$\sigma_{v_\phi}^{\rm MP}$\ [km\,s$^{-1}$] & log-$\mathcal{U}$($1.000$, $3.000$) & $108.158$ & [$106.534$, $109.806$] \\
$v_r^{\rm MP}$\ [km\,s$^{-1}$] & $\mathcal{U}$($-100.000$, $300.000$) & $8.075$ & [$4.914$, $10.833$] \\
$\sigma_{v_r}^{\rm MP}$\ [km\,s$^{-1}$] & log-$\mathcal{U}$($1.000$, $3.000$) & $137.318$ & [$135.235$, $139.425$] \\
$a^{\rm MP}$\ [dex] & $\mathcal{U}$($-2.500$, $-0.500$) & $-1.719$ & [$-1.730$, $-1.707$] \\
$b^{\rm MP}$\ [dex\,kpc$^{-1}$] & $\mathcal{U}$($-0.100$, $0.005$) & $-0.010$ & [$-0.010$, $-0.009$] \\
$\sigma_{\rm [Fe/H]}^{\rm MP}$\ [dex] & $\mathcal{U}$($0.000$, $3.000$) & $0.474$ & [$0.467$, $0.480$] \\
\hline
\end{tabular}
\end{center}
\end{table}

Figure~\ref{fig:gsefrac} illustrates the fraction $f_{\rm GSE}$ of RRLs with $p_{\rm GSE}$ above three probability cuts (0.5, 0.7, and 0.9), as a function of $R_{\rm GC}$. 
These fractions are computed as the number of RRLs with $p_{\rm GSE}$ larger than the corresponding threshold with respect to the total number of field RRLs per distance bin.
The figure shows that, if the selected probability cut is $p_{\rm GSE}>0.5$ or $p_{\rm GSE}>0.7$, the fraction of RRLs with GSE-like chemodynamics varies between $\sim$0.3--0.5 within 40\,kpc. 
Beyond that radius, $f_{\rm GSE}$ quickly drops to $\leq0.20$.
We note that the our estimation of $f_{\rm GSE}$ lies on the lower side of the overall GSE RRL fraction reported by \citet{Iorio2021} at $5 < R_{\rm GC}\ {\rm [kpc]}<25$ (between 50 and 80\,per\,cent), but is consistent with their estimation for stars with Cartesian $Z$ coordinates (above and below the plane) between 15 and 25\,kpc. 
Furthermore, for the lower $p_{\rm GSE}$ threshold, the GSE fraction does not reach $f_{\rm GSE}\sim0$ beyond $\sim40$\,kpc and, instead, a sudden increase is observed at $\sim75$\,kpc. 
We attribute this increase to the underdensity of metal-poor stars seen at that distance in Figure~\ref{fig:MDF_GMM_wtracks}.

\begin{figure}
\begin{center}
\includegraphics[angle=0,scale=.42]{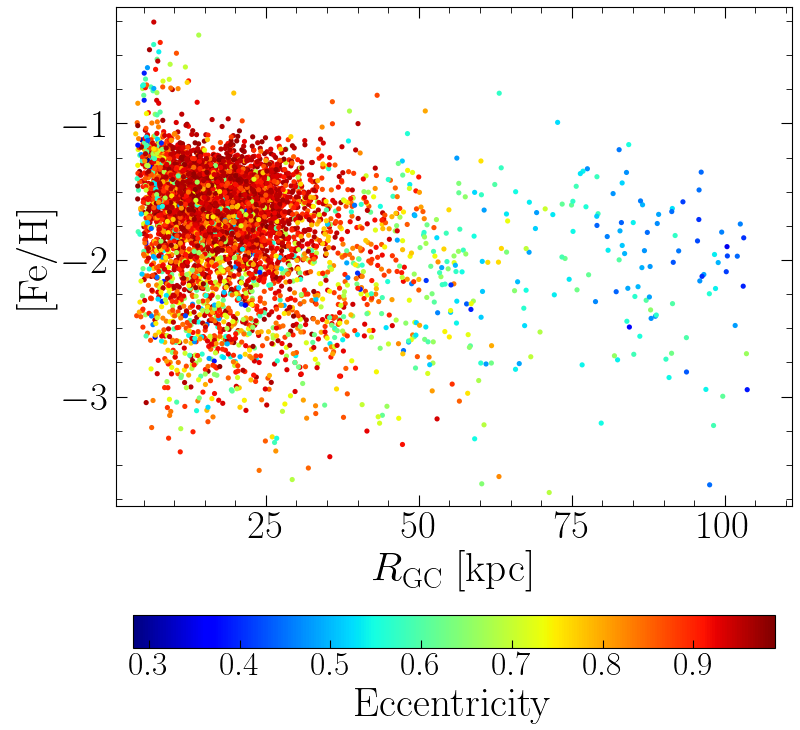}
\includegraphics[angle=0,scale=.58]{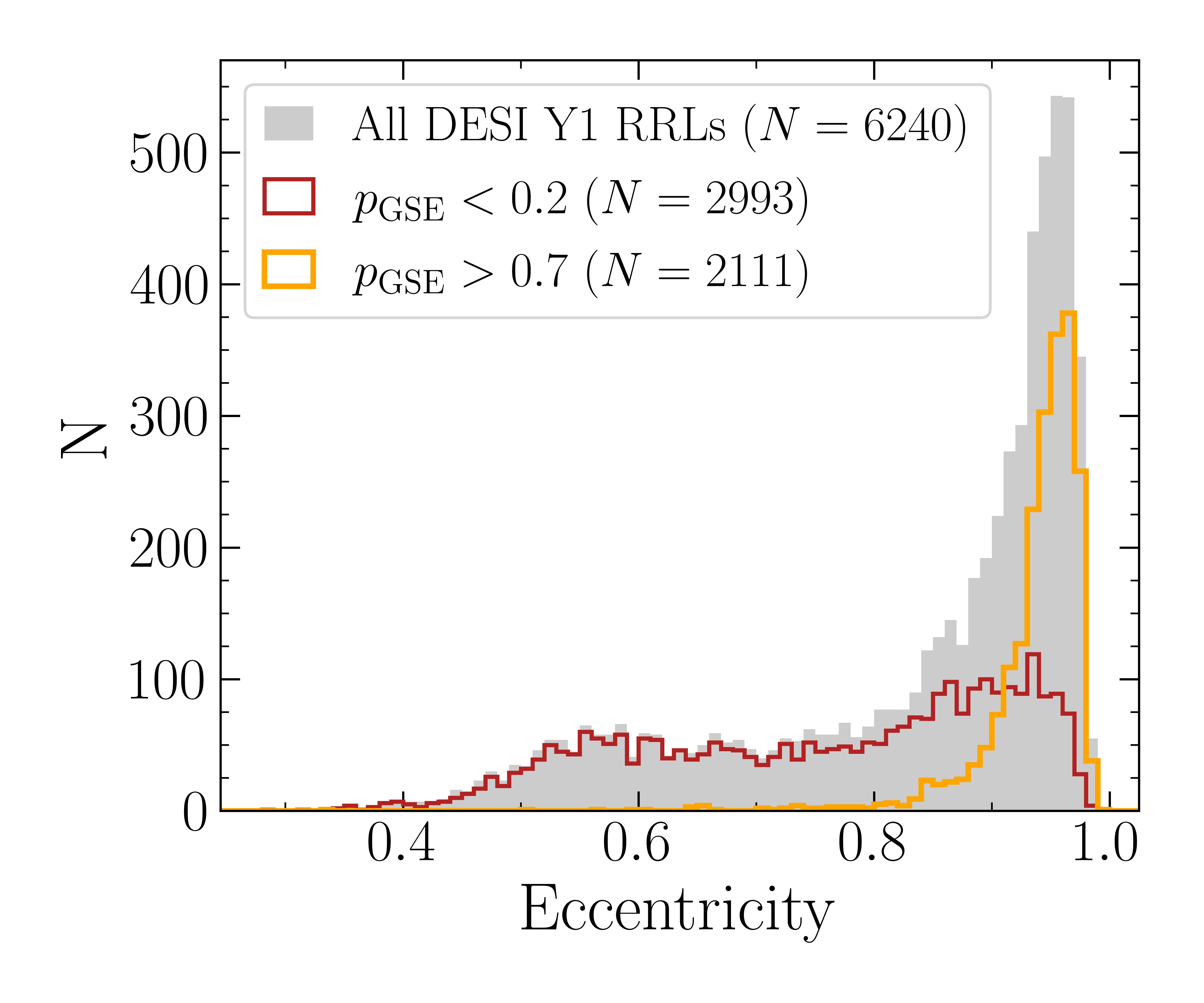}
\caption{
{\it Top}: Iron abundance as a function of Galactocentric distance and color coded by the eccentricity of the orbit of each star. 
This plot further demonstrates the presence of two populations in the data: a largely radial population, that dominates at $<45$\,kpc, and one with eccentricity $<0.5$, present at all radii. 
{\it Bottom}: Eccentricity distribution of the RRLs in our sample. 
Unfilled histograms represent the distribution of highly likely and unlikely GSE members, showing that we are able to distinguish the bulk of the predominantly radial orbits of the GSE stars with our mixture modeling approach. 
}
\label{fig:MDF_eccentricity}
\end{center}
\end{figure}

\subsection{The eccentricity of the metal-poor halo and GSE}
\label{sec:eccentricity}

In order to further inspect the overall kinematic properties of field and GSE RRLs, we compute the eccentricity of their orbits modeled using the Python module {\sc GALPY} \citep[][]{Bovy2015}\footnote{ \href{http://github.com/jobovy/galpy}{http://github.com/jobovy/galpy}}.
We integrate the orbits for 5\,Gyr 
backwards in time. 
We adopt the built-in {\it MWPotential2014} to model the Galactic potential, which consists of a spherical nucleus and bulge \citep[Hernquist potential;][]{Hernquist1990},
a Miyamoto-Nagai disk model \citep[][]{MiyamotoNagai1975}, and a spherical Navarro-Frenk-White dark matter halo \citep[][]{Navarro1997}.
Additionally, we considered the perturbation of the Milky Way potential caused by the Large Magellanic Cloud (LMC), as growing evidence indicates that its effects on the Milky Way potential can be significant \citep[especially for modeling orbits; see e.g.,][]{Erkal2018,Cunningham2020,Vasiliev2021,Medina2023}.
For the LMC current position, we adopted the central equatorial coordinates $\alpha=78.77$\,deg and $\delta=-69.01$\,deg, and a distance of $d_{\rm LMC}=49.6$\,kpc \citep[][]{Pietrzynski2019}. 
For its current motion, we assumed $\mu_\alpha=1.850$\,mas\,yr$^{-1}$ and $\mu_\delta= 0.234$\,mas\,yr$^{-1}$ as proper motions \citep[][]{GaiaCollaboration2018}, and a systemic line-of sight velocity of $262.2$\,km\,s$^{-1}$ \citep[][]{vanderMarel2002}.
We adopt an LMC mass of $1.88\times10^{11}$\,M$_\odot$ \citep[][]{Shipp2021} and a scale length that recovers the circular velocity at $8.7$\,kpc from the LMC center observed by \citet{vanderMarel2014} ($a_{\rm LMC}=20.22$\,kpc).  
To account for the fact that the LMC is not bound to the Milky Way in {\it MWPotential2014}, we multiply the halo mass by a factor of 1.5 (as suggested in {\sc GALPY}'s documentation). 
Lastly, we integrate the orbits in the presence of an LMC in motion, 
considering the effect of the Chandrasekhar dynamical friction in the orbit integration of the LMC, and ignoring the gravitational perturbations caused by other massive Milky Way  satellites.

To derive the orbital parameters of the stars in our sample and their uncertainties, 
we employ {\it Gaia} DR3 proper motions and DESI-derived systemic velocities.
We integrate orbits from 500 {\sc GALPY} realizations, varying the input parameters (systemic velocities, heliocentric distances, and proper motions) assuming that they follow Gaussian distributions built from the covariance matrices of the stars in  the {\it Gaia} DR3 catalog. 
The parameters and their uncertainties are then estimated as the median of the resulting distributions and their 16 and 84 percentiles, respectively. 
We note that a detailed analysis of the RRLs' orbital parameters is beyond the scope of this paper, and is presented in separate works \citep[][]{Medina2025b,Medina2025c}. 

Figure~\ref{fig:MDF_eccentricity} depicts the dependence of the iron abundance of our sample on Galactocentric distance, color-coded by orbital eccentricity. 
Additionally, Figure~\ref{fig:MDF_eccentricity} shows the distribution of eccentricities, for likely field and GSE RRLs. 
The figure clearly shows that the inner halo contains RRLs with eccentricities covering the entire [0,1] range, but it displays predominantly high eccentricities ($e>0.90$). 
As expected, a large number of  these stars are likely members of GSE, as $\sim35$\% (2,111 RRLs) of them display GSE membership probabilities $>0.70$. 
In contrast, the outer halo ($>45$\,kpc) is characterized by a dearth of stars in radial orbits, as most of the stars at these distances have eccentricities $<0.5$. 
We note, however, that despite their relatively lower $p_{\rm GSE}$, the non-association of these stars with GSE is not conclusive. 
Using simulations tailored to reproduce the merger between a MW-like galaxy and a (GSE-like) massive satellite, for instance, \citet{Koppelman2020} found that the debris from the dwarf galaxy's disruption could exhibit a rather intricate morphology.
This includes stars with a broad range of orbital eccentricities in both prograde and retrograde orbits, where a non-negligible fraction of accreted star particles could exhibit low eccentricities and be associated with material lost early on GSE's accretion.

\begin{figure}
\begin{center}
\includegraphics[angle=0,scale=.42]{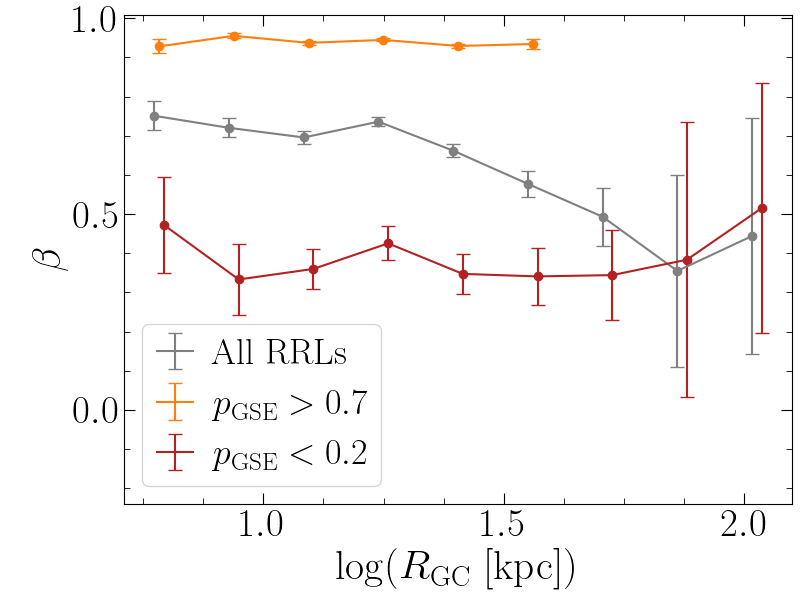}
\caption{
Velocity anisotropy parameter $\beta$ as a function of the logarithm of the Galactocentric distance $R_{\rm GC}$ for our entire sample and for the highly likely and unlikely GSE members shown in Figure~\ref{fig:MDF_eccentricity}.
A small offset in the x-axis is applied to the data for a better visualization.
}
\label{fig:beta_vs_dist}
\end{center}
\end{figure}

\subsection{The velocity anisotropy of the metal-poor halo and GSE}
\label{sec:velocity_anisotropy}

The data at hand enable us to investigate the overall velocity distribution of the metal-poor halo and GSE components.  
We estimate the velocity anisotropy parameter $\beta$ of each component adopting its definition in spherical coordinates, $\beta=1-(\sigma_\phi^2+\sigma_\theta^2)/(2\sigma_r^2)$ ($-\infty<\beta<1$), where $\sigma$ refers to the velocity dispersion, and the subindices $r$, $\phi$, and $\theta$ represent the Galactocentric distance, the azimuthal angle $0<\phi<2\pi$, and the polar angle $0<\theta<\pi$ \citep[][]{Binney2008,Bird2019}, respectively. 
This definition ensures that, for an isotropic system, 
$\sigma_\theta=\sigma_\phi=\sigma_r$ ($\beta=0$), whereas $\beta>0$ represents a system with orbits predominantly radial and $\beta<0$ one with mostly tangential orbits. 
If we define stars highly probable members of the GSE (metal-rich) component as those with membership probabilities $>0.7$ 
(2,111 RRLs), and those in the metal-poor component as stars with probabilities $<0.2$ (2,993 RRLs), we find $\beta_{\rm{\rm GSE}}\sim0.93$ and $\beta_{\rm{\rm MP}}\sim0.35$. 
With this result, we confirm the link between the bulk of the inner halo RRLs with the GSE merger event, taking into account that the stars in GSE display orbits that are largely radial \citep[][]{Belokurov2018}.

Similar to previous works \citep[e.g.,][]{Bird2019,Liu2022}, we measure the dependency of $\beta$ on Galactocentric radius for the stars likely associated to the GSE ($p_{\rm GSE}>0.7$), for the stars belonging to the metal-poor component of the halo ($p_{\rm GSE}<0.2$), and for the entire sample (irrespective of the stars' $p_{\rm GSE}$). 
We split the data into $\log(R_{\rm GC})$ bins and, in each bin, we follow the maximum likelihood methodology described in Section~\ref{sec:dracovlos}.
That is, we determine the velocity dispersion and its error (per velocity component) using Equation~\ref{eq:vdispmcmc} and from the 16th, 50th, and 84th percentiles of their posterior distributions. 
The $\beta$ parameter is then computed from its definition and its error is determined from the propagation of uncertainties.

The results of this analysis are displayed in Figure~\ref{fig:beta_vs_dist}. 
The figure shows that the stars likely associated with the GSE merger event exhibit predominantly radial orbits across the distance range they cover (out to $\log(R_{\rm GC}\ {\rm [kpc]})\sim1.60$, or $R_{\rm GC}\sim40$\,kpc), with $\beta=0.94\pm0.01$.
For the metal-poor component, we observe relatively constant $\beta$ values with distance, of $\sim 0.39\pm0.06$ (with increasing $\beta$ uncertainties with distance). 
We note that the flat behavior of $\beta$ for the GSE and the metal-poor component is not surprising, giving that our model to compute $p_{\rm GSE}$ assumes the variance of their radial and azimuthal velocity components distributions ($\sigma_{v_\phi}^{\rm GSE}$, $\sigma_{v_r}^{\rm GSE}$, $\sigma_{v_\phi}^{\rm MP}$, $\sigma_{v_r}^{\rm MP}$) to be independent of radius. 
When using the entire sample (i.e., without distinction between GSE and MP), the resulting $\beta$ values lie between those from the GSE and MP-like samples, with a trend of decreasing $\beta$ as a function of $\log (R_{\rm GC})$ across the explored distance range.
Using the Python package \code{scipy}, we find that this trend follows $\beta_{\rm }=0.91\ (\pm0.09) -  0.17\ (\pm0.07) \log(R_{\rm GC} {\rm \ [kpc]})$. 
A potential explanation for the observed decline in $\beta$ with distance is that the more metal-rich RRLs are only present within a certain radius (with a high $f_{\rm GSE}$; see Figure~\ref{fig:gsefrac}), affecting $\beta$ only within a constrained region and keeping $0.6<\beta<0.8$. 
Beyond that radius, the metal-poor component (with a more isotropic velocity distribution) dominates, resulting in the observed decreasing trend.

Our results are broadly consistent with several recent works that have measured the halo velocity anisotropy using different tracers. 
\citet{Bird2019}, for instance, analyzed the velocity distribution of  K giants observed by LAMOST and {\it Gaia} and found reasonably flat $\beta$ values for the metal-rich (${\rm [Fe/H]}<-1.8$\,dex) and the metal-poor (${\rm [Fe/H]}<-1.8$\,dex) end of the halo MDF out to $\sim25$\,kpc ($\beta\sim0.9$ and $0.6$, respectively). 
Beyond that radius, the reported $\beta$ of both subsamples declines to $\beta\sim0.3$--$0.5$ (for $25<R_{\rm GC}\ {\rm [kpc]}<60$).
In a similar study, \citet{Bird2021} combined LAMOST, SDSS/SEGUE, and {\it Gaia} data of K giants and blue horizontal-branch stars (BHBs) and found that, for $-1.7<{\rm [Fe/H]}<-1.0$, the halo exhibits a $\beta$ profile that begins to decline at $R_{\rm GC}\sim20$\,kpc, from $\beta~\sim0.9$ to $0.7$ for K giants, and $\beta~\sim0.8$ to $0.1$ for BHBs. 
For the metal-poor halo (${\rm [Fe/H]}<-1.7$\,dex), they find a relatively constant $\beta$ (between 0.2 and 0.7, depending on the metallicity range probed) across all distances, independent of star type.
Also using BHBs observed by LAMOST, \citet{Vickers2021} found that metal-rich BHBs have more radial velocity dispersion-dominated orbits at all radii within 25\,kpc ($\beta\sim0.70$) than metal-poor BHBs ($\beta\sim0.62$). 
\citet{Iorio2021}, imposing a fourfold symmetry and employing {\it Gaia} DR2 proper motions of RRLs, found that 50--80\,per\,cent of the halo out between 5--25\,kpc display radially-biased kinematics, where stars with $-1.7 <{\rm [Fe/H]}<-1.2$ have predominantly $\beta\sim0.9$ while for lower [Fe/H] stars $\beta$ drops to 0.2--0.4.
A similar trend is observed by \citet{Liu2022} using LAMOST RRL data, where they exhibit $\beta\sim0.8$ out to $\sim 20$--$30$\,kpc when using the entire sample, and $\sim0.7$ after the removal of GSE. 
Beyond this radius, these authors report $\beta$ to decline to 0.3--0.5 at $R_{\rm GC}\sim 45$\,kpc.
If we considered our entire sample (Figure~\ref{fig:beta_vs_dist}), we find a relatively constant $\beta$ value, with a decline starting at $\log(R_{\rm GC}\ {\rm [kpc]})\sim1.4$ ($\sim25$\,kpc).

\section{Conclusions}
\label{sec:conclusions}

Taking advantage of the rich and homogeneous dataset provided by DESI in its first year of operations, we built a catalog of 6,240 spectroscopically-characterized RRL based on a crossmatch with the {\it Gaia} DR3 RRL catalog.
With a total of over 12,000 individual epochs, this compilation is one of the largest homogeneous spectroscopic RRL catalogs existing to-date. 

The observed spectroscopic properties of our sample, namely line-of-sight velocities, effective temperatures, iron abundances, and surface gravities, are determined using the main pipelines developed for the DESI Milky Way survey (RVSpec and SP).
In addition, iron abundances were also derived using existing calibrations of the $\Delta$S method.
The value of [Fe/H] obtained with both methods cover a wide range of metallicities, from $\sim-3.8$\,dex to above solar, which makes this sample well-suited for a variety of stellar astrophysics and Galactic archaeology studies.  

We developed a novel Bayesian inference approach to build DESI-based radial velocity curves and to quantify the dependence between their shapes and the period of the RRLs' pulsation (and pulsation type). 
This method allows us to estimate the systemic (center of mass) radial velocity for each of the RRLs in our sample, correcting for the effect of radial pulsation regardless of the number of available epochs.  
Additionally, we utilize this approach and the observed (phase dependent) effective temperatures to compute the mean effective temperature of each RRL. 
The resulting corrected \teff\ significantly reduce the scatter of the \teff\ distribution of our RRLs, which enable us to recover and estimate empirically their Period-\teff\ relation. 
We note that the modeling of the variation of these spectroscopic properties has room for improvement, in particular for the short-period regime of RRab stars and for their phase of rapid contraction at the end of their pulsation cycle.  
Moreover, in this work we make no distinction between radial velocity curves obtained from different sets of absorption lines (as done by, e.g., \citealt{Sesar2012} and \citealt{Braga2021}) to correct for the Van Hoof effect. 
These improvements will be implemented with larger datasets at hand in future DESI data releases. 
We highlight that the methodology followed and the derived radial velocity and \teff\ curves, obtained from a single large and homogeneous database, can be applied to other homogeneous datasets to derive both systemic properties of RRLs from single-epoch and spectroscopic variation curves. 
The DESI Y1 RRL catalog, containing pulsation-corrected line-of-sight velocities and \teff, as well as the rest of the spectroscopic properties discussed in this work, will be made publicly available on the DESI data access website and via Zenodo (see Data Availability section)\footnote{Upon acceptance of publication of this manuscript.}.
The Stan and Python code required to derive the pulsational variation curves of RRLs based on single and multi-epoch spectra (for velocities and \teff) will be made available at \href{https://github.com/gmedinat/DESI-RRL-modeling}{https://github.com/gmedinat/DESI-RRL-modeling}.

We use the large number of RRLs observed by DESI in the Draco dwarf spheroidal galaxy to validate our estimated metallicities and velocities. 
In terms of [Fe/H], we find that our $\Delta$S results are the most similar with the expectations for Draco, considering its metallicity gradient.  
We find that estimating the RRLs systemic velocity using both our Bayesian method and the set of radial velocity curves from \citet{Braga2021} provide results consistent with the literature. 
Between these two methods, 
our Bayesian modeling results in a better agreement with the systemic velocity of Draco obtained from high-resolution spectroscopy. 
Moreover, the velocity corrections applied reduce the velocity dispersion of Draco RRLs to a value consistent with its known velocity dispersion and the typical uncertainty of our RRL corrections.  

To show the robustness of our data and as an example of usage of the catalog, 
we employ the available 7D information of our RRL sample (phase-space and [Fe/H]) to study the metallicity distribution of the halo. 
Our results suggest that, if the halo is composed of two populations that can be described by Gaussian distributions in chemodynamical space, the existence of a relatively metal-rich component (${\rm [Fe/H]}\sim-1.5$\,dex) is constrained to $<50$\,kpc in Galactocentric distance. 
We emphasize that assuming that the metallicities and velocities of the halo components follow Gaussian distributions is an oversimplification of the true nature of the halo \citep[see e.g.,][]{Leaman2012,Lancaster2019,Naidu2021,Iorio2021,Sanders2021}, and recognize this as a caveat of our analysis that leaves room 
for improvement in future studies. 
We associate the metal-rich component of the halo with the Gaia-Sausage-Enceladus (GSE) merger event, and confirm the slightly negative metallicity gradient detected in GSE RRLs by \citet{Liu2022}, although with a smaller decrease with radius ($-0.005$\,dex\,kpc$^{-1}$). 
Comparatively, we find a steeper metallicity gradient for the more metal-poor component (${\rm [Fe/H]}\sim-2.0$\,dex) of the halo ($-0.010$\,dex\,kpc$^{-1}$). 
Moreover, we find that the stars likely associated with the GSE are a dominant component in the inner halo, as $\sim$35\% have kinematics and [Fe/H] consistent with this merger. 
Indeed, the orbits of the stars in this component are predominantly radial, with a value of velocity anisotropy of $\beta\sim 0.94$, (in contrast to the anisotropy of the field sample, of $\beta \sim 0.39$). 
Our results indicate that  $\beta$ remains relatively constant as a function of distance  (between 10 and 50\,kpc) for the metal-rich and metal-poor components, while the sample as a whole displays a decrease of $\beta$ with distance ($\beta\sim0.8$ to below $\beta<0.5$ beyond 50\,kpc). 
We highlight that larger samples of distant RRLs with various chemical abundance information (at least $\alpha-$elements), as those that will be provided by future DESI data releases, will be key in providing clearer picture of the structure of the RRL metallicity and $\beta$ distribution in the outer halo (including the presence of multiple distinct components).

The DESI survey provides an ideal dataset to comprehensively map the Galactic outer halo for studies involving high-quality chemodynamical information of millions of stars.
The RR Lyrae catalog presented in this work, with its homogeneously derived spectroscopic properties, represents an opportunity to advance in this direction (and to study the physics of their pulsation), making the most of the multiple benefits of using these stars as tracers of the halo in the era of large spectroscopic surveys that just began.  
In this paper, we employed this rich dataset for a low-hanging-fruit scientific application, and the results of a detailed characterization of the sample presented in this work, with additional applications for the study of stellar pulsations, the accretion history of the Galaxy, and its mass (the latter being one of DESI's key projects), are underway and are presented in a series of separate papers \citep[][]{Medina2025b,Medina2025c}. 
Currently, the DESI survey is in its fourth out of five years of operations, and the analysis of RRLs in upcoming data releases has began. 
These extended datasets roughly duplicate the number of RRLs in DR1 (more than doubling the number of epochs for RRLs in DESI Y1), and will enable studies of halo RRLs with larger number statistics than those presented in this work. 
Thus, joined endeavors exploiting the synergies between RRL samples in large existing databases, like those from DESI or LAMOST \citep{Wang2024}, with those from large upcoming spectroscopic surveys, like WEAVE in the northern hemisphere (through its dedicated Galactic Archaeology surveys; \citealt{Dalton2012,Jin2024}) and 4MOST in the southern hemisphere (through its dedicated 4MOST {\it Gaia} RR Lyrae Survey, 4GRoundS; \citealt{deJong2019,Ibata2023_4grounds}) 
and the upcoming fourth {\it Gaia} data release will be pivotal for the development of Galactic and stellar astrophysics studies with catalog samples of sizes never seen before.

\section*{Data availability}
The data shown in all the figures of this manuscript, including the full RR Lyrae catalog, will be made available in Zenodo (\href{https://doi.org/10.5281/zenodo.15122042}{https://doi.org/10.5281/zenodo.15122042}). 
The full catalog will also be made available through the data access portal of DESI DR1 
(\href{https://data.desi.lbl.gov/doc/releases/dr1/#milky-way-survey-mws}{https://data.desi.lbl.gov/doc/releases/dr1/\#milky-way-survey-mws}).

\acknowledgments
We thank the anonymous referee for providing insightful comments and a thorough feedback, which greatly improved the content of this manuscript.
G.E.M. acknowledges valuable feedback from Joshua S. Speagle (University of Toronto) and Vittorio Braga (Istituto Nazionale di Astrofisica, INAF, Osservatorio Astronomico di Roma). 
G.E.M. and T.S.L. acknowledge financial support from Natural Sciences and Engineering Research Council of Canada (NSERC) through grant RGPIN-2022-04794.
G.E.M. acknowledges support from an Arts \& Science Postdoctoral Fellowship at the University of Toronto.
The Dunlap Institute is funded through an endowment established by the David Dunlap family and the University of Toronto.
S.K. acknowledges support from Science \& Technology Facilities Council (STFC) (grant ST/Y001001/1).

This material is based upon work supported by the U.S. Department of Energy (DOE), Office of Science, Office of High-Energy Physics, under Contract No. DE–AC02–05CH11231, and by the National Energy Research Scientific Computing Center, a DOE Office of Science User Facility under the same contract. Additional support for DESI was provided by the U.S. National Science Foundation (NSF), Division of Astronomical Sciences under Contract No. AST-0950945 to the NSF’s National Optical-Infrared Astronomy Research Laboratory; the Science and Technology Facilities Council of the United Kingdom; the Gordon and Betty Moore Foundation; the Heising-Simons Foundation; the French Alternative Energies and Atomic Energy Commission (CEA); the National Council of Humanities, Science and Technology of Mexico (CONAHCYT); the Ministry of Science, Innovation and Universities of Spain (MICIU/AEI/10.13039/501100011033), and by the DESI Member Institutions: \url{https://www.desi.lbl.gov/collaborating-institutions}. Any opinions, findings, and conclusions or recommendations expressed in this material are those of the author(s) and do not necessarily reflect the views of the U. S. National Science Foundation, the U. S. Department of Energy, or any of the listed funding agencies.

The authors are honored to be permitted to conduct scientific research on Iolkam Du’ag (Kitt Peak), a mountain with particular significance to the Tohono O’odham Nation.

For the purpose of open access, the author has applied a Creative Commons Attribution (CC BY) licence to any Author Accepted Manuscript version arising from this submission.

This research has made use of the SIMBAD database, operated at CDS, Strasbourg, France \citep{Simbad}.
This research has made use of NASA’s Astrophysics Data System Bibliographic Services.

{\it Software:} 
{\code{numpy} \citep{numpy}, 
\code{scipy} \citep{2020SciPy-NMeth},
\code{matplotlib} \citep{matplotlib}, 
\code{astropy} \citep{astropy,astropy:2018, astropy:2022},
\code{galpy} \citep[][]{Bovy2015}, 
\code{emcee} \citep[][]{emcee}.
}



\bibliography{references} 
\bibliographystyle{aasjournal}



\appendix

\end{document}

%% file: authors.tex
\author[0000-0003-0105-9576]{Gustavo~E.~Medina}
\affiliation{David A. Dunlap Department of Astronomy \& Astrophysics, University of Toronto, 50 St George Street, Toronto ON M5S 3H4, Canada}
\affiliation{Dunlap Institute for Astronomy \& Astrophysics, University of Toronto, 50 St George Street, Toronto, ON M5S 3H4, Canada}

\author[0000-0002-9110-6163]{Ting~S.~Li}
\affiliation{David A. Dunlap Department of Astronomy \& Astrophysics, University of Toronto, 50 St George Street, Toronto ON M5S 3H4, Canada}
\affiliation{Dunlap Institute for Astronomy \& Astrophysics, University of Toronto, 50 St George Street, Toronto, ON M5S 3H4, Canada}

\author[0000-0002-2644-135X]{Sergey~E.~Koposov}
\affiliation{Institute for Astronomy, University of Edinburgh, Royal Observatory, Blackford Hill, Edinburgh EH9 3HJ, UK}
\affiliation{Institute of Astronomy, University of Cambridge, Madingley Road, Cambridge CB3 0HA, UK}
\affiliation{Kavli Institute for Cosmology, University of Cambridge, Madingley Road, Cambridge CB3 0HA, UK}

\author[0000-0001-5805-5766]{A.~H.~Riley}
\affiliation{Institute for Computational Cosmology, Department of Physics, Durham University, South Road, Durham DH1 3LE, UK}

\author[0000-0002-0740-1507]{L. {Beraldo e Silva}}
\affiliation{Steward Observatory, University of Arizona, 933 N, Cherry Ave, Tucson, AZ 85721, USA}
\affiliation{Observatório Nacional, Rio de Janeiro - RJ, 20921-400, Brasil}

\author[0000-0002-6257-2341]{M.~Valluri}
\affiliation{Department of Astronomy, University of Michigan, Ann Arbor, MI 48109, USA}
\affiliation{University of Michigan, Ann Arbor, MI 48109, USA}

\author[0000-0002-5762-7571]{W.~Wang}
\affiliation{Department of Astronomy, School of Physics and Astronomy, and Shanghai Key Laboratory for Particle Physics and Cosmology, Shanghai Jiao Tong University, Shanghai 200240, People’s Republic of China}
\affiliation{State Key Laboratory of Dark Matter Physics, School of Physics and Astronomy,Shanghai Jiao Tong University, Shanghai 200240, China}
\affiliation{Shanghai Key Laboratory for Particle Physics and Cosmology, Shanghai 200240, China}

\author[0000-0002-5689-8791]{A.~Bystr\"om}
\affiliation{Institute for Astronomy, University of Edinburgh, Royal Observatory, Blackford Hill, Edinburgh EH9 3HJ, UK}

\author[0000-0001-9852-9954]{O.~Y.~Gnedin}
\affiliation{University of Michigan, Ann Arbor, MI 48109, USA}

\author[0000-0002-7667-0081]{R.~G.~Carlberg}
\affiliation{Department of Astronomy \& Astrophysics, University of Toronto, Toronto, ON M5S 3H4, Canada}

\author[0000-0003-0853-8887]{N.~Kizhuprakkat}
\affiliation{Institute of Astronomy and Department of Physics, National Tsing Hua
University, Hsinchu 30013, Taiwan}
\affiliation{Center for Informatics and Computation in Astronomy, National Tsing Hua
University, Hsinchu 30013, Taiwan}

\author{B.~A.~Weaver}
\affiliation{NSF NOIRLab, 950 N. Cherry Ave., Tucson, AZ 85719, USA}

\author{J.~Aguilar}
\affiliation{Lawrence Berkeley National Laboratory, 1 Cyclotron Road, Berkeley, CA 94720, USA}
\author[0000-0001-6098-7247]{S.~Ahlen}
\affiliation{Physics Dept., Boston University, 590 Commonwealth Avenue, Boston, MA 02215, USA}
\author[0000-0001-9712-0006]{D.~Bianchi}
\affiliation{Dipartimento di Fisica ``Aldo Pontremoli'', Universit\`a degli Studi di Milano, Via Celoria 16, I-20133 Milano, Italy}
\affiliation{INAF-Osservatorio Astronomico di Brera, Via Brera 28, 20122 Milano, Italy}
\author{D.~Brooks}
\affiliation{Department of Physics \& Astronomy, University College London, Gower Street, London, WC1E 6BT, UK}
\author{T.~Claybaugh}
\affiliation{Lawrence Berkeley National Laboratory, 1 Cyclotron Road, Berkeley, CA 94720, USA}
\author[0000-0001-8274-158X]{A.~P.~Cooper}
\affiliation{Institute of Astronomy and Department of Physics, National Tsing Hua University, 101 Kuang-Fu Rd. Sec. 2, Hsinchu 30013, Taiwan}
\author[0000-0002-1769-1640]{A.~de la Macorra}
\affiliation{Instituto de F\'{\i}sica, Universidad Nacional Aut\'{o}noma de M\'{e}xico,  Circuito de la Investigaci\'{o}n Cient\'{\i}fica, Ciudad Universitaria, Cd. de M\'{e}xico  C.~P.~04510,  M\'{e}xico}
\author[0000-0002-4928-4003]{A.~Dey}
\affiliation{NSF NOIRLab, 950 N. Cherry Ave., Tucson, AZ 85719, USA}
\author{P.~Doel}
\affiliation{Department of Physics \& Astronomy, University College London, Gower Street, London, WC1E 6BT, UK}
\author[0000-0002-3033-7312]{A.~Font-Ribera}
\affiliation{Institut de F\'{i}sica d’Altes Energies (IFAE), The Barcelona Institute of Science and Technology, Edifici Cn, Campus UAB, 08193, Bellaterra (Barcelona), Spain}
\author[0000-0002-2890-3725]{J.~E.~Forero-Romero}
\affiliation{Departamento de F\'isica, Universidad de los Andes, Cra. 1 No. 18A-10, Edificio Ip, CP 111711, Bogot\'a, Colombia}
\affiliation{Observatorio Astron\'omico, Universidad de los Andes, Cra. 1 No. 18A-10, Edificio H, CP 111711 Bogot\'a, Colombia}
\author{E.~Gaztañaga}
\affiliation{Institut d'Estudis Espacials de Catalunya (IEEC), c/ Esteve Terradas 1, Edifici RDIT, Campus PMT-UPC, 08860 Castelldefels, Spain}
\affiliation{Institute of Cosmology and Gravitation, University of Portsmouth, Dennis Sciama Building, Portsmouth, PO1 3FX, UK}
\affiliation{Institute of Space Sciences, ICE-CSIC, Campus UAB, Carrer de Can Magrans s/n, 08913 Bellaterra, Barcelona, Spain}
\author[0000-0003-3142-233X]{S.~Gontcho A Gontcho}
\affiliation{Lawrence Berkeley National Laboratory, 1 Cyclotron Road, Berkeley, CA 94720, USA}
\author{G.~Gutierrez}
\affiliation{Fermi National Accelerator Laboratory, PO Box 500, Batavia, IL 60510, USA}
\author[0000-0001-9822-6793]{J.~Guy}
\affiliation{Lawrence Berkeley National Laboratory, 1 Cyclotron Road, Berkeley, CA 94720, USA}
\author[0000-0002-6550-2023]{K.~Honscheid}
\affiliation{Center for Cosmology and AstroParticle Physics, The Ohio State University, 191 West Woodruff Avenue, Columbus, OH 43210, USA}
\affiliation{Department of Physics, The Ohio State University, 191 West Woodruff Avenue, Columbus, OH 43210, USA}
\affiliation{The Ohio State University, Columbus, 43210 OH, USA}
\author[0000-0002-6024-466X]{M.~Ishak}
\affiliation{Department of Physics, The University of Texas at Dallas, 800 W. Campbell Rd., Richardson, TX 75080, USA}
\author[0000-0003-3510-7134]{T.~Kisner}
\affiliation{Lawrence Berkeley National Laboratory, 1 Cyclotron Road, Berkeley, CA 94720, USA}
\author[0000-0003-1838-8528]{M.~Landriau}
\affiliation{Lawrence Berkeley National Laboratory, 1 Cyclotron Road, Berkeley, CA 94720, USA}
\author[0000-0001-7178-8868]{L.~Le~Guillou}
\affiliation{Sorbonne Universit\'{e}, CNRS/IN2P3, Laboratoire de Physique Nucl\'{e}aire et de Hautes Energies (LPNHE), FR-75005 Paris, France}
\author[0000-0002-1125-7384]{A.~Meisner}
\affiliation{NSF NOIRLab, 950 N. Cherry Ave., Tucson, AZ 85719, USA}
\author{R.~Miquel}
\affiliation{Instituci\'{o} Catalana de Recerca i Estudis Avan\c{c}ats, Passeig de Llu\'{\i}s Companys, 23, 08010 Barcelona, Spain}
\affiliation{Institut de F\'{i}sica d’Altes Energies (IFAE), The Barcelona Institute of Science and Technology, Edifici Cn, Campus UAB, 08193, Bellaterra (Barcelona), Spain}
\author{A.~D.~Myers}
\affiliation{Department of Physics \& Astronomy, University  of Wyoming, 1000 E. University, Dept.~3905, Laramie, WY 82071, USA}
\author[0000-0001-9070-3102]{S.~Nadathur}
\affiliation{Institute of Cosmology and Gravitation, University of Portsmouth, Dennis Sciama Building, Portsmouth, PO1 3FX, UK}
\author{C.~Poppett}
\affiliation{Lawrence Berkeley National Laboratory, 1 Cyclotron Road, Berkeley, CA 94720, USA}
\affiliation{Space Sciences Laboratory, University of California, Berkeley, 7 Gauss Way, Berkeley, CA  94720, USA}
\affiliation{University of California, Berkeley, 110 Sproul Hall \#5800 Berkeley, CA 94720, USA}
\author[0000-0001-7145-8674]{F.~Prada}
\affiliation{Instituto de Astrof\'{i}sica de Andaluc\'{i}a (CSIC), Glorieta de la Astronom\'{i}a, s/n, E-18008 Granada, Spain}
\author[0000-0001-6979-0125]{I.~P\'erez-R\`afols}
\affiliation{Departament de F\'isica, EEBE, Universitat Polit\`ecnica de Catalunya, c/Eduard Maristany 10, 08930 Barcelona, Spain}
\author{G.~Rossi}
\affiliation{Department of Physics and Astronomy, Sejong University, 209 Neungdong-ro, Gwangjin-gu, Seoul 05006, Republic of Korea}
\author[0000-0002-9646-8198]{E.~Sanchez}
\affiliation{CIEMAT, Avenida Complutense 40, E-28040 Madrid, Spain}
\author[0000-0002-6588-3508]{H.~Seo}
\affiliation{Department of Physics \& Astronomy, Ohio University, 139 University Terrace, Athens, OH 45701, USA}
\author{D.~Sprayberry}
\affiliation{NSF NOIRLab, 950 N. Cherry Ave., Tucson, AZ 85719, USA}
\author[0000-0003-1704-0781]{G.~Tarl\'{e}}
\affiliation{University of Michigan, 500 S. State Street, Ann Arbor, MI 48109, USA}
\author[0000-0003-2229-011X]{R.~H.~Wechsler}
\affiliation{Kavli Institute for Particle Astrophysics and Cosmology, Stanford University, Menlo Park, CA 94305, USA}
\affiliation{Physics Department, Stanford University, Stanford, CA 93405, USA}
\affiliation{SLAC National Accelerator Laboratory, 2575 Sand Hill Road, Menlo Park, CA 94025, USA}
\author[0000-0001-5381-4372]{R.~Zhou}
\affiliation{Lawrence Berkeley National Laboratory, 1 Cyclotron Road, Berkeley, CA 94720, USA}
\author[0000-0002-6684-3997]{H.~Zou}
\affiliation{National Astronomical Observatories, Chinese Academy of Sciences, A20 Datun Road, Chaoyang District, Beijing, 100101, P.~R.~China}

\correspondingauthor{Gustavo E. Medina}
\email{gustavo.medina@utoronto.ca}